\documentclass[12pt]{report}
\usepackage{amssymb}
\usepackage{amsmath}

\usepackage{graphicx}
\begin{document}
\voffset = -2.9cm
\hoffset = -1.3cm
\def\itm{\newline \makebox[8mm]{}}
\def\ls{\makebox[8mm]{}}
\def\fra#1#2{\frac{#1}{#2}}
\def\fr#1#2{#1/#2}
\def\frl#1#2{\mbox{\large $\frac{#1}{\rule[-0mm]{0mm}{3.15mm} #2}$}}
\def\frn#1#2{\mbox{\normalsize $\frac{#1}{\rule[-0mm]{0mm}{3.15mm} #2}$}}
\def\frm#1#2{\mbox{\normalsize $\frac{#1}{\rule[-0mm]{0mm}{2.85mm} #2}$}}
\def\frn#1#2{\mbox{\normalsize $\frac{#1}{\rule[-0mm]{0mm}{3.15mm} #2}$}}
\def\hs#1{\mbox{\hspace{#1}}}
\def\b{\begin{equation}}
\def\e{\end{equation}}
\def\arccot{\mbox{arccot}}
\vspace*{6mm}
\makebox[\textwidth][c]
{\large \bf{FRW Universe Models in Conformally Flat Spacetime Coordinates}}
\vspace*{-1.0mm} \newline
{\large \bf{II: Universe models with negative
and vanishing spatial \mbox{curvature}}}
\vspace{4mm} \newline
\makebox[\textwidth][c]
{\normalsize \O yvind Gr\o n$^{*}$ and Steinar Johannesen$^*$}
\vspace{1mm} \newline
\makebox[\textwidth][c]
{\scriptsize $*$ Oslo University College, Faculty of Engineering,
P.O.Box 4 St.Olavs Plass, N-0130 Oslo, Norway}
%
\vspace{6mm} \newline
{\bf \small Abstract}
{\small We deduce general expressions for the line element of universe
models with negative and vanishing spatial curvature described by
conformally flat spacetime coordinates. The empty Milne universe model
and models with dust, radiation and vacuum energy are exhibited.
Discussing the existence of particle horizons we show that there is
continual creation of space,matter and energy when conformal time is used
in Friedmann-Robertson-Walker models with negative spatial curvature.}
%
%
\vspace{10mm} \newline
{\bf 1. Introduction.}
\vspace{3mm} \newline
In the first paper [1] of this series we have developed a general formalism
for describing Friedmann-Robertson-Walker (FRW) universe models using
conformally flat spacetime coordinates (CFS) [2-9]. The line element is
written as a conformal factor times the Minkowski line element,
\begin{equation} \label{e_401}
ds^2 = A(T,R)^2 (-dT^2 + dR^2 + R^2 d \Omega^2)
\mbox{ ,}
\end{equation}
in such coordinates. Using standard cosmic coordinates the line element
has the form
\begin{equation} \label{e_402}
ds^2 = - dt^2 + a(t)^2 [\hs{0.6mm} d \chi^2 + S_k(\chi)^2 d \Omega^2]
\mbox{ ,}
\end{equation}
where
\begin{equation} \label{e_403}
S_k(\chi) =
\left\{ \begin{array}{lclcl}
\sin \chi  & \mbox{for} & k = 1  & , & 0 < \chi < \pi    \\
\chi       & \mbox{for} & k = 0  & , & 0 < \chi < \infty \\
\sinh \chi & \mbox{for} & k = -1 & , & 0 < \chi < \infty
\end{array} \right.
\mbox{ .}
\end{equation}
Introducing parametric time
\begin{equation} \label{e_404}
\eta = \int_{t_1}^t \frac{dt}{a(t)}
\end{equation}
where $t_1$ is an arbitrary constant, the line element takes the form
\begin{equation} \label{e_405}
ds^2 = a(\eta)^2 \hs{0.6mm} [\hs{0.3mm} - d{\eta}^2 + d \chi^2
+ S_k(\chi)^2 d \Omega^2]
\mbox{ .}
\end{equation}
\itm In reference [1] it is shown that the transformation
\begin{equation} \label{e_406}
T = \frl{1}{2} \hs{0.6mm} [\hs{0.3mm} f(\eta + \chi) + f(\eta - \chi)
\hs{0.3mm}]
\mbox{\hspace{2mm} , \hspace{3mm}}
R = \frl{1}{2} \hs{0.6mm} [\hs{0.3mm} f(\eta + \chi) - f(\eta - \chi)
\hs{0.3mm}]
\end{equation}
with
\begin{equation} \label{e_407}
f(x) =  c \left[ b + I_k \left( \frl{x - a}{2} \right) \right]^{-1} + d
\mbox{ ,}
\end{equation}
where $a$, $b$, $c$, $d$ are arbitrary constants and
\begin{equation} \label{e_408}
I_k(x) =
\left\{ \begin{array}{lcl}
\cot x  & \mbox{for} & k = 1  \\
1 / x   & \mbox{for} & k = 0  \\
\coth x & \mbox{for} & k = -1
\end{array} \right.
\mbox{ ,}
\end{equation}
leads from \eqref{e_405} to \eqref{e_401} with
\begin{equation} \label{e_409}
A(T,R) = \frl{a(\eta(T,R)) \hs{0.5mm} S_k(\chi(T,R))}{|R|}
\mbox{ .}
\end{equation}
We shall require that the CFS time coordinate $T$ is an increasing
function of the cosmic time $t$, and hence of the parametric time $\eta$,
for every value of $\chi$. Then it follows from equation \eqref{e_406}
that $f$ must be an increasing function. Hence equation (54) in
reference [1] implies that $c > 0$.
\itm In the present paper this formalism shall be applied to
FRW universe models with $k = -1$ and $k = 0$.
%
%
%
\vspace{5mm} \newline
{\bf 2. CFS coordinates in negatively curved universe models}
\vspace{3mm} \newline
M.J.Chodorowski [10] has recently presented an interesting
discussion of the concept \textit{space} in a cosmological context. He
has deduced the form of the Robertson-Walker line element for the case
of a negatively curved space as expressed in terms of conformal time,
and pointed out that space is a coordinate dependent concept.
\itm In this case $S_{-1} (\chi) = \sinh \chi$. The CFS coordinates
used by Chodorowski may be obtained by choosing the values $a = 0$,
$b = -1$, $c = 2 T_i$ and $d = T_i$ in equation \eqref{e_407}, where $T_i$
is the conformal time corresponding to $t = \eta = 0$ at $\chi = 0$.
This gives the generating function
\begin{equation} \label{e_203}
f(x) = T_i e^x
\mbox{ .}
\end{equation}
Hence the transformation \eqref{e_406} between the coordinates
$(\eta,\chi)$ and the conformal coordinates $(T,R)$ is
\begin{equation} \label{e_33}
T = T_i e^{\eta} \cosh \chi
\mbox{\hspace{2mm} , \hspace{3mm}}
R = T_i e^{\eta} \sinh \chi \; ,
\end{equation}
The inverse transformation is
\begin{equation} \label{e_34}
T_i e^{\eta} = \sqrt{T^{2} - R^{2}}
\mbox{\hspace{2mm} , \hspace{3mm}}
\tanh \chi = \frl{R}{T}
\mbox{ ,}
\end{equation}
where $R > 0$ and $T^2 > R^2$. At $\chi = R = 0$, the clocks
showing conformal time goe exponentially faster than the clocks
showing parametric time.
Equations \eqref{e_403}, \eqref{e_33} and \eqref{e_34} lead to
\begin{equation} \label{e_173}
S_{-1} (\chi (T,R)) = \frl{R^2}{T^2 - R^2}
\mbox{ .}
\end{equation}
It follows from equations \eqref{e_173} and \eqref{e_409} that
the line element for a universe model with negative spatial curvature,
as expressed in terms of the conformal coordinates $(T,R)$, takes the
conformally flat form
\begin{equation} \label{e_32}
ds^{2} = \frac{a(\eta (T,R))^{2}}
{T^{2} - R^{2}} \hs{1.0mm} ds_{M}^{2}
\mbox{ .}
\end{equation}
\vspace{-2mm}
\itm In the $(T,R)$-system, each reference particle with
$\chi = \mbox{constant}$ in the cosmic coordinate system has
a constant recession velocity
\begin{equation} \label{e_65}
V = \frl{R}{T} = \tanh \chi
\end{equation}
which is less than $1$. According to this equation $\chi$ is the
rapidity of a reference particle with radial coordinate $\chi$.
The question of superluminal expansion of space using the line
element \eqref{e_32} has been discussed by Lewis et al.[11].
\itm Figure 1 shows the cosmic coordinate system $(\eta,\chi)$ in a Minkowski
diagram referring to the conformal coordinate system of the observer at
$\chi = 0$. It follows from equations \eqref{e_34} that the world lines of
the reference particles with $\chi = \mbox{constant}$ are straight lines,
and the curves of the cosmic space $\eta = \mbox{constant}$ are hyperbolae
with centre at the origin as show in the diagram in \mbox{Figure 1.}
\vspace*{5mm} \newline
\begin{picture}(50,192)(-96,-182)
\qbezier(125.7955, -87.7955)(127.3448, -87.7955)(128.8996, -87.6648)
\qbezier(128.8996, -87.6648)(130.4544, -87.5342)(132.0258, -87.2720)
\qbezier(132.0258, -87.2720)(133.5971, -87.0099)(135.1961, -86.6143)
\qbezier(135.1961, -86.6143)(136.7952, -86.2187)(138.4332, -85.6869)
\qbezier(138.4332, -85.6869)(140.0713, -85.1551)(141.7600, -84.4833)
\qbezier(141.7600, -84.4833)(143.4487, -83.8115)(145.2000, -82.9949)
\qbezier(145.2000, -82.9949)(146.9513, -82.1784)(148.7777, -81.2113)
\qbezier(148.7777, -81.2113)(150.6041, -80.2442)(152.5185, -79.1197)
\qbezier(152.5185, -79.1197)(154.4329, -77.9952)(156.4489, -76.7052)
\qbezier(156.4489, -76.7052)(158.4649, -75.4153)(160.5968, -73.9509)
\qbezier(160.5968, -73.9509)(162.7288, -72.4864)(164.9917, -70.8370)
\qbezier(164.9917, -70.8370)(167.2546, -69.1876)(169.6647, -67.3416)
\qbezier(169.6647, -67.3416)(172.0747, -65.4956)(174.6489, -63.4398)
\qbezier(174.6489, -63.4398)(177.2231, -61.3840)(179.9798, -59.1039)
\qbezier(179.9798, -59.1039)(182.7365, -56.8238)(185.6952, -54.3032)
\qbezier(185.6952, -54.3032)(188.6539, -51.7826)(191.8356, -49.0036)
\qbezier(191.8356, -49.0036)(195.0173, -46.2246)(198.4446, -43.1675)
\qbezier(198.4446, -43.1675)(201.8718, -40.1104)(205.5690, -36.7535)
\qbezier(205.5690, -36.7535)(209.2662, -33.3966)(213.2595, -29.7161)
\put(125.7955, -124.6136){\line(1, 1){ 95.1136}}
\put(125.7955, -44.8409){\line(1, -1){ 39.8864}}
\qbezier(125.7955, -124.6136)(156.2436, -77.6823)(186.6918, -30.7509)
\put( 88.9773, -124.6136){\vector(1, 0){128.8636}}
\put(125.7955, -161.4318){\vector(0, 1){144.2045}}
\put(227.0455, -129.2159){\makebox(0,0)[]{\footnotesize{$R$}}}
\put(118.1250, -15.6932){\makebox(0,0)[]{\footnotesize{$T$}}}
\put(118.1250, -46.3750){\makebox(0,0)[]{\footnotesize{$T_0$}}}
\put(119.0455, -89.3295){\makebox(0,0)[]{\footnotesize{$T_i$}}}
\put(219.3750, -20.2955){\makebox(0,0)[]{\footnotesize{$t = 0$}}}
\put(185.6250, -20.2955){\makebox(0,0)[]{\footnotesize{$\chi = {\chi}_H$}}}
\put(135.0000, -132.2841){\makebox(0,0)[]{\footnotesize{$O$}}}
\put(135.0000, -40.2386){\makebox(0,0)[]{\footnotesize{$P$}}}
\put(156.4773, -64.7841){\makebox(0,0)[]{\footnotesize{$H$}}}
\end{picture}
\vspace{0mm} \newline
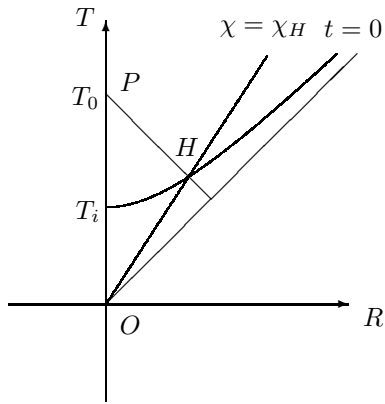
{\footnotesize \sf Figure 1. Minkowski diagram for universe models with
negative spatial curvature with reference to the conformal coordinates
$(T,R)$. Here the line $OH$ is the world line of a reference particle with
$\chi = \mbox{constant}$. The line $PH$ represents the backwards light cone.
The hyperbola represents the cosmic space at cosmic time $t=0$.}
\vspace{0mm} \newline
\itm The hyperbola shown in Figure 1 represents the Big Bang singularity
of the universe, i.e. the spacetime boundary of the universe. Conformal
space is represented by a horizontal line in this figure from the time
axis to the hyperbola. Hence the conformal space of the universe has
finite extension even if the cosmic space has infinite extension.
\itm Equation \eqref{e_33} shows that the velocity of the
$(T,R)$-system relative to the $(\eta,\chi)$-system, i.e. of a particle
with $R = \mbox{constant}$, is given by
\begin{equation} \label{e_106}
\frl{d \chi}{d \eta} = - \tanh \chi
\mbox{ .}
\end{equation}
Hence the $(T,R)$-system contracts relative to the $(\eta,\chi)$-system,
i.e. relative to the Hubble flow.
In a $(\eta,\chi)$-diagram the world lines of the reference particles
with $R = R_1$ are given by
\begin{equation} \label{e_108}
\sinh \chi = \frl{R_1}{T_i} \hs{1.0mm} e^{- \eta}
\mbox{ ,}
\end{equation}
and the simultaneity curves $T = T_1$ of the conformal space by
\begin{equation} \label{e_109}
\cosh \chi = \frl{T_1}{T_i} \hs{1.0mm} e^{- \eta}
\mbox{ .}
\end{equation}
These curves are shown in Figure 2.
\vspace*{5mm} \newline
\begin{picture}(50,192)(-96,-182)
\qbezier(160.7108, -124.6136)(158.6593, -122.0912)(156.8792, -119.7691)
\qbezier(156.8792, -119.7691)(154.9654, -117.2728)(153.2886, -114.9246)
\qbezier(153.2886, -114.9246)(151.5253, -112.4552)(149.9650, -110.0801)
\qbezier(149.9650, -110.0801)(148.3600, -107.6371)(146.9261, -105.2356)
\qbezier(146.9261, -105.2356)(145.4821, -102.8173)(144.1803, -100.3911)
\qbezier(144.1803, -100.3911)(142.8946, -97.9950)(141.7260, -95.5467)
\qbezier(141.7260, -95.5467)(140.5915, -93.1696)(139.5531, -90.7022)
\qbezier(139.5531, -90.7022)(138.5593, -88.3409)(137.6445, -85.8577)
\qbezier(137.6445, -85.8577)(136.7794, -83.5092)(135.9792, -81.0132)
\qbezier(135.9792, -81.0132)(135.2296, -78.6748)(134.5337, -76.1687)
\qbezier(134.5337, -76.1687)(133.8866, -73.8381)(133.2841, -71.3242)
\qbezier(133.2841, -71.3242)(132.7270, -68.9995)(132.2072, -66.4797)
\qbezier(132.2072, -66.4797)(131.7287, -64.1595)(131.2814, -61.6352)
\qbezier(131.2814, -61.6352)(130.8709, -59.3184)(130.4868, -56.7907)
\qbezier(130.4868, -56.7907)(130.1351, -54.4763)(129.8058, -51.9462)
\qbezier(129.8058, -51.9462)(129.5048, -49.6337)(129.2227, -47.1017)
\qbezier(129.2227, -47.1017)(128.9652, -44.7905)(128.7237, -42.2572)
\qbezier(128.7237, -42.2572)(128.5036, -39.9470)(128.2971, -37.4127)
\qbezier(128.2971, -37.4127)(128.1089, -35.1033)(127.9323, -32.5682)
\qbezier(183.4731, -124.6136)(180.9771, -121.9979)(178.8847, -119.7691)
\qbezier(178.8847, -119.7691)(176.4398, -117.1648)(174.3816, -114.9246)
\qbezier(174.3816, -114.9246)(172.0019, -112.3346)(169.9878, -110.0801)
\qbezier(169.9878, -110.0801)(167.6897, -107.5076)(165.7319, -105.2356)
\qbezier(165.7319, -105.2356)(163.5330, -102.6837)(161.6455, -100.3911)
\qbezier(161.6455, -100.3911)(159.5638, -97.8628)(157.7614, -95.5467)
\qbezier(157.7614, -95.5467)(155.8139, -93.0440)(154.1115, -90.7022)
\qbezier(154.1115, -90.7022)(152.3117, -88.2263)(150.7230, -85.8577)
\qbezier(150.7230, -85.8577)(149.0803, -83.4085)(147.6159, -81.0132)
\qbezier(147.6159, -81.0132)(146.1341, -78.5892)(144.8008, -76.1687)
\qbezier(144.8008, -76.1687)(143.4783, -73.7676)(142.2785, -71.3242)
\qbezier(142.2785, -71.3242)(141.1092, -68.9429)(140.0405, -66.4797)
\qbezier(140.0405, -66.4797)(139.0147, -64.1150)(138.0715, -61.6352)
\qbezier(138.0715, -61.6352)(137.1772, -59.2840)(136.3509, -56.7907)
\qbezier(136.3509, -56.7907)(135.5753, -54.4502)(134.8558, -51.9462)
\qbezier(134.8558, -51.9462)(134.1857, -49.6140)(133.5622, -47.1017)
\qbezier(133.5622, -47.1017)(132.9849, -44.7758)(132.4466, -42.2572)
\qbezier(132.4466, -42.2572)(131.9505, -39.9361)(131.4870, -37.4127)
\qbezier(131.4870, -37.4127)(131.0614, -35.0952)(130.6632, -32.5682)
\qbezier(169.5298, -124.6136)(168.8013, -123.9647)(168.0926, -123.3406)
\qbezier(168.0926, -123.3406)(167.3552, -122.6912)(166.6379, -122.0675)
\qbezier(166.6379, -122.0675)(165.8906, -121.4177)(165.1637, -120.7945)
\qbezier(165.1637, -120.7945)(164.4050, -120.1441)(163.6671, -119.5214)
\qbezier(163.6671, -119.5214)(162.8956, -118.8704)(162.1452, -118.2484)
\qbezier(162.1452, -118.2484)(161.3590, -117.5967)(160.5942, -116.9753)
\qbezier(160.5942, -116.9753)(159.7910, -116.3228)(159.0097, -115.7023)
\qbezier(159.0097, -115.7023)(158.1867, -115.0487)(157.3861, -114.4292)
\qbezier(157.3861, -114.4292)(156.5398, -113.7744)(155.7166, -113.1562)
\qbezier(155.7166, -113.1562)(154.8426, -112.4999)(153.9925, -111.8831)
\qbezier(153.9925, -111.8831)(153.0851, -111.2249)(152.2025, -110.6101)
\qbezier(152.2025, -110.6101)(151.2542, -109.9495)(150.3317, -109.3370)
\qbezier(150.3317, -109.3370)(149.3318, -108.6732)(148.3591, -108.0640)
\qbezier(148.3591, -108.0640)(147.2923, -107.3958)(146.2546, -106.7910)
\qbezier(146.2546, -106.7910)(145.0973, -106.1164)(143.9716, -105.5179)
\qbezier(143.9716, -105.5179)(142.6836, -104.8332)(141.4308, -104.2449)
\qbezier(141.4308, -104.2449)(139.9336, -103.5417)(138.4772, -102.9718)
\qbezier(138.4772, -102.9718)(136.5661, -102.2240)(134.7071, -101.6988)
\qbezier(134.7071, -101.6988)(130.3842, -100.4773)(126.2505, -100.4257)
\qbezier(194.1912, -124.6136)(192.8751, -123.3277)(191.6285, -122.1148)
\qbezier(191.6285, -122.1148)(190.3064, -120.8283)(189.0541, -119.6159)
\qbezier(189.0541, -119.6159)(187.7247, -118.3288)(186.4655, -117.1170)
\qbezier(186.4655, -117.1170)(185.1274, -115.8292)(183.8600, -114.6181)
\qbezier(183.8600, -114.6181)(182.5115, -113.3295)(181.2342, -112.1193)
\qbezier(181.2342, -112.1193)(179.8731, -110.8296)(178.5839, -109.6204)
\qbezier(178.5839, -109.6204)(177.2074, -108.3294)(175.9037, -107.1215)
\qbezier(175.9037, -107.1215)(174.5086, -105.8289)(173.1872, -104.6226)
\qbezier(173.1872, -104.6226)(171.7691, -103.3280)(170.4259, -102.1237)
\qbezier(170.4259, -102.1237)(168.9792, -100.8266)(167.6090, -99.6249)
\qbezier(167.6090, -99.6249)(166.1263, -98.3245)(164.7220, -97.1260)
\qbezier(164.7220, -97.1260)(163.1934, -95.8214)(161.7455, -94.6271)
\qbezier(161.7455, -94.6271)(160.1567, -93.3167)(158.6520, -92.1282)
\qbezier(158.6520, -92.1282)(156.9825, -90.8097)(155.4012, -89.6294)
\qbezier(155.4012, -89.6294)(153.6188, -88.2988)(151.9305, -87.1305)
\qbezier(151.9305, -87.1305)(149.9798, -85.7806)(148.1321, -84.6316)
\qbezier(148.1321, -84.6316)(145.9038, -83.2460)(143.7932, -82.1327)
\qbezier(143.7932, -82.1327)(141.0009, -80.6599)(138.3560, -79.6339)
\qbezier(138.3560, -79.6339)(132.1030, -77.2080)(126.2505, -77.1350)
\qbezier(215.0606, -124.6136)(213.1867, -122.7508)(211.4531, -121.0302)
\qbezier(211.4531, -121.0302)(209.5758, -119.1669)(207.8391, -117.4467)
\qbezier(207.8391, -117.4467)(205.9576, -115.5830)(204.2170, -113.8632)
\qbezier(204.2170, -113.8632)(202.3301, -111.9989)(200.5845, -110.2798)
\qbezier(200.5845, -110.2798)(198.6907, -108.4147)(196.9387, -106.6963)
\qbezier(196.9387, -106.6963)(195.0361, -104.8302)(193.2760, -103.1129)
\qbezier(193.2760, -103.1129)(191.3621, -101.2455)(189.5916, -99.5294)
\qbezier(189.5916, -99.5294)(187.6631, -97.6603)(185.8791, -95.9459)
\qbezier(185.8791, -95.9459)(183.9318, -94.0746)(182.1303, -92.3625)
\qbezier(182.1303, -92.3625)(180.1583, -90.4883)(178.3340, -88.7790)
\qbezier(178.3340, -88.7790)(176.3295, -86.9009)(174.4751, -85.1955)
\qbezier(174.4751, -85.1955)(172.4271, -83.3121)(170.5324, -81.6121)
\qbezier(170.5324, -81.6121)(168.4251, -79.7212)(166.4756, -78.0286)
\qbezier(166.4756, -78.0286)(164.2855, -76.1270)(162.2593, -74.4451)
\qbezier(162.2593, -74.4451)(159.9489, -72.5273)(157.8115, -70.8617)
\qbezier(157.8115, -70.8617)(155.3160, -68.9170)(153.0073, -67.2782)
\qbezier(153.0073, -67.2782)(150.1972, -65.2834)(147.5973, -63.6947)
\qbezier(147.5973, -63.6947)(144.1299, -61.5761)(140.9220, -60.1113)
\qbezier(140.9220, -60.1113)(133.2656, -56.6154)(126.2505, -56.5278)
\put( 88.9773, -124.6136){\vector(1, 0){144.2045}}
\put(125.7955, -161.4318){\vector(0, 1){144.2045}}
\put(242.3864, -129.2159){\makebox(0,0)[]{\footnotesize{$\chi$}}}
\put(116.5909, -15.6932){\makebox(0,0)[]{\footnotesize{$\eta$}}}
\put(159.5455, -37.1705){\makebox(0,0)[]{\footnotesize{$R = \mbox{const}$}}}
\put(220.9091, -98.5341){\makebox(0,0)[]{\footnotesize{$T = \mbox{const}$}}}
\end{picture}
\vspace{0mm} \newline
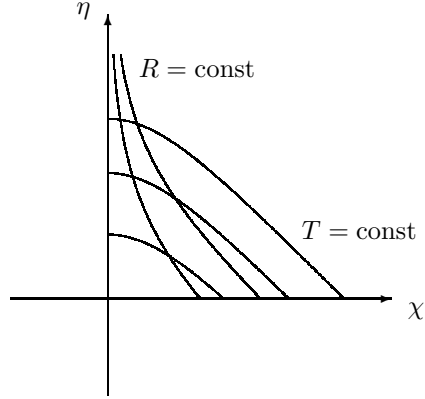
{\footnotesize \sf Figure 2. Minkowski diagram for universe models with
negative spatial curvature with reference to the $(\eta,\chi)$-coordinate
system. The diagram shows world lines $R = \mbox{constant}$ and
simultaneity curves $T = \mbox{constant} \,$. We see that the
$(T,R)$-system contracts relative to the $(\eta,\chi)$-system,
i.e. relative to the Hubble flow.}
%
%
\vspace{5mm} \newline
{\bf 3. The Milne universe model}
\vspace{3mm} \newline
The Milne universe is a model of an empty universe with $\Omega_0 = 0$.
This model has negative spatial curvature. The scale factor is
\begin{equation} \label{e_35}
a(t) = t
\mbox{ .}
\end{equation}
The normalisation condition $a(t_0) = 1$ is fullfilled since the unit of
time is chosen to be equal to the present age $t_0$ of the Milne universe.
Equation \eqref{e_404} then gives
\begin{equation} \label{e_36}
e^{\eta} = t
\end{equation}
with
$\eta \in \hs{-1.5mm} < \hs{-1.4mm} \mbox{$-\infty$},\infty \hs{-0.5mm} >$
when $t \in \hs{-1.5mm} < \hs{-0.5mm}  0,\infty \hs{-0.5mm} >$.
\itm For the Milne universe the transformation between the cosmic
coordinates and the conformal coordinates is
\begin{equation} \label{e_1}
T = t \cosh \chi
\mbox{\hspace{2mm} , \hspace{3mm}}
R = t \sinh \chi \; ,
\end{equation}
with inverse transformation
\begin{equation} \label{e_2}
t = \sqrt{T^{2} - R^{2}}
\mbox{\hspace{2mm} , \hspace{3mm}}
\tanh \chi = \frl{R}{T}
\mbox{ .}
\end{equation}
In this case $T = t$ at $\chi = 0$, i.e. the conformal clock at
$\chi = 0$ goes at the same rate as the cosmic clocks. However, while the
cosmic clocks go at the same rate at all positions, the conformal clocks
go faster for larger values of $\chi$. Inserting equation \eqref{e_2}
into equation \eqref{e_32} gives
\begin{equation} \label{e_38}
ds^{2} = ds_{M}^{2}
\mbox{ ,}
\end{equation}
which shows that in the case of the Milne universe model the
conformal coordinates are the same as the coordinates of flat
spacetime in a static reference frame in which the line element
takes the Minkowski form. Hence the Hubble parameter $H_R$ in the
conformal coordinate system vanishes in this case.
\itm According to equation \eqref{e_65} the conformal coordinate system
is comoving with an observer at $\chi = 0$. In the Milne universe the
conformal space is then identical to the private space of this observer.
Furthermore the public space at $t = 0$ is not represented by a hyperbola
in Figure 1, but by the light cone $R = T$. The private space still has
a finite extension at the time $T$. However, the public space at $t = t_1$
is represented by the hyperbola $T^2 - R^2 = t_1^2$ in the $(T,R)$-diagram,
and hence has an infinite extension.
%
%
\vspace{5mm} \newline
{\bf 4. Particle horizon in negatively curved universe models using
conformal time.}
\vspace{-2mm} \newline
We here consider universe models with negative spatial curvature,
and where the parametric time $\eta \rightarrow 0$ when the cosmic time
$t \rightarrow 0$. As noted above, the universe as described in conformally
flat coordinates extends only out to the hyperbola in Figure 1, and not
out to the light cone.
\itm As an interesting application of the conformal coordinates, we will
discuss the existence of a particle horizon using these coordinates.
In Figure 1 we have drawn a Minkowski diagram where the hyperbola
represents the simultaneity space at constant cosmic time $t = 0$
measured by clocks moving along the straight world lines from the
origin.
\itm The standard radial coordinate of the particle horizon is defined by
\begin{equation}\label{e_410}
\chi_H = \int_{t_i}^t \frac{dt}{a(t)}
\mbox{ .}
\end{equation}
From equations \eqref{e_410} and \eqref{e_404} it follows that
\begin{equation}\label{e_411}
\chi_H = \eta
\mbox{ .}
\end{equation}
We shall now show how this equation can be deduced directly from
the figure.
\itm Consider an observer at the point $P$ having coordinates $(0,T_0)$.
The line $PH$ given by $R = T_0 - T$ represents the backwards light cone
of this observer. A straight line $\chi = \mbox{constant}$ through the
origin represents a spherical surface in space at different times.
The particle horizon of this observer is the spherical surface within
which he may receive information emitted after cosmic time $t = 0$.
The space at $t = 0$ is represented by the hyperbola given
by $T^2 - R^2 = T_i^2$. The position of the horizon is given by
the intersection between $PH$ and the hyperbola which has coordinates
\begin{equation} \label{e_60}
T_{H}=\frac{T_0^2 + T_i^2}{2 T_0} \hs{4mm} , \hs{4mm}
R_{H}=\frac{T_0^2 - T_i^2}{2 T_0}
\mbox{ .}
\end{equation}
The time coordinate $\eta_P$ of the observer is found by inserting
$R = 0$ in the first of the equations \eqref{e_34} giving
\begin{equation} \label{e_61}
e^{\eta_P} = \frac{T_0}{T_i}
\mbox{ .}
\end{equation}
The radial coordinate of the horizon $\chi_H$
is found from the transformation \eqref{e_33} which leads to
\begin{equation} \label{e_62}
e^{\chi_H} = \cosh \chi_H + \sinh \chi_H
= \frac{T + R}{T_i e^{\eta_H}} = \frac{T + R}{T_i}
\mbox{ ,}
\end{equation}
since the time coordinate $\eta_H = 0$ at the point $H$ in Figure 1.
Inserting the coordinates of the point $H$ from
equation \eqref{e_60} into this equation, we arrive at
\begin{equation} \label{e_63}
e^{\chi_H} = \frl{T_0}{T_i}
\mbox{ .}
\end{equation}
Combining equations \eqref{e_61} and \eqref{e_63}, we
obtain equation \eqref{e_411}.
\itm A comoving object in the universe has a position given by a
fixed coordinate $\chi$. In the present case this corresponds to
a straight worldline in Figure 1. The physical significance of the
coordinate ${\chi}_H$ is that an observer at $R = 0$ cannot observe
objects with $\chi > {\chi}_H$ at the point of time $T_0$.
%
%
\vspace{5mm} \newline
{\bf 5. Continual creation in negatively curved universe models with
dust and radiation using conformal time.}
\vspace{3mm} \newline
The scale factor of a universe model with dust and radiation and with
negative spatial curvature, is given parametrically by [12,13]
\begin{equation} \label{e_72}
a = \alpha (\cosh \eta - 1) + \beta \sinh \eta
\end{equation}
and
\begin{equation} \label{e_73}
t = \alpha (\sinh \eta - \eta)
+ \beta (\cosh \eta - 1)
\mbox{ ,}
\end{equation}
where
\begin{equation} \label{e_74}
\alpha = \frl{{\Omega}_{m0}}{2(1 - {\Omega}_{0})}
\mbox{\hspace{2mm} and \hspace{2mm}}
\beta = \sqrt{\frl{{\Omega}_{\gamma 0}}{1 - {\Omega}_{0}}}
\mbox{ .}
\end{equation}
Here ${\Omega}_{m0}$ and ${\Omega}_{\gamma 0}$ are the present values
of the density parameters of dust and radiation, respectively, and
${\Omega}_{0} = {\Omega}_{m0} + {\Omega}_{\gamma 0}$. We have that
$\eta \in \hs{-2.3mm} < \hs{-0.5mm} 0,\infty \hs{-0.5mm} >$
when $t \in \hs{-2.3mm} < \hs{-0.5mm}  0,\infty \hs{-0.5mm} >$.
Transforming to conformal coordinates by means of
equation \eqref{e_33} and choosing
\begin{equation} \label{e_69}
T_i = \frn{1}{2} (\alpha + \beta)
\mbox{ ,}
\end{equation}
the line element \eqref{e_32} takes the form
\begin{equation} \label{e_75}
ds^2 = \left( 1 - \frac{\alpha + \beta}{2 \sqrt{T^2 - R^2}}
\right)^{\hs{-0.7mm} 2}
\left( 1 - \frac{\alpha - \beta}{2 \sqrt{T^2 - R^2}}
\right)^{\hs{-0.7mm} 2} ds_M^2
\mbox{ ,}
\end{equation}
where $T^2 - R^2 > T_i^2$.
Note that when $\beta = 0$, we obtain the line element of a dust dominated
universe
\begin{equation} \label{e_76}
ds^2 = \left( 1 - \frac{\alpha}{2 \sqrt{T^2 - R^2}}
\right)^{\hs{-0.7mm} 4} ds_M^2
\end{equation}
in accordance with [6]. Putting $\alpha = 0$ we obtain the line element
of a radiation dominated universe with negative spatial curvature in
conformal coordinates
\begin{equation} \label{e_77}
ds^2 = \left( 1 - \frac{{\beta}^2}{4 (T^2 - R^2)}
\right)^{\hs{-0.7mm} 2} ds_M^2
\mbox{ .}
\end{equation}
The relationship between the cosmic time and the conformal time at
$R = \chi = 0$ is
\begin{equation} \label{e_150}
dt = A(T,0) dT
\mbox{ .}
\end{equation}
From equation \eqref{e_75} it then follows that
\begin{equation} \label{e_151}
dt = \left( 1 - \frl{\alpha + \beta}{2T} \right)
\left( 1 - \frl{\alpha - \beta}{2T} \right) dT
\mbox{ ,}
\end{equation}
which shows that $dT / dt > 1$, i.e. the clocks showing conformal time
go at a faster rate than those showing cosmic time. The expression
\eqref{e_151} may be integrated with an integration constant determined
by the condition $t = 0$ for $T = T_i$. The result is, however, already
contained in the transformation equation \eqref{e_406}. With $R = 0$ we
have from equation \eqref{e_34}
\begin{equation} \label{e_152}
T = T_i e^{\eta}
\mbox{ .}
\end{equation}
Then, using equation \eqref{e_73} we obtain
\begin{equation} \label{e_153}
t = T - \alpha \ln \frl{T}{T_i} - \frl{1}{2} (\alpha - \beta) \frl{T_i}{T}
- \beta
\mbox{ ,}
\end{equation}
showing that $t = 0$ for $T = T_i = (\alpha + \beta) / 2$.
\itm In a universe model dominated by dust equation \eqref{e_153}
reduces to
\begin{equation} \label{e_154}
t = \frl{T^2 - T_i^2}{T} - 2 T_i \ln \frl{T}{T_i}
\mbox{ ,}
\end{equation}
and in a radiation dominated universe model,
\begin{equation} \label{e_155}
t = \frl{(T - T_i)^2}{T} < T
\mbox{ .}
\end{equation}
Hence the conformal age of the universe at $R = 0$ is larger than the
cosmic age, which is a coordinate effect.
\itm An interesting phenomenon taking place in the case where the
universe extends only out to the hyperbola in Figure 1, is continual
creation of new space, matter and radiation (Figure 3).
\itm At the local time $T_1$ the whole of the universe is represented
by the horizontal line segment $P_1 Q_1$. At the time $T_2$ this space
has expanded so that it is now represented by the line segment $P_2 Q_2$.
This space may be identified by reference particles enumerated by $\chi$
with $0 < \chi < {\chi}_1$.
The part of the line $T=T_2$ outside $P_2 Q_2$ and inside the hyperbola
represents new space which has been created after the time $T_1$.
This space contains new reference particles with $\chi > {\chi}_1$ that
did not exist at the point of time $T = T_1$. The enlargement of conformal
space is therefore due partly to expansion and partly to creation of new
space, matter and radiation.
\itm The usual picture of Big Bang using cosmic time $t$
and standard radial coordinate $\chi$ is that it happened
everywhere at a certain point of time $t = 0$. The
conformally flat picture is radically different.
The hyperbola $t = 0$ in figures 3a and 3b represents the
creation of the universe. In the conformally flat coordinate
system this creation appears as a spherical front expanding from
the position $R = 0$ of the observer.
According to the conformally flat picture of the universe an
arbitrary observer finds that the universe is isotropic, but
not homogeneous. He also finds that he is positioned at its center.
Particles and radiation are emitted at a spherical front with infinitely
high energy density and temperature, representing the boundary of the
universe, moving radially outwards with superluminal velocity $dR/dT$.
Hence there is continual creation of space, matter and energy in the
universe as depicted in conformally flat coordinates.
\vspace{2mm} \newline
\begin{picture}(50,189)(44,-181)
\qbezier(125.7955, -87.7955)(127.3448, -87.7955)(128.8996, -87.6648)
\qbezier(128.8996, -87.6648)(130.4544, -87.5342)(132.0258, -87.2720)
\qbezier(132.0258, -87.2720)(133.5971, -87.0099)(135.1961, -86.6143)
\qbezier(135.1961, -86.6143)(136.7952, -86.2187)(138.4332, -85.6869)
\qbezier(138.4332, -85.6869)(140.0713, -85.1551)(141.7600, -84.4833)
\qbezier(141.7600, -84.4833)(143.4487, -83.8115)(145.2000, -82.9949)
\qbezier(145.2000, -82.9949)(146.9513, -82.1784)(148.7777, -81.2113)
\qbezier(148.7777, -81.2113)(150.6041, -80.2442)(152.5185, -79.1197)
\qbezier(152.5185, -79.1197)(154.4329, -77.9952)(156.4489, -76.7052)
\qbezier(156.4489, -76.7052)(158.4649, -75.4153)(160.5968, -73.9509)
\qbezier(160.5968, -73.9509)(162.7288, -72.4864)(164.9917, -70.8370)
\qbezier(164.9917, -70.8370)(167.2546, -69.1876)(169.6647, -67.3416)
\qbezier(169.6647, -67.3416)(172.0747, -65.4956)(174.6489, -63.4398)
\qbezier(174.6489, -63.4398)(177.2231, -61.3840)(179.9798, -59.1039)
\qbezier(179.9798, -59.1039)(182.7365, -56.8238)(185.6952, -54.3032)
\qbezier(185.6952, -54.3032)(188.6539, -51.7826)(191.8356, -49.0036)
\qbezier(191.8356, -49.0036)(195.0173, -46.2246)(198.4446, -43.1675)
\qbezier(198.4446, -43.1675)(201.8718, -40.1104)(205.5690, -36.7535)
\qbezier(205.5690, -36.7535)(209.2662, -33.3966)(213.2595, -29.7161)
\put(125.7955, -124.6136){\line(1, 1){ 95.1136}}
\put(125.7955, -78.5909){\line(1, 0){ 55.2273}}
\qbezier(125.7955, -124.6136)(154.2375, -77.2102)(182.6795, -29.8068)
\put(125.7955, -50.9773){\line(1, 0){ 88.3636}}
\put( 88.9773, -124.6136){\vector(1, 0){128.8636}}
\put(125.7955, -161.4318){\vector(0, 1){144.2045}}
\put(227.0455, -129.2159){\makebox(0,0)[]{\footnotesize{$R$}}}
\put(118.1250, -15.6932){\makebox(0,0)[]{\footnotesize{$T$}}}
\put(216.3068, -20.2955){\makebox(0,0)[]{\footnotesize{$t = 0$}}}
\put(181.0227, -20.2955){\makebox(0,0)[]{\footnotesize{$\chi = {\chi}_1$}}}
\put(208.6364, -80.1250){\makebox(0,0)[]{\footnotesize{$T = T_1$}}}
\put(245.4545, -52.5114){\makebox(0,0)[]{\footnotesize{$T = T_2$}}}
\put(135.0000, -132.2841){\makebox(0,0)[]{\footnotesize{$O$}}}
\put(118.1250, -72.4545){\makebox(0,0)[]{\footnotesize{$P_1$}}}
\put(118.1250, -44.8409){\makebox(0,0)[]{\footnotesize{$P_2$}}}
\put(147.2727, -72.4545){\makebox(0,0)[]{\footnotesize{$Q_1$}}}
\put(165.6818, -44.8409){\makebox(0,0)[]{\footnotesize{$Q_2$}}}
\end{picture}
\begin{picture}(50,189)(-106,-181)
\qbezier(125.7955, -87.7955)(127.3448, -87.7955)(128.8996, -87.6648)
\qbezier(128.8996, -87.6648)(130.4544, -87.5342)(132.0258, -87.2720)
\qbezier(132.0258, -87.2720)(133.5971, -87.0099)(135.1961, -86.6143)
\qbezier(135.1961, -86.6143)(136.7952, -86.2187)(138.4332, -85.6869)
\qbezier(138.4332, -85.6869)(140.0713, -85.1551)(141.7600, -84.4833)
\qbezier(141.7600, -84.4833)(143.4487, -83.8115)(145.2000, -82.9949)
\qbezier(145.2000, -82.9949)(146.9513, -82.1784)(148.7777, -81.2113)
\qbezier(148.7777, -81.2113)(150.6041, -80.2442)(152.5185, -79.1197)
\qbezier(152.5185, -79.1197)(154.4329, -77.9952)(156.4489, -76.7052)
\qbezier(156.4489, -76.7052)(158.4649, -75.4153)(160.5968, -73.9509)
\qbezier(160.5968, -73.9509)(162.7288, -72.4864)(164.9917, -70.8370)
\qbezier(164.9917, -70.8370)(167.2546, -69.1876)(169.6647, -67.3416)
\qbezier(169.6647, -67.3416)(172.0747, -65.4956)(174.6489, -63.4398)
\qbezier(174.6489, -63.4398)(177.2231, -61.3840)(179.9798, -59.1039)
\qbezier(179.9798, -59.1039)(182.7365, -56.8238)(185.6952, -54.3032)
\qbezier(185.6952, -54.3032)(188.6539, -51.7826)(191.8356, -49.0036)
\qbezier(191.8356, -49.0036)(195.0173, -46.2246)(198.4446, -43.1675)
\qbezier(198.4446, -43.1675)(201.8718, -40.1104)(205.5690, -36.7535)
\qbezier(205.5690, -36.7535)(209.2662, -33.3966)(213.2595, -29.7161)
\put(125.7955, -124.6136){\line(1, 1){ 95.1136}}
\put(125.7955, -78.5909){\line(1, 0){ 55.2273}}
\qbezier(153.4091, -78.5909)(168.0443, -54.1989)(182.6795, -29.8068)
\qbezier(162.7413, -72.4545)(177.8456, -51.1307)(192.9499, -29.8068)
\qbezier(174.8864, -63.2500)(188.2636, -46.5284)(201.6409, -29.8068)
\qbezier(189.5664, -50.9773)(198.7335, -40.3920)(207.9006, -29.8068)
\put(125.7955, -50.9773){\line(1, 0){ 88.3636}}
\put( 88.9773, -124.6136){\vector(1, 0){128.8636}}
\put(125.7955, -161.4318){\vector(0, 1){144.2045}}
\put(227.0455, -129.2159){\makebox(0,0)[]{\footnotesize{$R$}}}
\put(118.1250, -15.6932){\makebox(0,0)[]{\footnotesize{$T$}}}
\put(208.6364, -80.1250){\makebox(0,0)[]{\footnotesize{$T = T_1$}}}
\put(245.4545, -52.5114){\makebox(0,0)[]{\footnotesize{$T = T_2$}}}
\end{picture}
%
\vspace*{-4mm} \newline
\hspace*{38.0mm} (a) \hspace{63.9mm} (b)
\vspace*{4mm} \newline
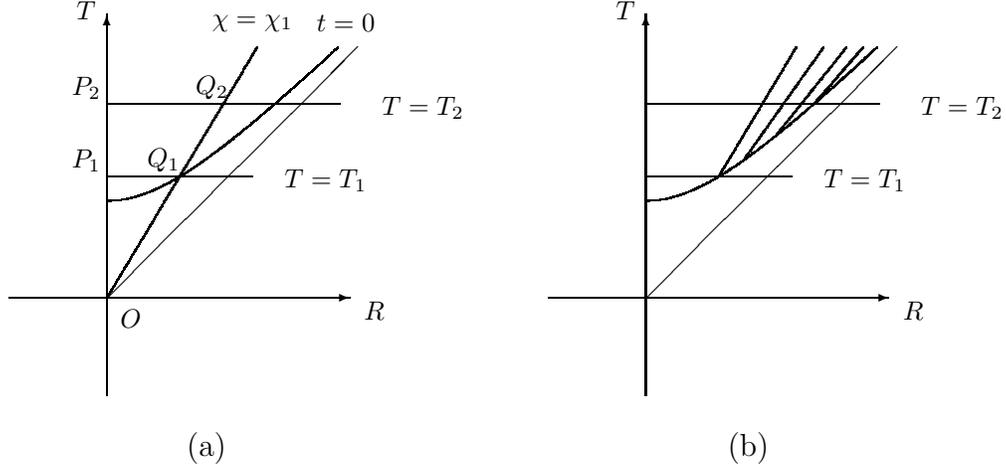
{\footnotesize \sf Figure 3. Minkowski diagram of a universe model with
radiation and dust using conformal coordinates of type I. The horizontal
line segment $P_1 Q_1$ in (a) represents the conformal space at the
conformal time $T_1$. At the conformal time $T = T_2$ the space $P_1 Q_1$
that existed at $T = T_1$ has expanded to $P_2 Q_2$. The figures show that
in addition, new conformal space has been created on the line $T = T_2$
outside the line segment $P_2 Q_2$ and inside the hyperbola, as indicated
by the world lines in (b).}
\vspace{0mm} \newline
\itm Although the conformal space is finite, a space traveller can never
reach the boundary of the universe since it moves outwards with a velocity
greater than the velocity of light. Hence no observer will ever risk
arriving at this terrible inferno.
\itm Differentiation of equation \eqref{e_34} while keeping $\eta = 0$
gives the the following expression for the velocity of the front
\begin{equation} \label{e_66}
\left( \frl{dR}{dT} \right)_{\eta = 0} = \frl{1}{\sqrt{1 - T_i^2/T^2}}
\end{equation}
at an arbitrary time $T > T_i$. On the other hand, the expansion of the
conformal space represented by the velocity of the Hubble, flow, is
\begin{equation} \label{e_71}
\left( \frl{dR}{dT} \right)_{\chi = {\chi}_1} = \tanh {\chi}_1
= \sqrt{1 - T_i^2/T_1^2}
\mbox{ .}
\end{equation}
Note that $\left( dR/dT \right)_{\eta = 0} > 1$ and
$\left( dR/dT \right)_{\chi = {\chi}_1} < 1$.
This means that the boundary of the universe moves faster than the
inertial flow representing the expansion of space. The initial velocity
of the boundary of the universe is infinitely great and then decreases
and approaches the velocity of light in an infinitely remote future. On
the other hand the conformally flat Hubble flow has a time independent
velocity less than that of light.
\itm In the conformally flat picture new space is continually created
at the boundary. However, using cosmic time no new space is created.
This apparent contradiction is solved by noting that constant cosmic
time and constant local time represent different simultaneities. So
the conformal space defined at constant conformal time is different
from the cosmic space defined at constant cosmic time.
\itm An exception among the negatively curved universe models
is the Milne universe. Due to the special form of the scale factor in
this universe model, the parametric time $\eta \rightarrow -\infty$
when the cosmic time $t \rightarrow 0 \,$. Hence the space $t=0$ is not
represented by a hyperbola, but by the light cone in Figure 1. This means
that there is no new space created using local time in this universe model.
\itm The Kretschmann curvature scalar for the models with dust and
with radiation are respectively
\begin{equation} \label{e_21}
K_d = \frac{245760 \hs{0.5mm} (T^2 - R^2)^3 \alpha^2}
{[2 \hs{0.5mm} \sqrt{T^2 - R^2} - \alpha]^{12}}
\mbox{\hspace{2mm}, \hspace{2mm}}
K_r = \frac{1572864 \hs{0.5mm} (T^2 - R^2)^4 \beta^4}
{[4 (T^2 - R^2) - \beta^2]^8}
\mbox{ .}
\end{equation}
These expressions show that there is a physical singularity with
infinitely great spacetime curvature at the boundary $T^2 - R^2 = T_i^2$
with continual creation.
%
%
%
\vspace{5mm} \newline
\textbf{6. Negatively curved universe models with vacuum energy}
\vspace{3mm} \newline
The inflationary era is a brief period dominated by vacuum energy
with accelerated expansion at the beginning of the universe. This
era is often said to make space flat. That cannot, however, be the
case. The inflationary era cannot change a universe model with curved
space, $k \ne 0 \,$, to a model with flat space, $k = 0 \,$. It can
only make space {\it approximately} flat. The curvature decreases
exponentially. Such a space is still curved, although
the curvature may be so small that we are not able to measure it.
So if our universe entered the inflationary era with curved
space, the space will still be curved today.
\itm The region of initial conditions for universe models with curved
space is much larger than that for flat space. Hence, if the
universe entered the inflationary era by some sort of quantum processes,
the probability that the space is curved is much larger than that
it is flat. One may therefore conclude that we probably live in a
universe with curved space, but that the space was inflated
so much in the inflationary era that we are not able to measure the
curvature.
\itm Although the difference between such a curved universe and a flat
one is negligible small as far as observed properties of the universe
is concerned, there are some very interesting conceptual differences
between curved universe models and flat ones.
For example a flat universe dominated by Lorentz Invariant Vacuum
Energy (LIVE) has a steady state character and may be infinitely old,
while a corresponding universe with negatively curved space is
evolving and has a finite age. In this case the scale factor is
\begin{equation} \label{e_79}
a(t) = \frl{1}{\widehat{H}_{\Lambda} \rule[-0mm]{0mm}{4.25mm}}
\sinh (\widehat{H}_{\Lambda} t)
\mbox{ ,}
\end{equation}
for $t > 0$, where
\begin{equation} \label{e_290}
\widehat{H}_{\Lambda} = l_0 \sqrt{\Lambda / 3}
\mbox{\hspace{2mm}, \hspace{2mm}}
l_0 = \frl{1}{H_0} \hs{0.7mm} \sqrt{\frl{1}{1 - \Omega_0}}
\mbox{ ,}
\end{equation}
and $\Lambda$ is the cosmological constant. The dimensionless
Hubble parameter, $\widehat{H} = l_0 \dot{a} / a$, is
\begin{equation} \label{e_5}
\widehat{H} = \widehat{H}_{\Lambda} \coth (\widehat{H}_{\Lambda} t)
\mbox{ .}
\end{equation}
\vspace{-2mm}
\itm Using equation \eqref{e_404}, the parametric time $\eta$ is [14]
\begin{equation} \label{e_80}
\eta = - \mbox{arccoth} (\cosh (\widehat{H}_{\Lambda} t))
= \ln (\tanh \frl{\widehat{H}_{\Lambda} t}{2})
\mbox{ .}
\end{equation}
Note that $\eta \rightarrow - \infty$ when $t \rightarrow 0 \,$,
and that $\eta \rightarrow 0$ when $t \rightarrow \infty \,$.
This means that the conformal space extends out to the light cone
$R = T$. Hence, there is no continual creation in such a universe
model. For this universe model the limit $t \rightarrow \infty$ at
$\chi = 0$ corresponds to a conformal time $T = T_i$ according to
equation \eqref{e_38}. It is therefore natural in this case to replace
the initial time $T_i$ with a final time $T_f$. Thus the conformal time
at $R = 0$ is
\begin{equation} \label{e_156}
T = T_f e^{\eta} = T_f \tanh \frl{\widehat{H}_{\Lambda} t}{2} < T_f
\mbox{ ,}
\end{equation}
where $T_f$ is the final conformal time. It follows that the LIVE
dominated universe with $k = -1$ has a finite conformal age. This is
a coordinate effect since
\begin{equation} \label{e_157}
dT = \frl{\widehat{H}_{\Lambda} T_f / 2}
{\cosh^2 \left( \widehat{H}_{\Lambda} t / \hs{0.5mm} 2 \right)
\rule[-0mm]{0mm}{4.25mm}} dt
\mbox{ ,}
\end{equation}
which shows that for $\widehat{H}_{\Lambda} t >> 1$ the rate of the
conformal time decreases exponentially compared to the rate of the
cosmic time.
\itm This model is, however, not realistic since the general theory
of relativity is valid only after the Planck time,
$t_{Pl} = 5.4 \cdot 10^{-44} s$.
Before this time the universe may have existed in a quantum era which
cannot be described without a quantum theory of gravity. We here assume
that the universe entered a vacuum dominated inflationary era at the
Planck time. This implies that the conformal space only extends out to
a hyperbola given by $t = t_{Pl}$, which represents the frontier against
a quantum era with properties that we cannot describe with the present
theories. At this frontier there is continual creation of new conformal
space.
\itm Equations \eqref{e_79} and \eqref{e_80} imply that the scale factor
is given in terms of the parametric time as
\begin{equation} \label{e_138}
a(\eta) = - \frl{1}{\widehat{H}_{\Lambda} \sinh \eta \rule[-0mm]{0mm}{4.25mm}}
\mbox{ .}
\end{equation}
\itm Together with equation \eqref{e_32} this relation implies that
that the line element for the present universe model, as expressed
by conformal coordinates, takes the form
\begin{equation} \label{e_142}
ds^2 = \frl{1}{{\Omega}_0 \widehat{H}_0^2 \rule[-0mm]{0mm}{4.25mm}}
\left[ \frl{2 T_f}{T_f^2 - (T^2 - R^2)} \right]^2 ds_M^2
\mbox{ .}
\end{equation}
This expression shows that the relationship between cosmic time and
conformal time is
\begin{equation} \label{e_178}
dt = \frl{1}{\widehat{H}_0 \sqrt{{\Omega}_0} \rule[-0mm]{0mm}{4.25mm}}
\hs{1.5mm} \frl{2 T_f}{T_f^2 - (T^2 - R^2)} \hs{1.5mm} dT
\mbox{ .}
\end{equation}
The coordinate $t$ is the proper time of the freely moving observers.
The expression shows that the clocks showing conformal time slows down
towards a vanishing rate of time as we approach the boundary
$T^2 - R^2 = T_f^2$ of the conformal space.
\itm The conformal space has an interesting behaviour in this universe
model. Since the parametric time $\eta = 0$ corresponds to cosmic time
$t \rightarrow \infty$, the hyperbola $\eta = 0$ in Figure \nolinebreak 4
represents a future limit beyond which there exists no space. Hence the
figure shows that conformal space is annihilated in the LIVE dominated
universe. At the point of time $T = T_f$, conformal space starts vanishing
at $R = 0$. Then a spherical hole develops which does not belong to the
conformal space. Note that space consists of simultaneous events
in spacetime, and the points inside the hole do not correspond to
events in the FRW-universe. An observer can never reach the boundary
of the hole since the cosmic time approaches infinity at this boundary.
At the local time $T_1$ the hole in the universe is represented
by the horizontal line segment $P_1 Q_1$. At the time $T_2$ this hole
has expanded so that it is now represented by the line segment $P_2 Q_2$.
The part of the line $T=T_2$ outside $P_2 Q_2$ and inside the hyperbola
represents new emptiness which has appeared after the time $T_1$. The
enlargement of the hole in conformal space is therefore due partly to
expansion and partly to annihilation of space.
\vspace{2mm} \newline
\begin{picture}(50,189)(-96,-181)
\qbezier(125.7955, -87.7955)(127.3448, -87.7955)(128.8996, -87.6648)
\qbezier(128.8996, -87.6648)(130.4544, -87.5342)(132.0258, -87.2720)
\qbezier(132.0258, -87.2720)(133.5971, -87.0099)(135.1961, -86.6143)
\qbezier(135.1961, -86.6143)(136.7952, -86.2187)(138.4332, -85.6869)
\qbezier(138.4332, -85.6869)(140.0713, -85.1551)(141.7600, -84.4833)
\qbezier(141.7600, -84.4833)(143.4487, -83.8115)(145.2000, -82.9949)
\qbezier(145.2000, -82.9949)(146.9513, -82.1784)(148.7777, -81.2113)
\qbezier(148.7777, -81.2113)(150.6041, -80.2442)(152.5185, -79.1197)
\qbezier(152.5185, -79.1197)(154.4329, -77.9952)(156.4489, -76.7052)
\qbezier(156.4489, -76.7052)(158.4649, -75.4153)(160.5968, -73.9509)
\qbezier(160.5968, -73.9509)(162.7288, -72.4864)(164.9917, -70.8370)
\qbezier(164.9917, -70.8370)(167.2546, -69.1876)(169.6647, -67.3416)
\qbezier(169.6647, -67.3416)(172.0747, -65.4956)(174.6489, -63.4398)
\qbezier(174.6489, -63.4398)(177.2231, -61.3840)(179.9798, -59.1039)
\qbezier(179.9798, -59.1039)(182.7365, -56.8238)(185.6952, -54.3032)
\qbezier(185.6952, -54.3032)(188.6539, -51.7826)(191.8356, -49.0036)
\qbezier(191.8356, -49.0036)(195.0173, -46.2246)(198.4446, -43.1675)
\qbezier(198.4446, -43.1675)(201.8718, -40.1104)(205.5690, -36.7535)
\qbezier(205.5690, -36.7535)(209.2662, -33.3966)(213.2595, -29.7161)
\put(125.7955, -124.6136){\line(1, 1){ 95.1136}}
\put(125.7955, -78.5909){\line(1, 0){ 55.2273}}
\qbezier(125.7955, -124.6136)(154.2375, -77.2102)(182.6795, -29.8068)
\put(125.7955, -50.9773){\line(1, 0){ 88.3636}}
\qbezier(125.7955, -118.4773)(127.3295, -120.0114)(128.8636, -121.5455)
\qbezier(125.7955, -112.3409)(128.8636, -115.4091)(131.9318, -118.4773)
\qbezier(125.7955, -106.2045)(130.3977, -110.8068)(135.0000, -115.4091)
\qbezier(125.7955, -100.0682)(131.9318, -106.2045)(138.0682, -112.3409)
\qbezier(125.7955, -93.9318)(133.4659, -101.6023)(141.1364, -109.2727)
\qbezier(125.7955, -87.7955)(135.0000, -97.0000)(144.2045, -106.2045)
\qbezier(131.4935, -87.3571)(139.3831, -95.2468)(147.2727, -103.1364)
\qbezier(136.5341, -86.2614)(143.4375, -93.1648)(150.3409, -100.0682)
\qbezier(141.1364, -84.7273)(147.2727, -90.8636)(153.4091, -97.0000)
\qbezier(145.4318, -82.8864)(150.9545, -88.4091)(156.4773, -93.9318)
\qbezier(149.5041, -80.8223)(154.5248, -85.8430)(159.5455, -90.8636)
\qbezier(153.4091, -78.5909)(158.0114, -83.1932)(162.6136, -87.7955)
\qbezier(157.1853, -76.2308)(161.4336, -80.4790)(165.6818, -84.7273)
\qbezier(160.8604, -73.7695)(164.8052, -77.7143)(168.7500, -81.6591)
\qbezier(164.4545, -71.2273)(168.1364, -74.9091)(171.8182, -78.5909)
\qbezier(167.9830, -68.6193)(171.4347, -72.0710)(174.8864, -75.5227)
\qbezier(171.4572, -65.9572)(174.7059, -69.2059)(177.9545, -72.4545)
\qbezier(174.8864, -63.2500)(177.9545, -66.3182)(181.0227, -69.3864)
\qbezier(178.2775, -60.5048)(181.1842, -63.4115)(184.0909, -66.3182)
\qbezier(181.6364, -57.7273)(184.3977, -60.4886)(187.1591, -63.2500)
\qbezier(184.9675, -54.9221)(187.5974, -57.5519)(190.2273, -60.1818)
\qbezier(188.2748, -52.0930)(190.7851, -54.6033)(193.2955, -57.1136)
\qbezier(191.5613, -49.2431)(193.9625, -51.6443)(196.3636, -54.0455)
\qbezier(194.8295, -46.3750)(197.1307, -48.6761)(199.4318, -50.9773)
\qbezier(198.0818, -43.4909)(200.2909, -45.7000)(202.5000, -47.9091)
\qbezier(201.3199, -40.5927)(203.4441, -42.7168)(205.5682, -44.8409)
\qbezier(204.5455, -37.6818)(206.5909, -39.7273)(208.6364, -41.7727)
\qbezier(207.7597, -34.7597)(209.7321, -36.7321)(211.7045, -38.7045)
\qbezier(210.9639, -31.8276)(212.8683, -33.7320)(214.7727, -35.6364)
\qbezier(214.1591, -28.8864)(216.0000, -30.7273)(217.8409, -32.5682)
\put( 88.9773, -124.6136){\vector(1, 0){128.8636}}
\put(125.7955, -161.4318){\vector(0, 1){144.2045}}
\put(227.0455, -129.2159){\makebox(0,0)[]{\footnotesize{$R$}}}
\put(118.1250, -15.6932){\makebox(0,0)[]{\footnotesize{$T$}}}
\put(216.3068, -20.2955){\makebox(0,0)[]{\footnotesize{$\eta = 0$}}}
\put(181.0227, -20.2955){\makebox(0,0)[]{\footnotesize{$\chi = {\chi}_1$}}}
\put(208.6364, -80.1250){\makebox(0,0)[]{\footnotesize{$T = T_1$}}}
\put(245.4545, -52.5114){\makebox(0,0)[]{\footnotesize{$T = T_2$}}}
\put(135.0000, -132.2841){\makebox(0,0)[]{\footnotesize{$O$}}}
\put(118.1250, -72.4545){\makebox(0,0)[]{\footnotesize{$P_1$}}}
\put(118.1250, -44.8409){\makebox(0,0)[]{\footnotesize{$P_2$}}}
\put(147.2727, -72.4545){\makebox(0,0)[]{\footnotesize{$Q_1$}}}
\put(165.6818, -44.8409){\makebox(0,0)[]{\footnotesize{$Q_2$}}}
\end{picture}
\vspace{0mm} \newline
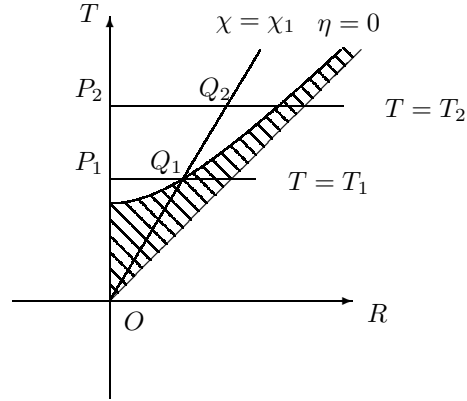
{\footnotesize \sf Figure 4. The hatched region represents a succession
of conformal spaces in a LIVE dominated universe at different points of
time. The hyperbola $\eta = 0$ corresponds to cosmic time
$t \rightarrow \infty$. The horizontal line segment $P_1 Q_1$ represents
a hole in the conformal space at the conformal time $T_1$.
At the conformal time $T = T_2$ the hole $P_1 Q_1$ that existed at
$T = T_1$ has expanded to $P_2 Q_2$. In addition the figure shows
that conformal space has been annihilated on the line $T = T_2$
outside the line segment $P_2 Q_2$ and inside the hyperbola.}
\vspace{0mm} \newline
\itm At the present time the universe is filled with radiation, matter
and vacuum energy. If the vacuum energy is of the LIVE type, the energy
density will remain constant in the future. But the density of radiation
and matter will decrease. Hence the universe will approach a vacuum
dominated state. This means that in conformal coordinates the final
destiny of our universe will be as described above.
%
%
%
\vspace{5mm} \newline
{\bf 7. A second type of conformal coordinates for universe models
with negative spatial curvature}
\vspace{3mm} \newline
For universe models with negative spatial curvature one may introduce
a second type of conformal coordinates $(\widehat{T},\widehat{R})$ by
choosing $a = 0$, $b = 0$, $c = 1$ and $d = 0$ in equation \eqref{e_407}.
This gives the generating function
\begin{equation} \label{e_204}
f(x) = \tanh (x / 2)
\mbox{ .}
\end{equation}
The transformation \eqref{e_406} then takes the form
\begin{equation} \label{e_17}
\widehat{T} = \frac{\sinh \eta}{\cosh \eta + \cosh \chi}
\mbox{\hspace{2mm} , \hspace{3mm}}
\widehat{R} = \frac{\sinh \chi}{\cosh \eta + \cosh \chi}
\mbox{ .}
\end{equation}
If the universe model begins at $\eta = 0$, this transformation
maps the first quadrant $\chi > 0$, $\eta > 0$ onto the triangle
$0 < \widehat{T} < 1$ and $0 < \widehat{R} < 1 - \widehat{T}$.
On the other hand, if the universe is infinitely old, the fourth
quadrant $\chi > 0$, $\eta < 0$ is mapped onto the triangle
$-1 < \widehat{T} < 0$ and $0 < \widehat{R} < 1 + \widehat{T}$.
The inverse transformation is
\begin{equation} \label{e_18}
\coth \eta = \frl{1 +
\left(\widehat{T}^{2} - \widehat{R}^{2} \right)}{2 \widehat{T}
\rule[-0mm]{0mm}{4.25mm}}
\mbox{\hspace{2mm} , \hspace{3mm}}
\coth \chi = \frl{1 -
\left(\widehat{T}^{2} - \widehat{R}^{2} \right)}{2 \widehat{R}
\rule[-0mm]{0mm}{4.25mm}}
\mbox{ .}
\end{equation}
From equation \eqref{e_409} it then follows that the line element now
takes the form
\begin{equation} \label{e_20}
ds^2 = \frac{4 \hs{0.9mm} a(\eta(\widehat{T},\widehat{R}))^2}
{[1 - (\widehat{T}^2 - \widehat{R}^2)]^2 - 4 \widehat{R}^2}
ds_M^2
\mbox{ .}
\end{equation}
As drawn in the $(\widehat{T},\widehat{R})$ spacetime diagrams in Figures 5
and 6, both the world lines of points with $\chi = {\chi}_0$ and the
simultaneity curves $\eta = {\eta}_0$ are hyperbolae, given respectively
by
\begin{equation} \label{e_31}
(\widehat{R} - a_1)^2 - \widehat{T}^2 = a_1^2 - 1
\mbox{\hspace{2mm} and \hspace{3mm}}
(\widehat{T} - b_1)^2 - \widehat{R}^2 = b_1^2 - 1
\mbox{ ,}
\end{equation}
where $a_1 = \coth {\chi}_0$ and $b_1 = \coth {\eta}_0$.
These equations are valid both for universe models dominated by radiation
and dust and by LIVE, the only difference being that ${\eta}_0 > 0$ with
radiation and dust, and ${\eta}_0 < 0$ with LIVE.
The velocity in the $(\widehat{T},\widehat{R})$-system of a particle with
$\chi = \mbox{constant}$ is given by
\begin{equation} \label{e_14}
V = \left( \frl{d\widehat{R}}{d\widehat{T} \rule[-0mm]{0mm}{4.25mm}}
\right)_{\chi = \mbox{\footnotesize constant}}
= - \frl{\sinh \eta \sinh \chi}{1 + \cosh \eta \cosh \chi}
= \frl{2 \widehat{T} \widehat{R}}
{-1 + (\widehat{T}^2 + \widehat{R}^2) \rule[-0mm]{0mm}{4.25mm}}
\mbox{ .}
\end{equation}
In this coordinate system the initial recession velocity vanishes.
Note that $V < 0$ for $\widehat{T} \widehat{R} > 0$ and $V > 0$ for
$\widehat{T} \widehat{R} < 0$ since $\widehat{T}^2 + \widehat{R}^2 < 1$.
\itm The reference particles of the $(\widehat{T},\widehat{R})$-system
and the $(T,R)$-system have different motions. The velocity in the
$(\eta,\chi)$-system of a particle with $\widehat{R} = \mbox{constant}$ is
\begin{equation} \label{e_15}
\frl{d \chi}{d \eta} = \frl{\sinh \eta \sinh \chi}
{1 + \cosh \eta \cosh \chi}
\mbox{ .}
\end{equation}
Hence the $(\widehat{T},\widehat{R})$-system expands relative to the
$(\eta,\chi)$-system while the $(T,R)$-system contracts for
$\widehat{T} > 0$.
\itm World lines of free particles defining the inertial Hubble flow
and simultaneity curves $\eta = \mbox{constant}$ in a universe with
negative spatial curvature containing dust and radiation are depicted
relative to the $(\widehat{T},\widehat{R})$-system in Figure 5.
This figure gives an illusion of a contracting universe of finite
extension with a finite age, since the world lines of particles with
$\chi = \mbox{constant}$, which define the Hubble flow, all terminate
at the point $(\widehat{T},\widehat{R}) = (1,0)$. However, according
to equation \eqref{e_18}, approaching this point means that the
parametic time $\eta$, and hence also the cosmic time $t$ and the
scale factor, approaches infinity. Furthemore, this equation also
implies that $\chi$ approaches infinity at the line
$\widehat{T} = 1 - \widehat{R}$.
\vspace*{5mm} \newline
\begin{picture}(50,192)(-96,-182)
\qbezier(120.0000, -32.0000)(124.0038, -36.4042)(127.6672, -40.5014)
\qbezier(127.6672, -40.5014)(131.3307, -44.5986)(134.6772, -48.4151)
\qbezier(134.6772, -48.4151)(138.0238, -52.2317)(141.0750, -55.7921)
\qbezier(141.0750, -55.7921)(144.1263, -59.3525)(146.9017, -62.6796)
\qbezier(146.9017, -62.6796)(149.6772, -66.0067)(152.1947, -69.1218)
\qbezier(152.1947, -69.1218)(154.7122, -72.2370)(156.9880, -75.1603)
\qbezier(156.9880, -75.1603)(159.2637, -78.0835)(161.3123, -80.8337)
\qbezier(161.3123, -80.8337)(163.3609, -83.5838)(165.1955, -86.1784)
\qbezier(165.1955, -86.1784)(167.0302, -88.7731)(168.6626, -91.2289)
\qbezier(168.6626, -91.2289)(170.2949, -93.6847)(171.7356, -96.0175)
\qbezier(171.7356, -96.0175)(173.1763, -98.3503)(174.4345, -100.5751)
\qbezier(174.4345, -100.5751)(175.6927, -102.7998)(176.7765, -104.9308)
\qbezier(176.7765, -104.9308)(177.8603, -107.0617)(178.7766, -109.1126)
\qbezier(178.7766, -109.1126)(179.6930, -111.1635)(180.4478, -113.1475)
\qbezier(180.4478, -113.1475)(181.2026, -115.1315)(181.8008, -117.0613)
\qbezier(181.8008, -117.0613)(182.3989, -118.9911)(182.8441, -120.8791)
\qbezier(182.8441, -120.8791)(183.2894, -122.7671)(183.5847, -124.6255)
\qbezier(183.5847, -124.6255)(183.8799, -126.4839)(184.0271, -128.3246)
\qbezier(184.0271, -128.3246)(184.1742, -130.1652)(184.1742, -132.0000)
\qbezier(120.0000, -32.0000)(121.9292, -35.0866)(123.7406, -38.1011)
\qbezier(123.7406, -38.1011)(125.5521, -41.1156)(127.2485, -44.0624)
\qbezier(127.2485, -44.0624)(128.9450, -47.0092)(130.5289, -49.8927)
\qbezier(130.5289, -49.8927)(132.1129, -52.7763)(133.5868, -55.6008)
\qbezier(133.5868, -55.6008)(135.0606, -58.4253)(136.4265, -61.1951)
\qbezier(136.4265, -61.1951)(137.7925, -63.9648)(139.0525, -66.6839)
\qbezier(139.0525, -66.6839)(140.3126, -69.4030)(141.4686, -72.0755)
\qbezier(141.4686, -72.0755)(142.6246, -74.7480)(143.6784, -77.3778)
\qbezier(143.6784, -77.3778)(144.7321, -80.0077)(145.6852, -82.5988)
\qbezier(145.6852, -82.5988)(146.6382, -85.1899)(147.4919, -87.7462)
\qbezier(147.4919, -87.7462)(148.3456, -90.3025)(149.1013, -92.8277)
\qbezier(149.1013, -92.8277)(149.8570, -95.3530)(150.5158, -97.8509)
\qbezier(150.5158, -97.8509)(151.1746, -100.3489)(151.7374, -102.8233)
\qbezier(151.7374, -102.8233)(152.3003, -105.2976)(152.7681, -107.7521)
\qbezier(152.7681, -107.7521)(153.2359, -110.2066)(153.6093, -112.6449)
\qbezier(153.6093, -112.6449)(153.9826, -115.0832)(154.2622, -117.5088)
\qbezier(154.2622, -117.5088)(154.5418, -119.9345)(154.7279, -122.3512)
\qbezier(154.7279, -122.3512)(154.9140, -124.7679)(155.0070, -127.1792)
\qbezier(155.0070, -127.1792)(155.1000, -129.5905)(155.1000, -132.0000)
\qbezier(120.0000, -32.0000)(120.6388, -34.6830)(121.2434, -37.3581)
\qbezier(121.2434, -37.3581)(121.8480, -40.0332)(122.4185, -42.7008)
\qbezier(122.4185, -42.7008)(122.9889, -45.3684)(123.5253, -48.0289)
\qbezier(123.5253, -48.0289)(124.0618, -50.6894)(124.5642, -53.3433)
\qbezier(124.5642, -53.3433)(125.0667, -55.9972)(125.5353, -58.6448)
\qbezier(125.5353, -58.6448)(126.0039, -61.2925)(126.4387, -63.9344)
\qbezier(126.4387, -63.9344)(126.8735, -66.5763)(127.2746, -69.2129)
\qbezier(127.2746, -69.2129)(127.6757, -71.8494)(128.0432, -74.4811)
\qbezier(128.0432, -74.4811)(128.4106, -77.1127)(128.7445, -79.7399)
\qbezier(128.7445, -79.7399)(129.0783, -82.3671)(129.3786, -84.9902)
\qbezier(129.3786, -84.9902)(129.6789, -87.6133)(129.9457, -90.2328)
\qbezier(129.9457, -90.2328)(130.2126, -92.8523)(130.4459, -95.4686)
\qbezier(130.4459, -95.4686)(130.6793, -98.0849)(130.8793, -100.6985)
\qbezier(130.8793, -100.6985)(131.0792, -103.3120)(131.2458, -105.9232)
\qbezier(131.2458, -105.9232)(131.4124, -108.5344)(131.5456, -111.1436)
\qbezier(131.5456, -111.1436)(131.6788, -113.7529)(131.7788, -116.3607)
\qbezier(131.7788, -116.3607)(131.8787, -118.9685)(131.9453, -121.5752)
\qbezier(131.9453, -121.5752)(132.0118, -124.1820)(132.0451, -126.7880)
\qbezier(132.0451, -126.7880)(132.0784, -129.3941)(132.0784, -132.0000)
\qbezier(120.0000, -67.8258)(121.8348, -67.8258)(123.6754, -67.9729)
\qbezier(123.6754, -67.9729)(125.5161, -68.1201)(127.3745, -68.4153)
\qbezier(127.3745, -68.4153)(129.2329, -68.7106)(131.1209, -69.1559)
\qbezier(131.1209, -69.1559)(133.0089, -69.6011)(134.9387, -70.1992)
\qbezier(134.9387, -70.1992)(136.8685, -70.7974)(138.8525, -71.5522)
\qbezier(138.8525, -71.5522)(140.8365, -72.3070)(142.8874, -73.2234)
\qbezier(142.8874, -73.2234)(144.9383, -74.1397)(147.0692, -75.2235)
\qbezier(147.0692, -75.2235)(149.2002, -76.3073)(151.4249, -77.5655)
\qbezier(151.4249, -77.5655)(153.6497, -78.8237)(155.9825, -80.2644)
\qbezier(155.9825, -80.2644)(158.3153, -81.7051)(160.7711, -83.3374)
\qbezier(160.7711, -83.3374)(163.2269, -84.9698)(165.8216, -86.8045)
\qbezier(165.8216, -86.8045)(168.4162, -88.6391)(171.1663, -90.6877)
\qbezier(171.1663, -90.6877)(173.9165, -92.7363)(176.8397, -95.0120)
\qbezier(176.8397, -95.0120)(179.7630, -97.2878)(182.8782, -99.8053)
\qbezier(182.8782, -99.8053)(185.9933, -102.3228)(189.3204, -105.0983)
\qbezier(189.3204, -105.0983)(192.6475, -107.8737)(196.2079, -110.9250)
\qbezier(196.2079, -110.9250)(199.7683, -113.9762)(203.5849, -117.3228)
\qbezier(203.5849, -117.3228)(207.4014, -120.6693)(211.4986, -124.3328)
\qbezier(211.4986, -124.3328)(215.5958, -127.9962)(220.0000, -132.0000)
\qbezier(120.0000, -96.9000)(122.4095, -96.9000)(124.8208, -96.9930)
\qbezier(124.8208, -96.9930)(127.2321, -97.0860)(129.6488, -97.2721)
\qbezier(129.6488, -97.2721)(132.0655, -97.4582)(134.4912, -97.7378)
\qbezier(134.4912, -97.7378)(136.9168, -98.0174)(139.3551, -98.3907)
\qbezier(139.3551, -98.3907)(141.7934, -98.7641)(144.2479, -99.2319)
\qbezier(144.2479, -99.2319)(146.7024, -99.6997)(149.1767, -100.2626)
\qbezier(149.1767, -100.2626)(151.6511, -100.8254)(154.1491, -101.4842)
\qbezier(154.1491, -101.4842)(156.6470, -102.1430)(159.1723, -102.8987)
\qbezier(159.1723, -102.8987)(161.6975, -103.6544)(164.2538, -104.5081)
\qbezier(164.2538, -104.5081)(166.8101, -105.3618)(169.4012, -106.3148)
\qbezier(169.4012, -106.3148)(171.9923, -107.2679)(174.6222, -108.3216)
\qbezier(174.6222, -108.3216)(177.2520, -109.3754)(179.9245, -110.5314)
\qbezier(179.9245, -110.5314)(182.5970, -111.6874)(185.3161, -112.9475)
\qbezier(185.3161, -112.9475)(188.0352, -114.2075)(190.8049, -115.5735)
\qbezier(190.8049, -115.5735)(193.5747, -116.9394)(196.3992, -118.4132)
\qbezier(196.3992, -118.4132)(199.2237, -119.8871)(202.1073, -121.4711)
\qbezier(202.1073, -121.4711)(204.9908, -123.0550)(207.9376, -124.7515)
\qbezier(207.9376, -124.7515)(210.8844, -126.4479)(213.8989, -128.2594)
\qbezier(213.8989, -128.2594)(216.9134, -130.0708)(220.0000, -132.0000)
\qbezier(120.0000, -119.9216)(122.6059, -119.9216)(125.2120, -119.9549)
\qbezier(125.2120, -119.9549)(127.8180, -119.9882)(130.4248, -120.0547)
\qbezier(130.4248, -120.0547)(133.0315, -120.1213)(135.6393, -120.2212)
\qbezier(135.6393, -120.2212)(138.2471, -120.3212)(140.8564, -120.4544)
\qbezier(140.8564, -120.4544)(143.4656, -120.5876)(146.0768, -120.7542)
\qbezier(146.0768, -120.7542)(148.6880, -120.9208)(151.3015, -121.1207)
\qbezier(151.3015, -121.1207)(153.9151, -121.3207)(156.5314, -121.5541)
\qbezier(156.5314, -121.5541)(159.1477, -121.7874)(161.7672, -122.0543)
\qbezier(161.7672, -122.0543)(164.3867, -122.3211)(167.0098, -122.6214)
\qbezier(167.0098, -122.6214)(169.6329, -122.9217)(172.2601, -123.2555)
\qbezier(172.2601, -123.2555)(174.8873, -123.5894)(177.5189, -123.9568)
\qbezier(177.5189, -123.9568)(180.1506, -124.3243)(182.7871, -124.7254)
\qbezier(182.7871, -124.7254)(185.4237, -125.1265)(188.0656, -125.5613)
\qbezier(188.0656, -125.5613)(190.7075, -125.9961)(193.3552, -126.4647)
\qbezier(193.3552, -126.4647)(196.0028, -126.9333)(198.6567, -127.4358)
\qbezier(198.6567, -127.4358)(201.3106, -127.9382)(203.9711, -128.4747)
\qbezier(203.9711, -128.4747)(206.6316, -129.0111)(209.2992, -129.5815)
\qbezier(209.2992, -129.5815)(211.9668, -130.1520)(214.6419, -130.7566)
\qbezier(214.6419, -130.7566)(217.3170, -131.3612)(220.0000, -132.0000)
\put(120.0000, -32.0000){\line(1, -1){100.0000}}
\put( 90.0000, -132.0000){\vector(1, 0){160.0000}}
\put(120.0000, -162.0000){\vector(0, 1){160.0000}}
\put(260.0000, -137.0000){\makebox(0,0)[]{\footnotesize{$\widehat{R}$}}}
\put(110.0000,  -2.0000){\makebox(0,0)[]{\footnotesize{$\widehat{T}$}}}
\put(170.0000, -142.0000){\makebox(0,0)[]{\footnotesize{$\chi = \mbox{const}$}}}
\put( 91.0000, -97.0000){\makebox(0,0)[]{\footnotesize{$\eta = \mbox{const}$}}}
\put(112.0000, -32.0000){\makebox(0,0)[]{\footnotesize{$1$}}}
\put(222.0000, -142.0000){\makebox(0,0)[]{\footnotesize{$1$}}}
\end{picture}
\vspace{0mm} \newline
{\footnotesize \sf Figure 5. Minkowski diagram for universe models with
negative spatial curvature containing dust and radiation with reference
to the conformal coordinate system of type II defined in
equation \eqref{e_17}. The diagram shows world lines
$\chi = \mbox{constant}$ and simultaneity curves
$\eta = \mbox{constant} \,$. We see that the $(\eta,\chi)$-system
contracts relative to the $(\widehat{T},\widehat{R})$-system.
The line $\widehat{T} + \widehat{R} = 1$ represents the limit
$\eta \rightarrow \infty$, $\chi \rightarrow \infty$.}
\vspace{0mm} \newline
\itm Figure 6 shows a corresponding Minkowski diagram as that of Figure 5,
but this time for a universe filled with vacuum energy. This figure provides
the picture of a universe of finite extension with a finite age, and with
attractive gravitation.
Since the world lines of the particles with $\chi = \mbox{constant}$
curve towards the $\widehat{T}$-axis, the coordinate velocity of
these particles decelerates. According to equations \eqref{e_80} and
\eqref{e_18}, the conformal time $\widehat{T} = - 1$ when the cosmic
time $t = 0 \,$, and $\widehat{T} \rightarrow 0$ when
$t \rightarrow \infty \,$, meaning that this universe will become
infinitely old. From equation \eqref{e_18} it also follows that
$\chi$ approaches infinity at the line $\widehat{T} = \widehat{R} - 1$.
Furthermore, equation \eqref{e_79} implies that $\ddot{a} > 0$, which
means that the Hubble flow has accelerated expansion.
\itm An important difference between the CFS systems of type I and II is
that in the $(T,R)$-system the free particles defining the Hubble flow
move with constant velocity along straight world lines, while in the
$(\widehat{T},\widehat{R})$-system their world lines are hyperbolae.
\itm For a negatively curved universe with radiation and dust
the line element with type II conformal coordinates takes the form
\begin{equation} \label{e_174}
ds^2 = \left[ \frl{2 \alpha \left( 1 + \widehat{T}^2 - \widehat{R}^2
- \sqrt{(1 + \widehat{T}^2 - \widehat{R}^2)^2 - 4 \widehat{T}^2}
\hs{2.0mm} \right) + 4 \beta \widehat{T}}
{(1 + \widehat{T}^2 - \widehat{R}^2)^2 - 4 \widehat{T}^2
\rule[-0mm]{0mm}{4.25mm}}
\hs{1.0mm} \right]^{\hs{-0.0mm} 2} ds_M^2
\mbox{ .}
\end{equation}
The type II conformal time at $\widehat{R} = 0$ is related to the
cosmic time by
\begin{equation} \label{e_158}
t = \alpha \left( \frl{2 \widehat{T}}{1 - \widehat{T}^2
\rule[-0mm]{0mm}{4.25mm}} - \arcsin \frl{2 \widehat{T}}{1 - \widehat{T}^2
\rule[-0mm]{0mm}{4.25mm}} \right)
+ \beta \frl{2 \widehat{T}^2}{1 - \widehat{T}^2 \rule[-0mm]{0mm}{4.25mm}}
\mbox{ .}
\end{equation}
The corresponding relationship for a LIVE dominated universe is
\begin{equation} \label{e_172}
\widehat{H}_{\Lambda} t = \ln \widehat{T}
\mbox{ .}
\end{equation}
The $\widehat{T}$-clocks goe faster than the cosmic clocks in this universe
model.
\vspace*{5mm} \newline
\begin{picture}(50,192)(-96,-182)
\qbezier(120.0000, -132.0000)(124.0038, -127.5958)(127.6672, -123.4986)
\qbezier(127.6672, -123.4986)(131.3307, -119.4014)(134.6772, -115.5849)
\qbezier(134.6772, -115.5849)(138.0238, -111.7683)(141.0750, -108.2079)
\qbezier(141.0750, -108.2079)(144.1263, -104.6475)(146.9017, -101.3204)
\qbezier(146.9017, -101.3204)(149.6772, -97.9933)(152.1947, -94.8782)
\qbezier(152.1947, -94.8782)(154.7122, -91.7630)(156.9880, -88.8397)
\qbezier(156.9880, -88.8397)(159.2637, -85.9165)(161.3123, -83.1663)
\qbezier(161.3123, -83.1663)(163.3609, -80.4162)(165.1955, -77.8216)
\qbezier(165.1955, -77.8216)(167.0302, -75.2269)(168.6626, -72.7711)
\qbezier(168.6626, -72.7711)(170.2949, -70.3153)(171.7356, -67.9825)
\qbezier(171.7356, -67.9825)(173.1763, -65.6497)(174.4345, -63.4249)
\qbezier(174.4345, -63.4249)(175.6927, -61.2002)(176.7765, -59.0692)
\qbezier(176.7765, -59.0692)(177.8603, -56.9383)(178.7766, -54.8874)
\qbezier(178.7766, -54.8874)(179.6930, -52.8365)(180.4478, -50.8525)
\qbezier(180.4478, -50.8525)(181.2026, -48.8685)(181.8008, -46.9387)
\qbezier(181.8008, -46.9387)(182.3989, -45.0089)(182.8441, -43.1209)
\qbezier(182.8441, -43.1209)(183.2894, -41.2329)(183.5847, -39.3745)
\qbezier(183.5847, -39.3745)(183.8799, -37.5161)(184.0271, -35.6754)
\qbezier(184.0271, -35.6754)(184.1742, -33.8348)(184.1742, -32.0000)
\qbezier(120.0000, -132.0000)(121.9292, -128.9134)(123.7406, -125.8989)
\qbezier(123.7406, -125.8989)(125.5521, -122.8844)(127.2485, -119.9376)
\qbezier(127.2485, -119.9376)(128.9450, -116.9908)(130.5289, -114.1073)
\qbezier(130.5289, -114.1073)(132.1129, -111.2237)(133.5868, -108.3992)
\qbezier(133.5868, -108.3992)(135.0606, -105.5747)(136.4265, -102.8049)
\qbezier(136.4265, -102.8049)(137.7925, -100.0352)(139.0525, -97.3161)
\qbezier(139.0525, -97.3161)(140.3126, -94.5970)(141.4686, -91.9245)
\qbezier(141.4686, -91.9245)(142.6246, -89.2520)(143.6784, -86.6222)
\qbezier(143.6784, -86.6222)(144.7321, -83.9923)(145.6852, -81.4012)
\qbezier(145.6852, -81.4012)(146.6382, -78.8101)(147.4919, -76.2538)
\qbezier(147.4919, -76.2538)(148.3456, -73.6975)(149.1013, -71.1723)
\qbezier(149.1013, -71.1723)(149.8570, -68.6470)(150.5158, -66.1491)
\qbezier(150.5158, -66.1491)(151.1746, -63.6511)(151.7374, -61.1767)
\qbezier(151.7374, -61.1767)(152.3003, -58.7024)(152.7681, -56.2479)
\qbezier(152.7681, -56.2479)(153.2359, -53.7934)(153.6093, -51.3551)
\qbezier(153.6093, -51.3551)(153.9826, -48.9168)(154.2622, -46.4912)
\qbezier(154.2622, -46.4912)(154.5418, -44.0655)(154.7279, -41.6488)
\qbezier(154.7279, -41.6488)(154.9140, -39.2321)(155.0070, -36.8208)
\qbezier(155.0070, -36.8208)(155.1000, -34.4095)(155.1000, -32.0000)
\qbezier(120.0000, -132.0000)(120.6388, -129.3170)(121.2434, -126.6419)
\qbezier(121.2434, -126.6419)(121.8480, -123.9668)(122.4185, -121.2992)
\qbezier(122.4185, -121.2992)(122.9889, -118.6316)(123.5253, -115.9711)
\qbezier(123.5253, -115.9711)(124.0618, -113.3106)(124.5642, -110.6567)
\qbezier(124.5642, -110.6567)(125.0667, -108.0028)(125.5353, -105.3552)
\qbezier(125.5353, -105.3552)(126.0039, -102.7075)(126.4387, -100.0656)
\qbezier(126.4387, -100.0656)(126.8735, -97.4237)(127.2746, -94.7871)
\qbezier(127.2746, -94.7871)(127.6757, -92.1506)(128.0432, -89.5189)
\qbezier(128.0432, -89.5189)(128.4106, -86.8873)(128.7445, -84.2601)
\qbezier(128.7445, -84.2601)(129.0783, -81.6329)(129.3786, -79.0098)
\qbezier(129.3786, -79.0098)(129.6789, -76.3867)(129.9457, -73.7672)
\qbezier(129.9457, -73.7672)(130.2126, -71.1477)(130.4459, -68.5314)
\qbezier(130.4459, -68.5314)(130.6793, -65.9151)(130.8793, -63.3015)
\qbezier(130.8793, -63.3015)(131.0792, -60.6880)(131.2458, -58.0768)
\qbezier(131.2458, -58.0768)(131.4124, -55.4656)(131.5456, -52.8564)
\qbezier(131.5456, -52.8564)(131.6788, -50.2471)(131.7788, -47.6393)
\qbezier(131.7788, -47.6393)(131.8787, -45.0315)(131.9453, -42.4248)
\qbezier(131.9453, -42.4248)(132.0118, -39.8180)(132.0451, -37.2120)
\qbezier(132.0451, -37.2120)(132.0784, -34.6059)(132.0784, -32.0000)
\qbezier(120.0000, -96.1742)(121.8348, -96.1742)(123.6754, -96.0271)
\qbezier(123.6754, -96.0271)(125.5161, -95.8799)(127.3745, -95.5847)
\qbezier(127.3745, -95.5847)(129.2329, -95.2894)(131.1209, -94.8441)
\qbezier(131.1209, -94.8441)(133.0089, -94.3989)(134.9387, -93.8008)
\qbezier(134.9387, -93.8008)(136.8685, -93.2026)(138.8525, -92.4478)
\qbezier(138.8525, -92.4478)(140.8365, -91.6930)(142.8874, -90.7766)
\qbezier(142.8874, -90.7766)(144.9383, -89.8603)(147.0692, -88.7765)
\qbezier(147.0692, -88.7765)(149.2002, -87.6927)(151.4249, -86.4345)
\qbezier(151.4249, -86.4345)(153.6497, -85.1763)(155.9825, -83.7356)
\qbezier(155.9825, -83.7356)(158.3153, -82.2949)(160.7711, -80.6626)
\qbezier(160.7711, -80.6626)(163.2269, -79.0302)(165.8216, -77.1955)
\qbezier(165.8216, -77.1955)(168.4162, -75.3609)(171.1663, -73.3123)
\qbezier(171.1663, -73.3123)(173.9165, -71.2637)(176.8397, -68.9880)
\qbezier(176.8397, -68.9880)(179.7630, -66.7122)(182.8782, -64.1947)
\qbezier(182.8782, -64.1947)(185.9933, -61.6772)(189.3204, -58.9017)
\qbezier(189.3204, -58.9017)(192.6475, -56.1263)(196.2079, -53.0750)
\qbezier(196.2079, -53.0750)(199.7683, -50.0238)(203.5849, -46.6772)
\qbezier(203.5849, -46.6772)(207.4014, -43.3307)(211.4986, -39.6672)
\qbezier(211.4986, -39.6672)(215.5958, -36.0038)(220.0000, -32.0000)
\qbezier(120.0000, -67.1000)(122.4095, -67.1000)(124.8208, -67.0070)
\qbezier(124.8208, -67.0070)(127.2321, -66.9140)(129.6488, -66.7279)
\qbezier(129.6488, -66.7279)(132.0655, -66.5418)(134.4912, -66.2622)
\qbezier(134.4912, -66.2622)(136.9168, -65.9826)(139.3551, -65.6093)
\qbezier(139.3551, -65.6093)(141.7934, -65.2359)(144.2479, -64.7681)
\qbezier(144.2479, -64.7681)(146.7024, -64.3003)(149.1767, -63.7374)
\qbezier(149.1767, -63.7374)(151.6511, -63.1746)(154.1491, -62.5158)
\qbezier(154.1491, -62.5158)(156.6470, -61.8570)(159.1723, -61.1013)
\qbezier(159.1723, -61.1013)(161.6975, -60.3456)(164.2538, -59.4919)
\qbezier(164.2538, -59.4919)(166.8101, -58.6382)(169.4012, -57.6852)
\qbezier(169.4012, -57.6852)(171.9923, -56.7321)(174.6222, -55.6784)
\qbezier(174.6222, -55.6784)(177.2520, -54.6246)(179.9245, -53.4686)
\qbezier(179.9245, -53.4686)(182.5970, -52.3126)(185.3161, -51.0525)
\qbezier(185.3161, -51.0525)(188.0352, -49.7925)(190.8049, -48.4265)
\qbezier(190.8049, -48.4265)(193.5747, -47.0606)(196.3992, -45.5868)
\qbezier(196.3992, -45.5868)(199.2237, -44.1129)(202.1073, -42.5289)
\qbezier(202.1073, -42.5289)(204.9908, -40.9450)(207.9376, -39.2485)
\qbezier(207.9376, -39.2485)(210.8844, -37.5521)(213.8989, -35.7406)
\qbezier(213.8989, -35.7406)(216.9134, -33.9292)(220.0000, -32.0000)
\qbezier(120.0000, -44.0784)(122.6059, -44.0784)(125.2120, -44.0451)
\qbezier(125.2120, -44.0451)(127.8180, -44.0118)(130.4248, -43.9453)
\qbezier(130.4248, -43.9453)(133.0315, -43.8787)(135.6393, -43.7788)
\qbezier(135.6393, -43.7788)(138.2471, -43.6788)(140.8564, -43.5456)
\qbezier(140.8564, -43.5456)(143.4656, -43.4124)(146.0768, -43.2458)
\qbezier(146.0768, -43.2458)(148.6880, -43.0792)(151.3015, -42.8793)
\qbezier(151.3015, -42.8793)(153.9151, -42.6793)(156.5314, -42.4459)
\qbezier(156.5314, -42.4459)(159.1477, -42.2126)(161.7672, -41.9457)
\qbezier(161.7672, -41.9457)(164.3867, -41.6789)(167.0098, -41.3786)
\qbezier(167.0098, -41.3786)(169.6329, -41.0783)(172.2601, -40.7445)
\qbezier(172.2601, -40.7445)(174.8873, -40.4106)(177.5189, -40.0432)
\qbezier(177.5189, -40.0432)(180.1506, -39.6757)(182.7871, -39.2746)
\qbezier(182.7871, -39.2746)(185.4237, -38.8735)(188.0656, -38.4387)
\qbezier(188.0656, -38.4387)(190.7075, -38.0039)(193.3552, -37.5353)
\qbezier(193.3552, -37.5353)(196.0028, -37.0667)(198.6567, -36.5642)
\qbezier(198.6567, -36.5642)(201.3106, -36.0618)(203.9711, -35.5253)
\qbezier(203.9711, -35.5253)(206.6316, -34.9889)(209.2992, -34.4185)
\qbezier(209.2992, -34.4185)(211.9668, -33.8480)(214.6419, -33.2434)
\qbezier(214.6419, -33.2434)(217.3170, -32.6388)(220.0000, -32.0000)
\put(120.0000, -132.0000){\line(1, 1){100.0000}}
\put( 90.0000, -32.0000){\vector(1, 0){160.0000}}
\put(120.0000, -162.0000){\vector(0, 1){160.0000}}
\put(260.0000, -37.0000){\makebox(0,0)[]{\footnotesize{$\widehat{R}$}}}
\put(110.0000,  -2.0000){\makebox(0,0)[]{\footnotesize{$\widehat{T}$}}}
\put(170.0000, -22.0000){\makebox(0,0)[]{\footnotesize{$\chi = \mbox{const}$}}}
\put( 91.0000, -68.0000){\makebox(0,0)[]{\footnotesize{$\eta = \mbox{const}$}}}
\put(110.0000, -132.0000){\makebox(0,0)[]{\footnotesize{$-1$}}}
\put(220.0000, -24.0000){\makebox(0,0)[]{\footnotesize{$1$}}}
\end{picture}
\vspace{0mm} \newline
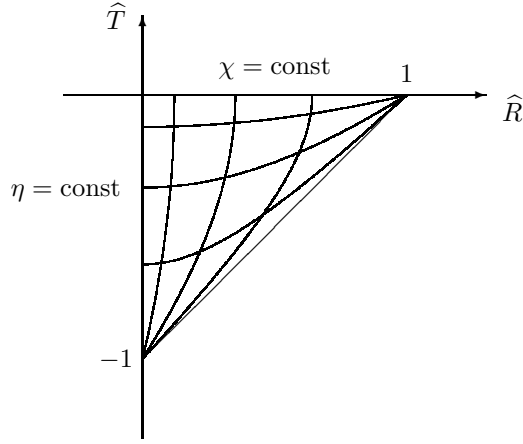
{\footnotesize \sf Figure 6. Minkowski diagram for universe models
with negative spatial curvature containing vacuum energy with reference
to the conformal coordinate system of type II defined in
equation \eqref{e_17}. The diagram shows world lines
$\chi = \mbox{constant}$ and simultaneity curves
$\eta = \mbox{constant} \,$. We see that in this case the
$(\eta,\chi)$-system expands relative to the
$(\widehat{T},\widehat{R})$-system.
The line $\widehat{T} + 1 = \widehat{R}$ represents the limit
$\eta \rightarrow 0$, $\chi \rightarrow \infty$, while $\chi = 0$
on the $\widehat{T}$-axis, and $\eta \rightarrow \infty$ on the
$\widehat{R}$-axis.}
\vspace{0mm} \newline
\itm From equations \eqref{e_80}, \eqref{e_18} and \eqref{e_20} it follows
that with the type II conformal coordinates, the line element of a
negatively curved universe model dominated by LIVE takes the form
\begin{equation} \label{e_137}
ds^2 = \frl{1}{\widehat{H}_{\Lambda}^2 \widehat{T}^2
\rule[-0mm]{0mm}{4.25mm}} ds_M^2
\mbox{ .}
\end{equation}
\vspace{-4.5mm}
\itm The coordinate transformation between the present conformal
coordinates and those introduced in equation \eqref{e_33} is
\begin{equation} \label{e_13}
\widehat{T} = \frl{(T^2 - T_i^2)
- R^2}{(T + T_i)^2 - R^2}
\mbox{\hspace{2mm} , \hspace{3mm}}
\widehat{R} = \frl{2 T_i R}{(T + T_i)^2 - R^2}
\mbox{ .}
\end{equation}
%
%
%
\vspace{5mm} \newline
{\bf 8. A third type of conformal coordinates for universe models
with negative spatial curvature}
\vspace{3mm} \newline
For universe models with negative spatial curvature one may also
introduce a third type of conformal coordinates
$(\widetilde{T},\widetilde{R})$, this time by choosing
$b = \coth (a / 2)$, $c = 1 / \sinh^{2} (a / 2)$ and
$d = \tanh (a / 2)$ in equation \eqref{e_407}. This gives the
generating function
\begin{equation} \label{e_205}
f(x) = - \coth (x / 2)
\mbox{ .}
\end{equation}
Then the transformation \eqref{e_406} takes the form
\begin{equation} \label{e_133}
\widetilde{T} = \frac{\sinh \eta}{\cosh \chi - \cosh \eta}
\mbox{\hspace{2mm} , \hspace{3mm}}
\widetilde{R} = \frac{\sinh \chi}{\cosh \eta - \cosh \chi}
\mbox{ .}
\end{equation}
If the universe model begins at $\eta = 0$, this transformation
maps the region $0 < \eta < \chi$ onto the region
$0 < \widetilde{T} < -1 - \widetilde{R}$, and the region
$0 < \chi < \eta$ onto the region
$0 < \widetilde{R} < -1 - \widetilde{T}$.
On the other hand, if the universe is infinitely old,
the region $- \chi < \eta < 0$ is mapped onto the
region $1 + \widetilde{R} < \widetilde{T} < 0$, and the region
$0 < \chi < -\eta$ onto the region
$0 < \widetilde{R} < \widetilde{T} - 1$.
\itm The inverse transformation is
\begin{equation} \label{e_134}
\coth \eta = \frl{\left(\widetilde{R}^{2} - \widetilde{T}^{2} \right) - 1}
{2 \widetilde{T}
\rule[-0mm]{0mm}{4.25mm}}
\mbox{\hspace{2mm} , \hspace{3mm}}
\coth \chi = \frl{\left(\widetilde{T}^{2} - \widetilde{R}^{2} \right) - 1}
{2 \widetilde{R}
\rule[-0mm]{0mm}{4.25mm}}
\mbox{ .}
\end{equation}
From equation \eqref{e_409} it follows that the line element again
takes the form \eqref{e_20} with $(\widehat{T},\widehat{R})$
replaced by $(\widetilde{T},\widetilde{R})$.
As drawn in the $(\widetilde{T},\widetilde{R})$ spacetime diagram in
Figure 7 and 8, both the world lines of points with $\chi = {\chi}_0$
and the simultaneity curves $\eta = {\eta}_0$ are hyperbolae, given
respectively by
\begin{equation} \label{e_179}
(\widetilde{R} + a_1)^2 - \widetilde{T}^2 = a_1^2 - 1
\mbox{\hspace{2mm} and \hspace{3mm}}
(\widetilde{T} + b_1)^2 - \widetilde{R}^2 = b_1^2 - 1
\mbox{ ,}
\end{equation}
where $a_1 = \coth {\chi}_0$ and $b_1 = \coth {\eta}_0$.
Again these equations are valid both for universe models dominated
by radiation and dust and by LIVE, the only difference being that
${\eta}_0 > 0$ with radiation and dust, and ${\eta}_0 < 0$ with LIVE.
In the present case the line element is still given by
equation \eqref{e_137}.
%
\vspace*{5mm} \newline
\begin{picture}(50,248)(-96,-195)
%
%
\qbezier(129.1304, -71.5652)(129.6814, -73.1631)(130.3234, -74.7946)
\qbezier(130.3234, -74.7946)(130.9653, -76.4261)(131.7002, -78.0963)
\qbezier(131.7002, -78.0963)(132.4352, -79.7666)(133.2654, -81.4809)
\qbezier(133.2654, -81.4809)(134.0956, -83.1953)(135.0238, -84.9591)
\qbezier(135.0238, -84.9591)(135.9520, -86.7230)(136.9811, -88.5419)
\qbezier(136.9811, -88.5419)(138.0102, -90.3609)(139.1435, -92.2408)
\qbezier(139.1435, -92.2408)(140.2767, -94.1207)(141.5178, -96.0674)
\qbezier(141.5178, -96.0674)(142.7588, -98.0142)(144.1115, -100.0339)
\qbezier(144.1115, -100.0339)(145.4643, -102.0537)(146.9330, -104.1530)
\qbezier(146.9330, -104.1530)(148.4017, -106.2522)(149.9911, -108.4376)
\qbezier(149.9911, -108.4376)(151.5805, -110.6229)(153.2956, -112.9014)
\qbezier(153.2956, -112.9014)(155.0107, -115.1798)(156.8570, -117.5585)
\qbezier(156.8570, -117.5585)(158.7032, -119.9372)(160.6865, -122.4238)
\qbezier(160.6865, -122.4238)(162.6698, -124.9104)(164.7964, -127.5127)
\qbezier(164.7964, -127.5127)(166.9230, -130.1150)(169.1996, -132.8413)
\qbezier(169.1996, -132.8413)(171.4763, -135.5676)(173.9102, -138.4266)
\qbezier(173.9102, -138.4266)(176.3441, -141.2856)(178.9431, -144.2863)
\qbezier(178.9431, -144.2863)(181.5421, -147.2870)(184.3143, -150.4390)
\qbezier(184.3143, -150.4390)(187.0866, -153.5909)(190.0409, -156.9042)
\qbezier(129.1304, -71.5652)(129.3968, -73.7497)(129.7228, -75.9422)
\qbezier(129.7228, -75.9422)(130.0488, -78.1348)(130.4347, -80.3370)
\qbezier(130.4347, -80.3370)(130.8205, -82.5392)(131.2665, -84.7528)
\qbezier(131.2665, -84.7528)(131.7124, -86.9664)(132.2188, -89.1929)
\qbezier(132.2188, -89.1929)(132.7253, -91.4194)(133.2925, -93.6606)
\qbezier(133.2925, -93.6606)(133.8598, -95.9017)(134.4883, -98.1591)
\qbezier(134.4883, -98.1591)(135.1168, -100.4166)(135.8070, -102.6919)
\qbezier(135.8070, -102.6919)(136.4972, -104.9673)(137.2496, -107.2623)
\qbezier(137.2496, -107.2623)(138.0021, -109.5574)(138.8173, -111.8737)
\qbezier(138.8173, -111.8737)(139.6326, -114.1901)(140.5112, -116.5296)
\qbezier(140.5112, -116.5296)(141.3898, -118.8690)(142.3325, -121.2332)
\qbezier(142.3325, -121.2332)(143.2752, -123.5975)(144.2826, -125.9882)
\qbezier(144.2826, -125.9882)(145.2900, -128.3790)(146.3629, -130.7981)
\qbezier(146.3629, -130.7981)(147.4359, -133.2173)(148.5751, -135.6665)
\qbezier(148.5751, -135.6665)(149.7143, -138.1157)(150.9206, -140.5969)
\qbezier(150.9206, -140.5969)(152.1269, -143.0781)(153.4013, -145.5930)
\qbezier(153.4013, -145.5930)(154.6756, -148.1079)(156.0190, -150.6585)
\qbezier(156.0190, -150.6585)(157.3623, -153.2091)(158.7756, -155.7972)
\qbezier(158.7756, -155.7972)(160.1889, -158.3854)(161.6732, -161.0129)
\qbezier(203.7678, -151.2356)(198.8902, -146.4301)(194.5934, -142.2153)
\qbezier(194.5934, -142.2153)(190.2966, -138.0005)(186.5090, -134.3062)
\qbezier(186.5090, -134.3062)(182.7213, -130.6119)(179.3799, -127.3766)
\qbezier(179.3799, -127.3766)(176.0384, -124.1413)(173.0875, -121.3112)
\qbezier(173.0875, -121.3112)(170.1366, -118.4810)(167.5270, -116.0089)
\qbezier(167.5270, -116.0089)(164.9175, -113.5367)(162.6059, -111.3814)
\qbezier(162.6059, -111.3814)(160.2943, -109.2261)(158.2421, -107.3518)
\qbezier(158.2421, -107.3518)(156.1900, -105.4774)(154.3631, -103.8529)
\qbezier(154.3631, -103.8529)(152.5362, -102.2283)(150.9042, -100.8264)
\qbezier(150.9042, -100.8264)(149.2721, -99.4245)(147.8078, -98.2219)
\qbezier(147.8078, -98.2219)(146.3435, -97.0194)(145.0224, -95.9962)
\qbezier(145.0224, -95.9962)(143.7014, -94.9730)(142.5017, -94.1121)
\qbezier(142.5017, -94.1121)(141.3019, -93.2512)(140.2035, -92.5383)
\qbezier(140.2035, -92.5383)(139.1051, -91.8253)(138.0898, -91.2485)
\qbezier(138.0898, -91.2485)(137.0744, -90.6717)(136.1252, -90.2213)
\qbezier(136.1252, -90.2213)(135.1760, -89.7710)(134.2771, -89.4396)
\qbezier(134.2771, -89.4396)(133.3782, -89.1082)(132.5146, -88.8903)
\qbezier(132.5146, -88.8903)(131.6511, -88.6724)(130.8086, -88.5644)
\qbezier(130.8086, -88.5644)(129.9660, -88.4564)(129.1304, -88.4564)
\qbezier(198.0474, -156.4256)(194.6632, -153.2359)(191.5603, -150.3465)
\qbezier(191.5603, -150.3465)(188.4574, -147.4571)(185.6092, -144.8431)
\qbezier(185.6092, -144.8431)(182.7610, -142.2291)(180.1431, -139.8681)
\qbezier(180.1431, -139.8681)(177.5252, -137.5071)(175.1150, -135.3788)
\qbezier(175.1150, -135.3788)(172.7049, -133.2505)(170.4818, -131.3367)
\qbezier(170.4818, -131.3367)(168.2588, -129.4228)(166.2037, -127.7069)
\qbezier(166.2037, -127.7069)(164.1486, -125.9911)(162.2439, -124.4585)
\qbezier(162.2439, -124.4585)(160.3392, -122.9259)(158.5684, -121.5635)
\qbezier(158.5684, -121.5635)(156.7977, -120.2010)(155.1457, -118.9969)
\qbezier(155.1457, -118.9969)(153.4938, -117.7929)(151.9465, -116.7369)
\qbezier(151.9465, -116.7369)(150.3991, -115.6809)(148.9431, -114.7639)
\qbezier(148.9431, -114.7639)(147.4870, -113.8469)(146.1098, -113.0611)
\qbezier(146.1098, -113.0611)(144.7326, -112.2752)(143.4224, -111.6138)
\qbezier(143.4224, -111.6138)(142.1121, -110.9523)(140.8576, -110.4095)
\qbezier(140.8576, -110.4095)(139.6031, -109.8668)(138.3936, -109.4380)
\qbezier(138.3936, -109.4380)(137.1840, -109.0093)(136.0091, -108.6910)
\qbezier(136.0091, -108.6910)(134.8341, -108.3726)(133.6836, -108.1619)
\qbezier(133.6836, -108.1619)(132.5332, -107.9511)(131.3973, -107.8462)
\qbezier(131.3973, -107.8462)(130.2614, -107.7413)(129.1304, -107.7413)
\qbezier(179.0697, -158.0028)(177.2808, -156.5838)(175.5697, -155.2643)
\qbezier(175.5697, -155.2643)(173.8586, -153.9448)(172.2198, -152.7205)
\qbezier(172.2198, -152.7205)(170.5810, -151.4962)(169.0092, -150.3631)
\qbezier(169.0092, -150.3631)(167.4374, -149.2300)(165.9275, -148.1845)
\qbezier(165.9275, -148.1845)(164.4176, -147.1390)(162.9647, -146.1776)
\qbezier(162.9647, -146.1776)(161.5118, -145.2163)(160.1112, -144.3360)
\qbezier(160.1112, -144.3360)(158.7106, -143.4557)(157.3579, -142.6537)
\qbezier(157.3579, -142.6537)(156.0051, -141.8516)(154.6958, -141.1252)
\qbezier(154.6958, -141.1252)(153.3864, -140.3988)(152.1163, -139.7457)
\qbezier(152.1163, -139.7457)(150.8461, -139.0926)(149.6111, -138.5107)
\qbezier(149.6111, -138.5107)(148.3760, -137.9288)(147.1721, -137.4162)
\qbezier(147.1721, -137.4162)(145.9681, -136.9035)(144.7914, -136.4586)
\qbezier(144.7914, -136.4586)(143.6146, -136.0136)(142.4613, -135.6348)
\qbezier(142.4613, -135.6348)(141.3080, -135.2560)(140.1743, -134.9422)
\qbezier(140.1743, -134.9422)(139.0407, -134.6284)(137.9230, -134.3786)
\qbezier(137.9230, -134.3786)(136.8054, -134.1288)(135.7001, -133.9421)
\qbezier(135.7001, -133.9421)(134.5949, -133.7554)(133.4985, -133.6313)
\qbezier(133.4985, -133.6313)(132.4021, -133.5072)(131.3109, -133.4453)
\qbezier(131.3109, -133.4453)(130.2198, -133.3833)(129.1304, -133.3833)
\qbezier( 24.2263,   8.9104)( 27.5395,   5.9561)( 30.6915,   3.1839)
\qbezier( 30.6915,   3.1839)( 33.8434,   0.4116)( 36.8442,  -2.1873)
\qbezier( 36.8442,  -2.1873)( 39.8449,  -4.7863)( 42.7038,  -7.2202)
\qbezier( 42.7038,  -7.2202)( 45.5628,  -9.6542)( 48.2891, -11.9308)
\qbezier( 48.2891, -11.9308)( 51.0154, -14.2075)( 53.6178, -16.3341)
\qbezier( 53.6178, -16.3341)( 56.2201, -18.4607)( 58.7066, -20.4439)
\qbezier( 58.7066, -20.4439)( 61.1932, -22.4272)( 63.5719, -24.2735)
\qbezier( 63.5719, -24.2735)( 65.9506, -26.1197)( 68.2291, -27.8348)
\qbezier( 68.2291, -27.8348)( 70.5075, -29.5499)( 72.6929, -31.1393)
\qbezier( 72.6929, -31.1393)( 74.8782, -32.7287)( 76.9775, -34.1974)
\qbezier( 76.9775, -34.1974)( 79.0767, -35.6662)( 81.0965, -37.0189)
\qbezier( 81.0965, -37.0189)( 83.1163, -38.3716)( 85.0630, -39.6127)
\qbezier( 85.0630, -39.6127)( 87.0098, -40.8537)( 88.8896, -41.9870)
\qbezier( 88.8896, -41.9870)( 90.7695, -43.1202)( 92.5885, -44.1493)
\qbezier( 92.5885, -44.1493)( 94.4075, -45.1784)( 96.1713, -46.1066)
\qbezier( 96.1713, -46.1066)( 97.9352, -47.0348)( 99.6495, -47.8650)
\qbezier( 99.6495, -47.8650)(101.3639, -48.6953)(103.0341, -49.4302)
\qbezier(103.0341, -49.4302)(104.7044, -50.1651)(106.3358, -50.8071)
\qbezier(106.3358, -50.8071)(107.9673, -51.4490)(109.5652, -52.0000)
\qbezier( 20.1175, -19.4572)( 22.7451, -20.9415)( 25.3332, -22.3548)
\qbezier( 25.3332, -22.3548)( 27.9213, -23.7681)( 30.4719, -25.1115)
\qbezier( 30.4719, -25.1115)( 33.0225, -26.4548)( 35.5374, -27.7291)
\qbezier( 35.5374, -27.7291)( 38.0524, -29.0035)( 40.5335, -30.2098)
\qbezier( 40.5335, -30.2098)( 43.0147, -31.4162)( 45.4640, -32.5554)
\qbezier( 45.4640, -32.5554)( 47.9132, -33.6946)( 50.3323, -34.7675)
\qbezier( 50.3323, -34.7675)( 52.7514, -35.8404)( 55.1422, -36.8478)
\qbezier( 55.1422, -36.8478)( 57.5330, -37.8552)( 59.8972, -38.7979)
\qbezier( 59.8972, -38.7979)( 62.2614, -39.7406)( 64.6009, -40.6192)
\qbezier( 64.6009, -40.6192)( 66.9403, -41.4979)( 69.2567, -42.3131)
\qbezier( 69.2567, -42.3131)( 71.5731, -43.1283)( 73.8681, -43.8808)
\qbezier( 73.8681, -43.8808)( 76.1631, -44.6332)( 78.4385, -45.3235)
\qbezier( 78.4385, -45.3235)( 80.7139, -46.0137)( 82.9713, -46.6422)
\qbezier( 82.9713, -46.6422)( 85.2287, -47.2707)( 87.4699, -47.8379)
\qbezier( 87.4699, -47.8379)( 89.7110, -48.4052)( 91.9375, -48.9116)
\qbezier( 91.9375, -48.9116)( 94.1641, -49.4180)( 96.3776, -49.8640)
\qbezier( 96.3776, -49.8640)( 98.5912, -50.3099)(100.7934, -50.6958)
\qbezier(100.7934, -50.6958)(102.9957, -51.0816)(105.1882, -51.4076)
\qbezier(105.1882, -51.4076)(107.3807, -51.7336)(109.5652, -52.0000)
\qbezier( 92.6741, -52.0000)( 92.6741, -51.1644)( 92.5660, -50.3219)
\qbezier( 92.5660, -50.3219)( 92.4580, -49.4793)( 92.2401, -48.6158)
\qbezier( 92.2401, -48.6158)( 92.0222, -47.7523)( 91.6908, -46.8534)
\qbezier( 91.6908, -46.8534)( 91.3595, -45.9545)( 90.9091, -45.0053)
\qbezier( 90.9091, -45.0053)( 90.4588, -44.0560)( 89.8819, -43.0407)
\qbezier( 89.8819, -43.0407)( 89.3051, -42.0253)( 88.5922, -40.9269)
\qbezier( 88.5922, -40.9269)( 87.8792, -39.8285)( 87.0183, -38.6288)
\qbezier( 87.0183, -38.6288)( 86.1574, -37.4290)( 85.1342, -36.1080)
\qbezier( 85.1342, -36.1080)( 84.1110, -34.7870)( 82.9085, -33.3226)
\qbezier( 82.9085, -33.3226)( 81.7060, -31.8583)( 80.3041, -30.2263)
\qbezier( 80.3041, -30.2263)( 78.9022, -28.5942)( 77.2776, -26.7674)
\qbezier( 77.2776, -26.7674)( 75.6530, -24.9405)( 73.7786, -22.8883)
\qbezier( 73.7786, -22.8883)( 71.9043, -20.8361)( 69.7490, -18.5245)
\qbezier( 69.7490, -18.5245)( 67.5937, -16.2129)( 65.1216, -13.6034)
\qbezier( 65.1216, -13.6034)( 62.6494, -10.9938)( 59.8193,  -8.0429)
\qbezier( 59.8193,  -8.0429)( 56.9891,  -5.0920)( 53.7538,  -1.7505)
\qbezier( 53.7538,  -1.7505)( 50.5185,   1.5909)( 46.8242,   5.3785)
\qbezier( 46.8242,   5.3785)( 43.1300,   9.1661)( 38.9152,  13.4630)
\qbezier( 38.9152,  13.4630)( 34.7004,  17.7598)( 29.8949,  22.6374)
\qbezier( 73.3891, -52.0000)( 73.3891, -50.8690)( 73.2842, -49.7331)
\qbezier( 73.2842, -49.7331)( 73.1793, -48.5973)( 72.9686, -47.4468)
\qbezier( 72.9686, -47.4468)( 72.7578, -46.2963)( 72.4395, -45.1214)
\qbezier( 72.4395, -45.1214)( 72.1211, -43.9464)( 71.6924, -42.7369)
\qbezier( 71.6924, -42.7369)( 71.2637, -41.5273)( 70.7209, -40.2728)
\qbezier( 70.7209, -40.2728)( 70.1781, -39.0183)( 69.5167, -37.7081)
\qbezier( 69.5167, -37.7081)( 68.8552, -36.3978)( 68.0693, -35.0206)
\qbezier( 68.0693, -35.0206)( 67.2835, -33.6434)( 66.3665, -32.1874)
\qbezier( 66.3665, -32.1874)( 65.4495, -30.7313)( 64.3935, -29.1840)
\qbezier( 64.3935, -29.1840)( 63.3376, -27.6366)( 62.1335, -25.9847)
\qbezier( 62.1335, -25.9847)( 60.9294, -24.3327)( 59.5670, -22.5620)
\qbezier( 59.5670, -22.5620)( 58.2045, -20.7913)( 56.6719, -18.8865)
\qbezier( 56.6719, -18.8865)( 55.1393, -16.9818)( 53.4235, -14.9267)
\qbezier( 53.4235, -14.9267)( 51.7076, -12.8717)( 49.7938, -10.6486)
\qbezier( 49.7938, -10.6486)( 47.8799,  -8.4255)( 45.7516,  -6.0154)
\qbezier( 45.7516,  -6.0154)( 43.6233,  -3.6052)( 41.2623,  -0.9873)
\qbezier( 41.2623,  -0.9873)( 38.9013,   1.6306)( 36.2873,   4.4788)
\qbezier( 36.2873,   4.4788)( 33.6734,   7.3270)( 30.7839,  10.4299)
\qbezier( 30.7839,  10.4299)( 27.8945,  13.5327)( 24.7048,  16.9170)
\qbezier( 47.7471, -52.0000)( 47.7471, -50.9106)( 47.6852, -49.8195)
\qbezier( 47.6852, -49.8195)( 47.6232, -48.7284)( 47.4991, -47.6320)
\qbezier( 47.4991, -47.6320)( 47.3750, -46.5356)( 47.1883, -45.4303)
\qbezier( 47.1883, -45.4303)( 47.0017, -44.3251)( 46.7518, -43.2074)
\qbezier( 46.7518, -43.2074)( 46.5020, -42.0898)( 46.1882, -40.9561)
\qbezier( 46.1882, -40.9561)( 45.8744, -39.8225)( 45.4956, -38.6691)
\qbezier( 45.4956, -38.6691)( 45.1169, -37.5158)( 44.6719, -36.3390)
\qbezier( 44.6719, -36.3390)( 44.2269, -35.1623)( 43.7143, -33.9584)
\qbezier( 43.7143, -33.9584)( 43.2016, -32.7544)( 42.6197, -31.5193)
\qbezier( 42.6197, -31.5193)( 42.0378, -30.2843)( 41.3847, -29.0142)
\qbezier( 41.3847, -29.0142)( 40.7316, -27.7440)( 40.0052, -26.4347)
\qbezier( 40.0052, -26.4347)( 39.2788, -25.1253)( 38.4768, -23.7726)
\qbezier( 38.4768, -23.7726)( 37.6747, -22.4198)( 36.7944, -21.0192)
\qbezier( 36.7944, -21.0192)( 35.9142, -19.6187)( 34.9528, -18.1658)
\qbezier( 34.9528, -18.1658)( 33.9915, -16.7129)( 32.9460, -15.2030)
\qbezier( 32.9460, -15.2030)( 31.9004, -13.6930)( 30.7673, -12.1212)
\qbezier( 30.7673, -12.1212)( 29.6342, -10.5494)( 28.4099,  -8.9106)
\qbezier( 28.4099,  -8.9106)( 27.1856,  -7.2718)( 25.8661,  -5.5607)
\qbezier( 25.8661,  -5.5607)( 24.5466,  -3.8496)( 23.1277,  -2.0607)
\put( 29.3478,  28.2174){\line(1, -1){180.0000}}
\put( 21.5217, -52.0000){\vector(1, 0){195.6522}}
\put(129.1304, -159.6087){\vector(0, 1){195.6522}}
\put(223.0435, -54.9348){\makebox(0,0)[]{\footnotesize{$\widetilde{R}$}}}
\put(123.2609,  37.0217){\makebox(0,0)[]{\footnotesize{$\widetilde{T}$}}}
\put(160.4348, -172.3261){\makebox(0,0)[]{\footnotesize{$\chi = \mbox{const}$}}}
\put(219.1304, -166.4565){\makebox(0,0)[]{\footnotesize{$\eta = \mbox{const}$}}}
\put( -6.8478, -20.6957){\makebox(0,0)[]{\footnotesize{$\eta = \mbox{const}$}}}
\put( -2.9348,  18.4348){\makebox(0,0)[]{\footnotesize{$\chi = \mbox{const}$}}}
\put(140.8696, -71.5652){\makebox(0,0)[]{\footnotesize{$-1$}}}
\put(115.4348, -46.1304){\makebox(0,0)[]{\footnotesize{$-1$}}}
\end{picture}
\vspace{0mm} \newline
{\footnotesize \sf Figure 7. Minkowski diagram for universe models with
negative spatial curvature containing dust and radiation with reference
to the conformal coordinate system of type III defined in
equation \eqref{e_133}. The diagram shows world lines
$\chi = \mbox{constant}$ and simultaneity curves
$\eta = \mbox{constant} \,$. The line
$\widetilde{T} + \widetilde{R} + 1 = 0$ represents the limit
$\chi \rightarrow \infty$, $\eta \rightarrow \infty$.
Furthermore $\eta \rightarrow 0$ on the $\widetilde{R}$-axis,
and $\eta \rightarrow \infty$ at the point $\widetilde{T} = -1$
on the $\widetilde{T}$-axis.}
\vspace{0mm} \newline
\newpage
\begin{picture}(50,232)(-96,-177)
\qbezier(190.0409,  33.3390)(187.0866,  30.0257)(184.3143,  26.8737)
\qbezier(184.3143,  26.8737)(181.5421,  23.7218)(178.9431,  20.7211)
\qbezier(178.9431,  20.7211)(176.3441,  17.7203)(173.9102,  14.8614)
\qbezier(173.9102,  14.8614)(171.4763,  12.0024)(169.1996,   9.2761)
\qbezier(169.1996,   9.2761)(166.9230,   6.5498)(164.7964,   3.9475)
\qbezier(164.7964,   3.9475)(162.6698,   1.3451)(160.6865,  -1.1414)
\qbezier(160.6865,  -1.1414)(158.7032,  -3.6280)(156.8570,  -6.0067)
\qbezier(156.8570,  -6.0067)(155.0107,  -8.3854)(153.2956, -10.6639)
\qbezier(153.2956, -10.6639)(151.5805, -12.9423)(149.9911, -15.1276)
\qbezier(149.9911, -15.1276)(148.4017, -17.3130)(146.9330, -19.4123)
\qbezier(146.9330, -19.4123)(145.4643, -21.5115)(144.1115, -23.5313)
\qbezier(144.1115, -23.5313)(142.7588, -25.5511)(141.5178, -27.4978)
\qbezier(141.5178, -27.4978)(140.2767, -29.4445)(139.1435, -31.3244)
\qbezier(139.1435, -31.3244)(138.0102, -33.2043)(136.9811, -35.0233)
\qbezier(136.9811, -35.0233)(135.9520, -36.8422)(135.0238, -38.6061)
\qbezier(135.0238, -38.6061)(134.0956, -40.3700)(133.2654, -42.0843)
\qbezier(133.2654, -42.0843)(132.4352, -43.7986)(131.7002, -45.4689)
\qbezier(131.7002, -45.4689)(130.9653, -47.1391)(130.3234, -48.7706)
\qbezier(130.3234, -48.7706)(129.6814, -50.4021)(129.1304, -52.0000)
\qbezier(161.6732,  37.4477)(160.1889,  34.8202)(158.7756,  32.2320)
\qbezier(158.7756,  32.2320)(157.3623,  29.6439)(156.0190,  27.0933)
\qbezier(156.0190,  27.0933)(154.6756,  24.5427)(153.4013,  22.0278)
\qbezier(153.4013,  22.0278)(152.1269,  19.5128)(150.9206,  17.0317)
\qbezier(150.9206,  17.0317)(149.7143,  14.5505)(148.5751,  12.1013)
\qbezier(148.5751,  12.1013)(147.4359,   9.6520)(146.3629,   7.2329)
\qbezier(146.3629,   7.2329)(145.2900,   4.8138)(144.2826,   2.4230)
\qbezier(144.2826,   2.4230)(143.2752,   0.0322)(142.3325,  -2.3320)
\qbezier(142.3325,  -2.3320)(141.3898,  -4.6962)(140.5112,  -7.0357)
\qbezier(140.5112,  -7.0357)(139.6326,  -9.3751)(138.8173, -11.6915)
\qbezier(138.8173, -11.6915)(138.0021, -14.0078)(137.2496, -16.3029)
\qbezier(137.2496, -16.3029)(136.4972, -18.5979)(135.8070, -20.8733)
\qbezier(135.8070, -20.8733)(135.1168, -23.1487)(134.4883, -25.4061)
\qbezier(134.4883, -25.4061)(133.8598, -27.6635)(133.2925, -29.9047)
\qbezier(133.2925, -29.9047)(132.7253, -32.1458)(132.2188, -34.3723)
\qbezier(132.2188, -34.3723)(131.7124, -36.5989)(131.2665, -38.8124)
\qbezier(131.2665, -38.8124)(130.8205, -41.0260)(130.4347, -43.2282)
\qbezier(130.4347, -43.2282)(130.0488, -45.4304)(129.7228, -47.6230)
\qbezier(129.7228, -47.6230)(129.3968, -49.8155)(129.1304, -52.0000)
\qbezier(129.1304, -35.1089)(129.9660, -35.1089)(130.8086, -35.0008)
\qbezier(130.8086, -35.0008)(131.6511, -34.8928)(132.5146, -34.6749)
\qbezier(132.5146, -34.6749)(133.3782, -34.4570)(134.2771, -34.1256)
\qbezier(134.2771, -34.1256)(135.1760, -33.7943)(136.1252, -33.3439)
\qbezier(136.1252, -33.3439)(137.0744, -32.8936)(138.0898, -32.3167)
\qbezier(138.0898, -32.3167)(139.1051, -31.7399)(140.2035, -31.0269)
\qbezier(140.2035, -31.0269)(141.3019, -30.3140)(142.5017, -29.4531)
\qbezier(142.5017, -29.4531)(143.7014, -28.5922)(145.0224, -27.5690)
\qbezier(145.0224, -27.5690)(146.3435, -26.5458)(147.8078, -25.3433)
\qbezier(147.8078, -25.3433)(149.2721, -24.1407)(150.9042, -22.7388)
\qbezier(150.9042, -22.7388)(152.5362, -21.3370)(154.3631, -19.7124)
\qbezier(154.3631, -19.7124)(156.1900, -18.0878)(158.2421, -16.2134)
\qbezier(158.2421, -16.2134)(160.2943, -14.3391)(162.6059, -12.1838)
\qbezier(162.6059, -12.1838)(164.9175, -10.0285)(167.5270,  -7.5563)
\qbezier(167.5270,  -7.5563)(170.1366,  -5.0842)(173.0875,  -2.2540)
\qbezier(173.0875,  -2.2540)(176.0384,   0.5761)(179.3799,   3.8114)
\qbezier(179.3799,   3.8114)(182.7213,   7.0467)(186.5090,  10.7410)
\qbezier(186.5090,  10.7410)(190.2966,  14.4353)(194.5934,  18.6501)
\qbezier(194.5934,  18.6501)(198.8902,  22.8649)(203.7678,  27.6704)
\qbezier(129.1304, -15.8239)(130.2614, -15.8239)(131.3973, -15.7190)
\qbezier(131.3973, -15.7190)(132.5332, -15.6141)(133.6836, -15.4034)
\qbezier(133.6836, -15.4034)(134.8341, -15.1926)(136.0091, -14.8743)
\qbezier(136.0091, -14.8743)(137.1840, -14.5559)(138.3936, -14.1272)
\qbezier(138.3936, -14.1272)(139.6031, -13.6984)(140.8576, -13.1557)
\qbezier(140.8576, -13.1557)(142.1121, -12.6129)(143.4224, -11.9514)
\qbezier(143.4224, -11.9514)(144.7326, -11.2900)(146.1098, -10.5041)
\qbezier(146.1098, -10.5041)(147.4870,  -9.7183)(148.9431,  -8.8013)
\qbezier(148.9431,  -8.8013)(150.3991,  -7.8843)(151.9465,  -6.8283)
\qbezier(151.9465,  -6.8283)(153.4938,  -5.7723)(155.1457,  -4.5683)
\qbezier(155.1457,  -4.5683)(156.7977,  -3.3642)(158.5684,  -2.0017)
\qbezier(158.5684,  -2.0017)(160.3392,  -0.6393)(162.2439,   0.8933)
\qbezier(162.2439,   0.8933)(164.1486,   2.4259)(166.2037,   4.1417)
\qbezier(166.2037,   4.1417)(168.2588,   5.8576)(170.4818,   7.7714)
\qbezier(170.4818,   7.7714)(172.7049,   9.6853)(175.1150,  11.8136)
\qbezier(175.1150,  11.8136)(177.5252,  13.9419)(180.1431,  16.3029)
\qbezier(180.1431,  16.3029)(182.7610,  18.6639)(185.6092,  21.2779)
\qbezier(185.6092,  21.2779)(188.4574,  23.8919)(191.5603,  26.7813)
\qbezier(191.5603,  26.7813)(194.6632,  29.6707)(198.0474,  32.8604)
\qbezier(129.1304,   9.8181)(130.2198,   9.8181)(131.3109,   9.8801)
\qbezier(131.3109,   9.8801)(132.4021,   9.9420)(133.4985,  10.0661)
\qbezier(133.4985,  10.0661)(134.5949,  10.1902)(135.7001,  10.3769)
\qbezier(135.7001,  10.3769)(136.8054,  10.5636)(137.9230,  10.8134)
\qbezier(137.9230,  10.8134)(139.0407,  11.0632)(140.1743,  11.3770)
\qbezier(140.1743,  11.3770)(141.3080,  11.6908)(142.4613,  12.0696)
\qbezier(142.4613,  12.0696)(143.6146,  12.4484)(144.7914,  12.8933)
\qbezier(144.7914,  12.8933)(145.9681,  13.3383)(147.1721,  13.8509)
\qbezier(147.1721,  13.8509)(148.3760,  14.3636)(149.6111,  14.9455)
\qbezier(149.6111,  14.9455)(150.8461,  15.5274)(152.1163,  16.1805)
\qbezier(152.1163,  16.1805)(153.3864,  16.8336)(154.6958,  17.5600)
\qbezier(154.6958,  17.5600)(156.0051,  18.2864)(157.3579,  19.0885)
\qbezier(157.3579,  19.0885)(158.7106,  19.8905)(160.1112,  20.7708)
\qbezier(160.1112,  20.7708)(161.5118,  21.6510)(162.9647,  22.6124)
\qbezier(162.9647,  22.6124)(164.4176,  23.5737)(165.9275,  24.6193)
\qbezier(165.9275,  24.6193)(167.4374,  25.6648)(169.0092,  26.7979)
\qbezier(169.0092,  26.7979)(170.5810,  27.9310)(172.2198,  29.1553)
\qbezier(172.2198,  29.1553)(173.8586,  30.3796)(175.5697,  31.6991)
\qbezier(175.5697,  31.6991)(177.2808,  33.0186)(179.0697,  34.4376)
\qbezier(109.5652, -71.5652)(107.9673, -72.1162)(106.3358, -72.7582)
\qbezier(106.3358, -72.7582)(104.7044, -73.4001)(103.0341, -74.1350)
\qbezier(103.0341, -74.1350)(101.3639, -74.8699)( 99.6495, -75.7002)
\qbezier( 99.6495, -75.7002)( 97.9352, -76.5304)( 96.1713, -77.4586)
\qbezier( 96.1713, -77.4586)( 94.4075, -78.3868)( 92.5885, -79.4159)
\qbezier( 92.5885, -79.4159)( 90.7695, -80.4450)( 88.8896, -81.5783)
\qbezier( 88.8896, -81.5783)( 87.0098, -82.7115)( 85.0630, -83.9526)
\qbezier( 85.0630, -83.9526)( 83.1163, -85.1936)( 81.0965, -86.5463)
\qbezier( 81.0965, -86.5463)( 79.0767, -87.8990)( 76.9775, -89.3678)
\qbezier( 76.9775, -89.3678)( 74.8782, -90.8365)( 72.6929, -92.4259)
\qbezier( 72.6929, -92.4259)( 70.5075, -94.0153)( 68.2291, -95.7304)
\qbezier( 68.2291, -95.7304)( 65.9506, -97.4455)( 63.5719, -99.2918)
\qbezier( 63.5719, -99.2918)( 61.1932, -101.1380)( 58.7066, -103.1213)
\qbezier( 58.7066, -103.1213)( 56.2201, -105.1046)( 53.6178, -107.2311)
\qbezier( 53.6178, -107.2311)( 51.0154, -109.3577)( 48.2891, -111.6344)
\qbezier( 48.2891, -111.6344)( 45.5628, -113.9110)( 42.7038, -116.3450)
\qbezier( 42.7038, -116.3450)( 39.8449, -118.7789)( 36.8442, -121.3779)
\qbezier( 36.8442, -121.3779)( 33.8434, -123.9769)( 30.6915, -126.7491)
\qbezier( 30.6915, -126.7491)( 27.5395, -129.5213)( 24.2263, -132.4756)
\qbezier(109.5652, -71.5652)(107.3807, -71.8316)(105.1882, -72.1576)
\qbezier(105.1882, -72.1576)(102.9957, -72.4836)(100.7934, -72.8694)
\qbezier(100.7934, -72.8694)( 98.5912, -73.2553)( 96.3776, -73.7012)
\qbezier( 96.3776, -73.7012)( 94.1641, -74.1472)( 91.9375, -74.6536)
\qbezier( 91.9375, -74.6536)( 89.7110, -75.1600)( 87.4699, -75.7273)
\qbezier( 87.4699, -75.7273)( 85.2287, -76.2945)( 82.9713, -76.9230)
\qbezier( 82.9713, -76.9230)( 80.7139, -77.5515)( 78.4385, -78.2418)
\qbezier( 78.4385, -78.2418)( 76.1631, -78.9320)( 73.8681, -79.6844)
\qbezier( 73.8681, -79.6844)( 71.5731, -80.4369)( 69.2567, -81.2521)
\qbezier( 69.2567, -81.2521)( 66.9403, -82.0674)( 64.6009, -82.9460)
\qbezier( 64.6009, -82.9460)( 62.2614, -83.8246)( 59.8972, -84.7673)
\qbezier( 59.8972, -84.7673)( 57.5330, -85.7100)( 55.1422, -86.7174)
\qbezier( 55.1422, -86.7174)( 52.7514, -87.7248)( 50.3323, -88.7977)
\qbezier( 50.3323, -88.7977)( 47.9132, -89.8706)( 45.4640, -91.0098)
\qbezier( 45.4640, -91.0098)( 43.0147, -92.1490)( 40.5335, -93.3554)
\qbezier( 40.5335, -93.3554)( 38.0524, -94.5617)( 35.5374, -95.8361)
\qbezier( 35.5374, -95.8361)( 33.0225, -97.1104)( 30.4719, -98.4538)
\qbezier( 30.4719, -98.4538)( 27.9213, -99.7971)( 25.3332, -101.2104)
\qbezier( 25.3332, -101.2104)( 22.7451, -102.6237)( 20.1175, -104.1080)
\qbezier( 29.8949, -146.2026)( 34.7004, -141.3250)( 38.9152, -137.0282)
\qbezier( 38.9152, -137.0282)( 43.1300, -132.7313)( 46.8242, -128.9437)
\qbezier( 46.8242, -128.9437)( 50.5185, -125.1561)( 53.7538, -121.8147)
\qbezier( 53.7538, -121.8147)( 56.9891, -118.4732)( 59.8193, -115.5223)
\qbezier( 59.8193, -115.5223)( 62.6494, -112.5714)( 65.1216, -109.9618)
\qbezier( 65.1216, -109.9618)( 67.5937, -107.3523)( 69.7490, -105.0407)
\qbezier( 69.7490, -105.0407)( 71.9043, -102.7291)( 73.7786, -100.6769)
\qbezier( 73.7786, -100.6769)( 75.6530, -98.6247)( 77.2776, -96.7979)
\qbezier( 77.2776, -96.7979)( 78.9022, -94.9710)( 80.3041, -93.3390)
\qbezier( 80.3041, -93.3390)( 81.7060, -91.7069)( 82.9085, -90.2426)
\qbezier( 82.9085, -90.2426)( 84.1110, -88.7782)( 85.1342, -87.4572)
\qbezier( 85.1342, -87.4572)( 86.1574, -86.1362)( 87.0183, -84.9364)
\qbezier( 87.0183, -84.9364)( 87.8792, -83.7367)( 88.5922, -82.6383)
\qbezier( 88.5922, -82.6383)( 89.3051, -81.5399)( 89.8819, -80.5245)
\qbezier( 89.8819, -80.5245)( 90.4588, -79.5092)( 90.9091, -78.5600)
\qbezier( 90.9091, -78.5600)( 91.3595, -77.6107)( 91.6908, -76.7118)
\qbezier( 91.6908, -76.7118)( 92.0222, -75.8130)( 92.2401, -74.9494)
\qbezier( 92.2401, -74.9494)( 92.4580, -74.0859)( 92.5660, -73.2434)
\qbezier( 92.5660, -73.2434)( 92.6741, -72.4008)( 92.6741, -71.5652)
\qbezier( 24.7048, -140.4822)( 27.8945, -137.0980)( 30.7839, -133.9951)
\qbezier( 30.7839, -133.9951)( 33.6734, -130.8922)( 36.2873, -128.0440)
\qbezier( 36.2873, -128.0440)( 38.9013, -125.1958)( 41.2623, -122.5779)
\qbezier( 41.2623, -122.5779)( 43.6233, -119.9600)( 45.7516, -117.5498)
\qbezier( 45.7516, -117.5498)( 47.8799, -115.1397)( 49.7938, -112.9166)
\qbezier( 49.7938, -112.9166)( 51.7076, -110.6935)( 53.4235, -108.6385)
\qbezier( 53.4235, -108.6385)( 55.1393, -106.5834)( 56.6719, -104.6787)
\qbezier( 56.6719, -104.6787)( 58.2045, -102.7740)( 59.5670, -101.0032)
\qbezier( 59.5670, -101.0032)( 60.9294, -99.2325)( 62.1335, -97.5805)
\qbezier( 62.1335, -97.5805)( 63.3376, -95.9286)( 64.3935, -94.3812)
\qbezier( 64.3935, -94.3812)( 65.4495, -92.8339)( 66.3665, -91.3779)
\qbezier( 66.3665, -91.3779)( 67.2835, -89.9218)( 68.0693, -88.5446)
\qbezier( 68.0693, -88.5446)( 68.8552, -87.1674)( 69.5167, -85.8571)
\qbezier( 69.5167, -85.8571)( 70.1781, -84.5469)( 70.7209, -83.2924)
\qbezier( 70.7209, -83.2924)( 71.2637, -82.0379)( 71.6924, -80.8284)
\qbezier( 71.6924, -80.8284)( 72.1211, -79.6188)( 72.4395, -78.4439)
\qbezier( 72.4395, -78.4439)( 72.7578, -77.2689)( 72.9686, -76.1184)
\qbezier( 72.9686, -76.1184)( 73.1793, -74.9680)( 73.2842, -73.8321)
\qbezier( 73.2842, -73.8321)( 73.3891, -72.6962)( 73.3891, -71.5652)
\qbezier( 23.1277, -121.5045)( 24.5466, -119.7156)( 25.8661, -118.0045)
\qbezier( 25.8661, -118.0045)( 27.1856, -116.2934)( 28.4099, -114.6546)
\qbezier( 28.4099, -114.6546)( 29.6342, -113.0158)( 30.7673, -111.4440)
\qbezier( 30.7673, -111.4440)( 31.9004, -109.8722)( 32.9460, -108.3623)
\qbezier( 32.9460, -108.3623)( 33.9915, -106.8524)( 34.9528, -105.3995)
\qbezier( 34.9528, -105.3995)( 35.9142, -103.9466)( 36.7944, -102.5460)
\qbezier( 36.7944, -102.5460)( 37.6747, -101.1454)( 38.4768, -99.7927)
\qbezier( 38.4768, -99.7927)( 39.2788, -98.4399)( 40.0052, -97.1305)
\qbezier( 40.0052, -97.1305)( 40.7316, -95.8212)( 41.3847, -94.5511)
\qbezier( 41.3847, -94.5511)( 42.0378, -93.2809)( 42.6197, -92.0459)
\qbezier( 42.6197, -92.0459)( 43.2016, -90.8108)( 43.7143, -89.6069)
\qbezier( 43.7143, -89.6069)( 44.2269, -88.4029)( 44.6719, -87.2262)
\qbezier( 44.6719, -87.2262)( 45.1169, -86.0494)( 45.4956, -84.8961)
\qbezier( 45.4956, -84.8961)( 45.8744, -83.7428)( 46.1882, -82.6091)
\qbezier( 46.1882, -82.6091)( 46.5020, -81.4754)( 46.7518, -80.3578)
\qbezier( 46.7518, -80.3578)( 47.0017, -79.2401)( 47.1883, -78.1349)
\qbezier( 47.1883, -78.1349)( 47.3750, -77.0297)( 47.4991, -75.9332)
\qbezier( 47.4991, -75.9332)( 47.6232, -74.8368)( 47.6852, -73.7457)
\qbezier( 47.6852, -73.7457)( 47.7471, -72.6546)( 47.7471, -71.5652)
\put( 29.3478, -151.7826){\line(1, 1){180.0000}}
\put( 21.5217, -71.5652){\vector(1, 0){195.6522}}
\put(129.1304, -159.6087){\vector(0, 1){195.6522}}
\put(223.0435, -74.5000){\makebox(0,0)[]{\footnotesize{$\widetilde{R}$}}}
\put(123.2609,  37.0217){\makebox(0,0)[]{\footnotesize{$\widetilde{T}$}}}
\put(160.4348,  47.7826){\makebox(0,0)[]{\footnotesize{$\chi = \mbox{const}$}}}
\put(219.1304,  41.9130){\makebox(0,0)[]{\footnotesize{$\eta = \mbox{const}$}}}
\put( -6.8478, -102.8696){\makebox(0,0)[]{\footnotesize{$\eta = \mbox{const}$}}}
\put( -2.9348, -142.0000){\makebox(0,0)[]{\footnotesize{$\chi = \mbox{const}$}}}
\put(139.8913, -52.0000){\makebox(0,0)[]{\footnotesize{$1$}}}
\put(115.4348, -78.4130){\makebox(0,0)[]{\footnotesize{$-1$}}}
\end{picture}
\vspace{0mm} \newline
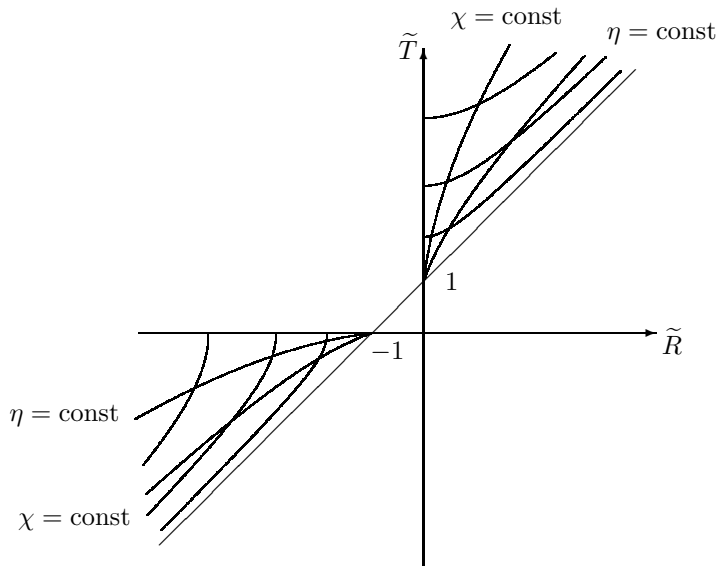
{\footnotesize \sf Figure 8. Minkowski diagram for universe models with
negative spatial curvature containing vacuum energy with reference to the
conformal coordinate system of type III defined in equation \eqref{e_133}.
The diagram shows world lines $\chi = \mbox{constant}$ and
simultaneity curves $\eta = \mbox{constant} \,$. The line
$\widetilde{T} = \widetilde{R} + 1$ represents the limit
$\chi \rightarrow \infty$, $\eta \rightarrow -\infty$.
Furthermore $\eta \rightarrow 0$ on the $\widetilde{R}$-axis,
and $\eta \rightarrow -\infty$ at the point $\widetilde{T} = 1$
on the $\widetilde{T}$-axis.}
\vspace{0mm} \newline
%
%
\vspace{5mm} \newline
{\bf 9. Flat universe models}
\vspace{3mm} \newline
%
%
{\bf 9.1. Conformal coordinates in flat universe models}
\vspace{3mm} \newline
In flat universe models $\eta$ and $\chi$ already are conformal
coordinates, corresponding to $a = 0$, $b = 0$, $c = 2$ and $d = 0$
in equation \eqref{e_407}. This gives the generating function
$f(x) = x$. However, $\eta = \mbox{constant}$ defines the same
cosmic space as that given by $t = \mbox{constant}$. But when $k = 0$,
we can introduce a second type of conformal coordinates defined by
$a = 1$, $b = 2$, $c = 2$ and $d = -1$ in equation \eqref{e_407}.
This gives the generating function
\begin{equation} \label{e_206}
f(x) = - 1 / x
\mbox{ .}
\end{equation}
The transformation \eqref{e_406} then takes the form
\begin{equation} \label{e_89}
T = \frl{\eta}{\chi^2 - \eta^2}
\mbox{\hspace{2mm} and \hspace{2mm}}
R = \frl{\chi}{\eta^2 - \chi^2}
\mbox{ .}
\end{equation}
If the universe model begins at $\eta = 0$, this transformation
maps the region $0 < \eta < \chi$ onto the region
$0 < T < - R$, and the region $0 < \chi < \eta$ onto the region
$0 < R < - T$. On the other hand, if the universe is infinitely old,
the region $- \chi < \eta < 0$ is mapped onto the region $R < T < 0$,
and the region $0 < \chi < - \eta$ onto the region $0 < R < T$.
The inverse transformation has the same form,
\begin{equation} \label{e_82}
\eta = \frl{T}{R^2 - T^2}
\mbox{\hspace{2mm} and \hspace{2mm}}
\chi = \frl{R}{T^2 - R^2}
\mbox{ .}
\end{equation}
Since ${\partial T}/{\partial \eta} > 0$, $T$ increases in the same
direction as $\eta$ and $t$.
\itm The line element takes the form
\begin{equation} \label{e_85}
ds^2 = \left[ \frl{a(\eta(T,R))}
{T^2 - R^2
\rule[-0mm]{0mm}{4.25mm}} \right]^{\hs{-0.0mm} 2}
(- dT^2 + dR^2 + R^2 d \Omega^2)
\mbox{ .}
\end{equation}
From equation (46) in reference [1] it follows that a particle with
$\chi = \mbox{constant}$ has a recession velocity in the
$(T,R)$-system given by
\begin{equation} \label{e_99}
V = \frl{2 T R}{T^2 + R^2
\rule[-0mm]{0mm}{4.25mm}}
\mbox{ ,}
\end{equation}
Again the initial recession velocity vanishes. Furthermore $V > 0$ for
$TR > 0$ and $V < 0$ for $TR < 0$. Note that $V < 0$ corresponds to
expansion and $V > 0$ to contraction when $R < 0$.
\itm A flat universe model with radiation and dust has [2]
\begin{equation} \label{e_113}
a = \frl{1}{2} \hs{0.5mm} \alpha \eta^2 + \beta \eta
\mbox{\hspace{2mm} , \hspace{3mm}}
t = \frl{1}{6} \hs{0.5mm} \alpha \eta^3
+ \frl{1}{2} \hs{0.5mm} \beta \eta^2
\end{equation}
where
\begin{equation} \label{e_277}
\alpha = {\Omega}_{m0} / 2
\mbox{\hspace{2mm} and \hspace{3mm}}
\beta = \sqrt{{\Omega}_{\gamma 0}}
\mbox{ .}
\end{equation}
The relationship between the conformal time at $R = 0$ and the
cosmic time is
\begin{equation} \label{e_159}
t = \frl{3 \beta T - \alpha}{6 T^3}
\mbox{ .}
\end{equation}
The conformal clocks go faster than the cosmic ones.
Here $\eta \in \hs{-2.3mm} < \hs{-0.5mm} 0,\infty \hs{-0.5mm} >$
when $t \in \hs{-2.3mm} < \hs{-0.5mm}  0,\infty \hs{-0.5mm} >$.
From equation \eqref{e_89} it then follows that the conformal time
$T$ is negative for these universe models, and $T \rightarrow 0$
when $\chi$ is constant and $\eta \rightarrow \infty$. The line
element as expressed by the $(T,R)$-coordinates takes the form
\begin{equation} \label{e_7}
ds^2 = \left[ \frac{({\Omega}_{m0} T - 4 \sqrt{{\Omega}_{\gamma 0}}
\hs{0.5mm} (T^2 - R^2)) \hs{0.5mm} T}
{4 (T^2 - R^2)^3} \right]^2 ds_M^2
\mbox{ .}
\end{equation}
\itm The Einstein-deSitter universe is a dust dominated, flat universe.
It has $\beta = {\Omega}_{\gamma} = 0$. The line element for this universe
in the $T,R$-coordinate system is
\begin{equation} \label{e_175}
ds^2 = \left[ \frac{{\Omega}_{m0} T^2}
{4 (T^2 - R^2)^3} \right]^2 ds_M^2
\mbox{ .}
\end{equation}
\itm The world lines of the reference particles in the cosmic coordinate
system, $\chi = \chi_0$, in the conformal reference frame are
given by
\begin{equation} \label{e_83}
(R + a_2)^2 - T^2 = a_2^2
\mbox{ ,}
\end{equation}
where $a_2 = (2 \hs{0.5mm} \chi_0)^{-1} > 0$.
The simultaneity curves of the cosmic space, $\eta = \eta_0$,
as given in the conformal system, are
\begin{equation} \label{e_84}
(T + b_2)^2 - R^2 = b_2^2
\mbox{ ,}
\end{equation}
where $b_2 = (2 \hs{0.5mm} \eta_0)^{-1}$.
These two sets of hyperbolae are drawn for a universe model with
matter and radiation in Figure 9.
\vspace*{6mm} \newline
\begin{picture}(50,232)(-96,-187)
\qbezier(126.7347, -61.1837)(126.7347, -62.7865)(126.8681, -64.3949)
\qbezier(126.8681, -64.3949)(127.0016, -66.0033)(127.2694, -67.6283)
\qbezier(127.2694, -67.6283)(127.5372, -69.2534)(127.9412, -70.9064)
\qbezier(127.9412, -70.9064)(128.3452, -72.5593)(128.8883, -74.2517)
\qbezier(128.8883, -74.2517)(129.4313, -75.9440)(130.1171, -77.6874)
\qbezier(130.1171, -77.6874)(130.8029, -79.4308)(131.6363, -81.2373)
\qbezier(131.6363, -81.2373)(132.4696, -83.0438)(133.4562, -84.9260)
\qbezier(133.4562, -84.9260)(134.4428, -86.8081)(135.5895, -88.7789)
\qbezier(135.5895, -88.7789)(136.7363, -90.7497)(138.0510, -92.8228)
\qbezier(138.0510, -92.8228)(139.3658, -94.8959)(140.8577, -97.0856)
\qbezier(140.8577, -97.0856)(142.3496, -99.2753)(144.0290, -101.5968)
\qbezier(144.0290, -101.5968)(145.7083, -103.9183)(147.5868, -106.3877)
\qbezier(147.5868, -106.3877)(149.4652, -108.8570)(151.5558, -111.4913)
\qbezier(151.5558, -111.4913)(153.6463, -114.1256)(155.9634, -116.9430)
\qbezier(155.9634, -116.9430)(158.2805, -119.7604)(160.8401, -122.7805)
\qbezier(160.8401, -122.7805)(163.3998, -125.8006)(166.2197, -129.0442)
\qbezier(166.2197, -129.0442)(169.0397, -132.2879)(172.1394, -135.7775)
\qbezier(172.1394, -135.7775)(175.2392, -139.2671)(178.6402, -143.0269)
\qbezier(178.6402, -143.0269)(182.0412, -146.7867)(185.7669, -150.8426)
\qbezier(126.7347, -61.1837)(126.7347, -63.4368)(126.8225, -65.6916)
\qbezier(126.8225, -65.6916)(126.9102, -67.9464)(127.0859, -70.2063)
\qbezier(127.0859, -70.2063)(127.2616, -72.4663)(127.5255, -74.7348)
\qbezier(127.5255, -74.7348)(127.7893, -77.0033)(128.1417, -79.2838)
\qbezier(128.1417, -79.2838)(128.4942, -81.5642)(128.9357, -83.8602)
\qbezier(128.9357, -83.8602)(129.3773, -86.1562)(129.9086, -88.4711)
\qbezier(129.9086, -88.4711)(130.4399, -90.7860)(131.0618, -93.1233)
\qbezier(131.0618, -93.1233)(131.6837, -95.4607)(132.3971, -97.8240)
\qbezier(132.3971, -97.8240)(133.1106, -100.1874)(133.9166, -102.5803)
\qbezier(133.9166, -102.5803)(134.7227, -104.9732)(135.6225, -107.3994)
\qbezier(135.6225, -107.3994)(136.5224, -109.8256)(137.5175, -112.2886)
\qbezier(137.5175, -112.2886)(138.5126, -114.7517)(139.6043, -117.2553)
\qbezier(139.6043, -117.2553)(140.6961, -119.7590)(141.8863, -122.3072)
\qbezier(141.8863, -122.3072)(143.0764, -124.8553)(144.3667, -127.4517)
\qbezier(144.3667, -127.4517)(145.6571, -130.0481)(147.0495, -132.6968)
\qbezier(147.0495, -132.6968)(148.4420, -135.3454)(149.9386, -138.0503)
\qbezier(149.9386, -138.0503)(151.4353, -140.7552)(153.0385, -143.5205)
\qbezier(153.0385, -143.5205)(154.6417, -146.2858)(156.3539, -149.1156)
\qbezier(156.3539, -149.1156)(158.0660, -151.9454)(159.8897, -154.8441)
\qbezier(201.6147, -144.7838)(195.8766, -139.0803)(190.9455, -134.1895)
\qbezier(190.9455, -134.1895)(186.0143, -129.2987)(181.7750, -125.1064)
\qbezier(181.7750, -125.1064)(177.5357, -120.9141)(173.8893, -117.3224)
\qbezier(173.8893, -117.3224)(170.2428, -113.7308)(167.1042, -110.6559)
\qbezier(167.1042, -110.6559)(163.9655, -107.5811)(161.2614, -104.9512)
\qbezier(161.2614, -104.9512)(158.5573, -102.3214)(156.2245, -100.0752)
\qbezier(156.2245, -100.0752)(153.8918, -97.8290)(151.8760, -95.9141)
\qbezier(151.8760, -95.9141)(149.8602, -93.9991)(148.1143, -92.3707)
\qbezier(148.1143, -92.3707)(146.3684, -90.7422)(144.8517, -89.3623)
\qbezier(144.8517, -89.3623)(143.3349, -87.9824)(142.0119, -86.8187)
\qbezier(142.0119, -86.8187)(140.6889, -85.6551)(139.5287, -84.6806)
\qbezier(139.5287, -84.6806)(138.3685, -83.7061)(137.3442, -82.8980)
\qbezier(137.3442, -82.8980)(136.3198, -82.0899)(135.4073, -81.4293)
\qbezier(135.4073, -81.4293)(134.4947, -80.7688)(133.6728, -80.2403)
\qbezier(133.6728, -80.2403)(132.8508, -79.7118)(132.1003, -79.3031)
\qbezier(132.1003, -79.3031)(131.3497, -78.8945)(130.6530, -78.5960)
\qbezier(130.6530, -78.5960)(129.9563, -78.2976)(129.2972, -78.1024)
\qbezier(129.2972, -78.1024)(128.6380, -77.9072)(128.0011, -77.8107)
\qbezier(128.0011, -77.8107)(127.3642, -77.7143)(126.7347, -77.7143)
\qbezier(192.4512, -148.9573)(188.9578, -145.6053)(185.7889, -142.5930)
\qbezier(185.7889, -142.5930)(182.6201, -139.5807)(179.7429, -136.8769)
\qbezier(179.7429, -136.8769)(176.8656, -134.1730)(174.2499, -131.7494)
\qbezier(174.2499, -131.7494)(171.6342, -129.3257)(169.2528, -127.1569)
\qbezier(169.2528, -127.1569)(166.8713, -124.9881)(164.6993, -123.0516)
\qbezier(164.6993, -123.0516)(162.5272, -121.1151)(160.5419, -119.3907)
\qbezier(160.5419, -119.3907)(158.5566, -117.6662)(156.7373, -116.1358)
\qbezier(156.7373, -116.1358)(154.9180, -114.6054)(153.2458, -113.2532)
\qbezier(153.2458, -113.2532)(151.5735, -111.9009)(150.0309, -110.7126)
\qbezier(150.0309, -110.7126)(148.4882, -109.5243)(147.0591, -108.4876)
\qbezier(147.0591, -108.4876)(145.6299, -107.4509)(144.2993, -106.5549)
\qbezier(144.2993, -106.5549)(142.9687, -105.6590)(141.7229, -104.8945)
\qbezier(141.7229, -104.8945)(140.4770, -104.1299)(139.3028, -103.4889)
\qbezier(139.3028, -103.4889)(138.1286, -102.8478)(137.0139, -102.3235)
\qbezier(137.0139, -102.3235)(135.8991, -101.7991)(134.8322, -101.3861)
\qbezier(134.8322, -101.3861)(133.7653, -100.9730)(132.7350, -100.6670)
\qbezier(132.7350, -100.6670)(131.7048, -100.3609)(130.7005, -100.1586)
\qbezier(130.7005, -100.1586)(129.6962, -99.9563)(128.7073, -99.8557)
\qbezier(128.7073, -99.8557)(127.7184, -99.7551)(126.7347, -99.7551)
\qbezier(175.4520, -153.1212)(173.6243, -151.6089)(171.8871, -150.2072)
\qbezier(171.8871, -150.2072)(170.1499, -148.8056)(168.4964, -147.5092)
\qbezier(168.4964, -147.5092)(166.8429, -146.2128)(165.2668, -145.0167)
\qbezier(165.2668, -145.0167)(163.6908, -143.8205)(162.1860, -142.7200)
\qbezier(162.1860, -142.7200)(160.6812, -141.6195)(159.2419, -140.6104)
\qbezier(159.2419, -140.6104)(157.8026, -139.6013)(156.4232, -138.6797)
\qbezier(156.4232, -138.6797)(155.0438, -137.7581)(153.7190, -136.9204)
\qbezier(153.7190, -136.9204)(152.3943, -136.0827)(151.1190, -135.3258)
\qbezier(151.1190, -135.3258)(149.8438, -134.5688)(148.6131, -133.8897)
\qbezier(148.6131, -133.8897)(147.3824, -133.2105)(146.1916, -132.6065)
\qbezier(146.1916, -132.6065)(145.0007, -132.0025)(143.8451, -131.4714)
\qbezier(143.8451, -131.4714)(142.6895, -130.9402)(141.5647, -130.4799)
\qbezier(141.5647, -130.4799)(140.4398, -130.0195)(139.3414, -129.6282)
\qbezier(139.3414, -129.6282)(138.2431, -129.2368)(137.1669, -128.9130)
\qbezier(137.1669, -128.9130)(136.0907, -128.5891)(135.0325, -128.3315)
\qbezier(135.0325, -128.3315)(133.9744, -128.0739)(132.9302, -127.8816)
\qbezier(132.9302, -127.8816)(131.8860, -127.6893)(130.8518, -127.5615)
\qbezier(130.8518, -127.5615)(129.8176, -127.4337)(128.7893, -127.3699)
\qbezier(128.7893, -127.3699)(127.7610, -127.3061)(126.7347, -127.3061)
\qbezier( 37.0758,  -2.1514)( 41.1317,  -5.8772)( 44.8915,  -9.2782)
\qbezier( 44.8915,  -9.2782)( 48.6512, -12.6792)( 52.1409, -15.7789)
\qbezier( 52.1409, -15.7789)( 55.6305, -18.8787)( 58.8741, -21.6986)
\qbezier( 58.8741, -21.6986)( 62.1178, -24.5186)( 65.1379, -27.0782)
\qbezier( 65.1379, -27.0782)( 68.1579, -29.6379)( 70.9754, -31.9550)
\qbezier( 70.9754, -31.9550)( 73.7928, -34.2721)( 76.4271, -36.3626)
\qbezier( 76.4271, -36.3626)( 79.0614, -38.4531)( 81.5307, -40.3316)
\qbezier( 81.5307, -40.3316)( 84.0001, -42.2100)( 86.3216, -43.8894)
\qbezier( 86.3216, -43.8894)( 88.6431, -45.5688)( 90.8328, -47.0607)
\qbezier( 90.8328, -47.0607)( 93.0225, -48.5526)( 95.0956, -49.8673)
\qbezier( 95.0956, -49.8673)( 97.1687, -51.1821)( 99.1395, -52.3288)
\qbezier( 99.1395, -52.3288)(101.1103, -53.4755)(102.9924, -54.4622)
\qbezier(102.9924, -54.4622)(104.8746, -55.4488)(106.6811, -56.2821)
\qbezier(106.6811, -56.2821)(108.4876, -57.1154)(110.2310, -57.8012)
\qbezier(110.2310, -57.8012)(111.9744, -58.4870)(113.6667, -59.0301)
\qbezier(113.6667, -59.0301)(115.3590, -59.5731)(117.0120, -59.9771)
\qbezier(117.0120, -59.9771)(118.6650, -60.3812)(120.2900, -60.6490)
\qbezier(120.2900, -60.6490)(121.9151, -60.9168)(123.5235, -61.0502)
\qbezier(123.5235, -61.0502)(125.1319, -61.1837)(126.7347, -61.1837)
\qbezier( 33.0743, -28.0287)( 35.9730, -29.8524)( 38.8028, -31.5645)
\qbezier( 38.8028, -31.5645)( 41.6326, -33.2766)( 44.3979, -34.8798)
\qbezier( 44.3979, -34.8798)( 47.1631, -36.4830)( 49.8681, -37.9797)
\qbezier( 49.8681, -37.9797)( 52.5730, -39.4764)( 55.2216, -40.8689)
\qbezier( 55.2216, -40.8689)( 57.8703, -42.2613)( 60.4667, -43.5516)
\qbezier( 60.4667, -43.5516)( 63.0631, -44.8419)( 65.6112, -46.0321)
\qbezier( 65.6112, -46.0321)( 68.1593, -47.2222)( 70.6630, -48.3140)
\qbezier( 70.6630, -48.3140)( 73.1667, -49.4058)( 75.6298, -50.4009)
\qbezier( 75.6298, -50.4009)( 78.0928, -51.3960)( 80.5190, -52.2958)
\qbezier( 80.5190, -52.2958)( 82.9451, -53.1957)( 85.3381, -54.0018)
\qbezier( 85.3381, -54.0018)( 87.7310, -54.8078)( 90.0944, -55.5212)
\qbezier( 90.0944, -55.5212)( 92.4577, -56.2347)( 94.7951, -56.8566)
\qbezier( 94.7951, -56.8566)( 97.1324, -57.4785)( 99.4473, -58.0098)
\qbezier( 99.4473, -58.0098)(101.7622, -58.5411)(104.0582, -58.9826)
\qbezier(104.0582, -58.9826)(106.3541, -59.4242)(108.6346, -59.7766)
\qbezier(108.6346, -59.7766)(110.9151, -60.1290)(113.1836, -60.3929)
\qbezier(113.1836, -60.3929)(115.4521, -60.6568)(117.7120, -60.8324)
\qbezier(117.7120, -60.8324)(119.9720, -61.0081)(122.2268, -61.0959)
\qbezier(122.2268, -61.0959)(124.4816, -61.1837)(126.7347, -61.1837)
\qbezier(110.2041, -61.1837)(110.2041, -60.5541)(110.1076, -59.9172)
\qbezier(110.1076, -59.9172)(110.0112, -59.2803)(109.8160, -58.6212)
\qbezier(109.8160, -58.6212)(109.6208, -57.9621)(109.3224, -57.2654)
\qbezier(109.3224, -57.2654)(109.0239, -56.5687)(108.6152, -55.8181)
\qbezier(108.6152, -55.8181)(108.2065, -55.0675)(107.6781, -54.2456)
\qbezier(107.6781, -54.2456)(107.1496, -53.4236)(106.4890, -52.5111)
\qbezier(106.4890, -52.5111)(105.8285, -51.5986)(105.0204, -50.5742)
\qbezier(105.0204, -50.5742)(104.2123, -49.5498)(103.2378, -48.3897)
\qbezier(103.2378, -48.3897)(102.2633, -47.2295)(101.0996, -45.9065)
\qbezier(101.0996, -45.9065)( 99.9360, -44.5835)( 98.5561, -43.0667)
\qbezier( 98.5561, -43.0667)( 97.1761, -41.5499)( 95.5477, -39.8041)
\qbezier( 95.5477, -39.8041)( 93.9193, -38.0582)( 92.0043, -36.0424)
\qbezier( 92.0043, -36.0424)( 90.0893, -34.0266)( 87.8432, -31.6938)
\qbezier( 87.8432, -31.6938)( 85.5970, -29.3611)( 82.9671, -26.6570)
\qbezier( 82.9671, -26.6570)( 80.3373, -23.9528)( 77.2624, -20.8142)
\qbezier( 77.2624, -20.8142)( 74.1876, -17.6755)( 70.5959, -14.0291)
\qbezier( 70.5959, -14.0291)( 67.0043, -10.3827)( 62.8119,  -6.1434)
\qbezier( 62.8119,  -6.1434)( 58.6196,  -1.9040)( 53.7288,   3.0271)
\qbezier( 53.7288,   3.0271)( 48.8380,   7.9583)( 43.1346,  13.6964)
\qbezier( 88.1633, -61.1837)( 88.1633, -60.1999)( 88.0626, -59.2111)
\qbezier( 88.0626, -59.2111)( 87.9620, -58.2222)( 87.7597, -57.2179)
\qbezier( 87.7597, -57.2179)( 87.5575, -56.2136)( 87.2514, -55.1833)
\qbezier( 87.2514, -55.1833)( 86.9453, -54.1531)( 86.5323, -53.0862)
\qbezier( 86.5323, -53.0862)( 86.1192, -52.0192)( 85.5949, -50.9045)
\qbezier( 85.5949, -50.9045)( 85.0706, -49.7898)( 84.4295, -48.6156)
\qbezier( 84.4295, -48.6156)( 83.7884, -47.4414)( 83.0239, -46.1955)
\qbezier( 83.0239, -46.1955)( 82.2594, -44.9496)( 81.3635, -43.6191)
\qbezier( 81.3635, -43.6191)( 80.4675, -42.2885)( 79.4308, -40.8593)
\qbezier( 79.4308, -40.8593)( 78.3941, -39.4302)( 77.2058, -37.8875)
\qbezier( 77.2058, -37.8875)( 76.0175, -36.3448)( 74.6652, -34.6726)
\qbezier( 74.6652, -34.6726)( 73.3129, -33.0004)( 71.7825, -31.1811)
\qbezier( 71.7825, -31.1811)( 70.2522, -29.3618)( 68.5277, -27.3765)
\qbezier( 68.5277, -27.3765)( 66.8033, -25.3912)( 64.8667, -23.2191)
\qbezier( 64.8667, -23.2191)( 62.9302, -21.0471)( 60.7615, -18.6656)
\qbezier( 60.7615, -18.6656)( 58.5927, -16.2842)( 56.1690, -13.6685)
\qbezier( 56.1690, -13.6685)( 53.7453, -11.0528)( 51.0415,  -8.1755)
\qbezier( 51.0415,  -8.1755)( 48.3376,  -5.2982)( 45.3254,  -2.1294)
\qbezier( 45.3254,  -2.1294)( 42.3131,   1.0394)( 38.9610,   4.5329)
\qbezier( 60.6122, -61.1837)( 60.6122, -60.1574)( 60.5485, -59.1291)
\qbezier( 60.5485, -59.1291)( 60.4847, -58.1008)( 60.3569, -57.0666)
\qbezier( 60.3569, -57.0666)( 60.2291, -56.0323)( 60.0367, -54.9881)
\qbezier( 60.0367, -54.9881)( 59.8444, -53.9440)( 59.5868, -52.8858)
\qbezier( 59.5868, -52.8858)( 59.3292, -51.8277)( 59.0054, -50.7515)
\qbezier( 59.0054, -50.7515)( 58.6816, -49.6753)( 58.2902, -48.5769)
\qbezier( 58.2902, -48.5769)( 57.8989, -47.4785)( 57.4385, -46.3537)
\qbezier( 57.4385, -46.3537)( 56.9781, -45.2289)( 56.4470, -44.0733)
\qbezier( 56.4470, -44.0733)( 55.9158, -42.9177)( 55.3118, -41.7268)
\qbezier( 55.3118, -41.7268)( 54.7079, -40.5360)( 54.0287, -39.3053)
\qbezier( 54.0287, -39.3053)( 53.3495, -38.0746)( 52.5926, -36.7993)
\qbezier( 52.5926, -36.7993)( 51.8356, -35.5241)( 50.9980, -34.1993)
\qbezier( 50.9980, -34.1993)( 50.1603, -32.8746)( 49.2387, -31.4952)
\qbezier( 49.2387, -31.4952)( 48.3171, -30.1158)( 47.3080, -28.6765)
\qbezier( 47.3080, -28.6765)( 46.2989, -27.2372)( 45.1984, -25.7324)
\qbezier( 45.1984, -25.7324)( 44.0978, -24.2276)( 42.9017, -22.6515)
\qbezier( 42.9017, -22.6515)( 41.7056, -21.0754)( 40.4092, -19.4220)
\qbezier( 40.4092, -19.4220)( 39.1128, -17.7685)( 37.7111, -16.0313)
\qbezier( 37.7111, -16.0313)( 36.3095, -14.2941)( 34.7972, -12.4664)
\put( 41.3265,  24.2245){\line(1, -1){170.8163}}
\put( 30.3061, -61.1837){\vector(1, 0){192.8571}}
\put(126.7347, -157.6122){\vector(0, 1){192.8571}}
\put(231.4286, -65.3163){\makebox(0,0)[]{\footnotesize{$R$}}}
\put(118.4694,  36.6224){\makebox(0,0)[]{\footnotesize{$T$}}}
\put(162.5510, -164.5000){\makebox(0,0)[]{\footnotesize{$\chi = \mbox{const}$}}}
\put(217.6531, -156.2347){\makebox(0,0)[]{\footnotesize{$\eta = \mbox{const}$}}}
\put(  4.1327, -28.6735){\makebox(0,0)[]{\footnotesize{$\eta = \mbox{const}$}}}
\put( 12.3980,  13.2041){\makebox(0,0)[]{\footnotesize{$\chi = \mbox{const}$}}}
\end{picture}
\vspace{0mm} \newline
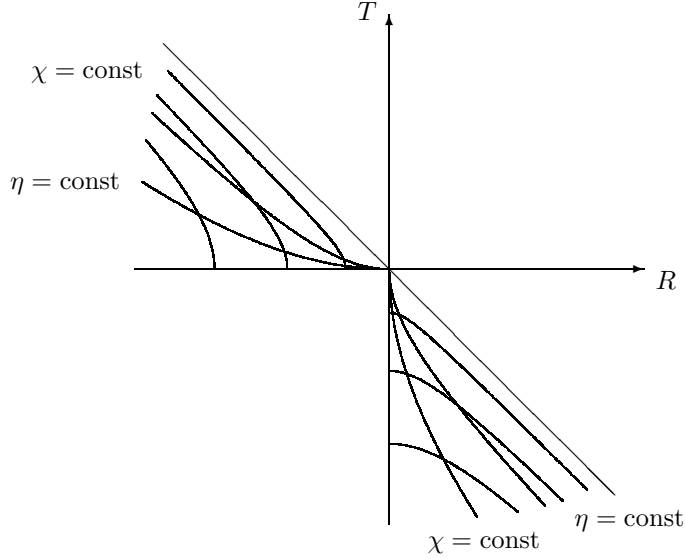
{\footnotesize \sf Figure 9. Minkowski diagram with reference to the
conformal coordinate system for a flat universe model dominated by matter
and radiation. The diagram shows world lines $\chi = \mbox{constant}$ and
simultaneity curves $\eta = \mbox{constant} \,$. For this universe model
the Hubble flow is expanding for $T > 0$ and contracting for $T < 0$
relative to the conformal system.}
\vspace{0mm} \newline
\itm The Minkowski diagram in Figure 9 shows that the universe
with radiation and dust appears at $T = 0$ having an infinitely
great extension. Then a hole develops expanding with the velocity
of light. For an infinitely large conformal time the Hubble flow
expands relative to the conformal frame. As $T$ approaches infinity,
the conformal clocks are reset to come from minus infinity, and $R$
changes from negative to positive values. The universe then contracts
relative to the conformal frame to vanishing extension at $T = 0$.
Hence during the period with positive time the conformal frame
contracts relative to the Hubble flow, and during the succeeding
period with negative time the conformal frame expands faster than
the Hubble flow.
\itm The world lines of the reference particles of the conformal
coordinate system, $R = R_0$, in the $(\eta,\chi)$-system are given
by (see Figure 10)
\begin{equation} \label{e_3}
(\chi + a_3)^2 - \eta^2 = a_3^2
\mbox{ ,}
\end{equation}
where $a_3 = (2 R_0)^{-1} \,$.
The simultaneity curves of the conformal space, $T = T_0$,
have the equation
\begin{equation} \label{e_4}
(\eta + b_3)^2 - \chi^2 = b_3^2
\mbox{ ,}
\end{equation}
where $b_3 = (2 T_0)^{-1}$.
\vspace*{5mm} \newline
\begin{picture}(50,192)(-96,-182)
\qbezier(191.5359, -24.7662)(187.3867, -29.2830)(183.5993, -33.4700)
\qbezier(183.5993, -33.4700)(179.8118, -37.6571)(176.3598, -41.5432)
\qbezier(176.3598, -41.5432)(172.9078, -45.4294)(169.7674, -49.0416)
\qbezier(169.7674, -49.0416)(166.6270, -52.6539)(163.7765, -56.0172)
\qbezier(163.7765, -56.0172)(160.9260, -59.3804)(158.3456, -62.5180)
\qbezier(158.3456, -62.5180)(155.7652, -65.6556)(153.4371, -68.5893)
\qbezier(153.4371, -68.5893)(151.1090, -71.5229)(149.0171, -74.2728)
\qbezier(149.0171, -74.2728)(146.9252, -77.0228)(145.0550, -79.6081)
\qbezier(145.0550, -79.6081)(143.1848, -82.1934)(141.5233, -84.6319)
\qbezier(141.5233, -84.6319)(139.8619, -87.0705)(138.3977, -89.3792)
\qbezier(138.3977, -89.3792)(136.9336, -91.6878)(135.6565, -93.8826)
\qbezier(135.6565, -93.8826)(134.3795, -96.0773)(133.2808, -98.1734)
\qbezier(133.2808, -98.1734)(132.1821, -100.2694)(131.2540, -102.2812)
\qbezier(131.2540, -102.2812)(130.3260, -104.2930)(129.5623, -106.2345)
\qbezier(129.5623, -106.2345)(128.7985, -108.1760)(128.1938, -110.0607)
\qbezier(128.1938, -110.0607)(127.5890, -111.9453)(127.1391, -113.7861)
\qbezier(127.1391, -113.7861)(126.6891, -115.6269)(126.3909, -117.4366)
\qbezier(126.3909, -117.4366)(126.0927, -119.2464)(125.9441, -121.0375)
\qbezier(125.9441, -121.0375)(125.7955, -122.8287)(125.7955, -124.6136)
\qbezier(162.7181, -20.3100)(160.6872, -23.5381)(158.7804, -26.6895)
\qbezier(158.7804, -26.6895)(156.8737, -29.8409)(155.0884, -32.9204)
\qbezier(155.0884, -32.9204)(153.3030, -35.9999)(151.6362, -39.0121)
\qbezier(151.6362, -39.0121)(149.9695, -42.0244)(148.4188, -44.9741)
\qbezier(148.4188, -44.9741)(146.8681, -47.9237)(145.4311, -50.8152)
\qbezier(145.4311, -50.8152)(143.9942, -53.7066)(142.6688, -56.5443)
\qbezier(142.6688, -56.5443)(141.3434, -59.3820)(140.1276, -62.1702)
\qbezier(140.1276, -62.1702)(138.9117, -64.9584)(137.8036, -67.7013)
\qbezier(137.8036, -67.7013)(136.6954, -70.4443)(135.6933, -73.1461)
\qbezier(135.6933, -73.1461)(134.6911, -75.8480)(133.7935, -78.5128)
\qbezier(133.7935, -78.5128)(132.8959, -81.1777)(132.1014, -83.8096)
\qbezier(132.1014, -83.8096)(131.3069, -86.4415)(130.6143, -89.0445)
\qbezier(130.6143, -89.0445)(129.9217, -91.6475)(129.3300, -94.2254)
\qbezier(129.3300, -94.2254)(128.7383, -96.8034)(128.2466, -99.3602)
\qbezier(128.2466, -99.3602)(127.7549, -101.9171)(127.3624, -104.4567)
\qbezier(127.3624, -104.4567)(126.9699, -106.9964)(126.6761, -109.5227)
\qbezier(126.6761, -109.5227)(126.3822, -112.0489)(126.1866, -114.5657)
\qbezier(126.1866, -114.5657)(125.9910, -117.0824)(125.8932, -119.5935)
\qbezier(125.8932, -119.5935)(125.7955, -122.1045)(125.7955, -124.6136)
\qbezier(125.7955, -106.2045)(126.4965, -106.2045)(127.2058, -106.0971)
\qbezier(127.2058, -106.0971)(127.9151, -105.9897)(128.6491, -105.7723)
\qbezier(128.6491, -105.7723)(129.3831, -105.5550)(130.1590, -105.2226)
\qbezier(130.1590, -105.2226)(130.9349, -104.8903)(131.7707, -104.4351)
\qbezier(131.7707, -104.4351)(132.6066, -103.9800)(133.5220, -103.3915)
\qbezier(133.5220, -103.3915)(134.4373, -102.8030)(135.4535, -102.0673)
\qbezier(135.4535, -102.0673)(136.4698, -101.3317)(137.6105, -100.4318)
\qbezier(137.6105, -100.4318)(138.7513, -99.5318)(140.0433, -98.4466)
\qbezier(140.0433, -98.4466)(141.3353, -97.3614)(142.8087, -96.0655)
\qbezier(142.8087, -96.0655)(144.2821, -94.7697)(145.9712, -93.2329)
\qbezier(145.9712, -93.2329)(147.6603, -91.6962)(149.6046, -89.8827)
\qbezier(149.6046, -89.8827)(151.5489, -88.0692)(153.7937, -85.9366)
\qbezier(153.7937, -85.9366)(156.0386, -83.8040)(158.6364, -81.3026)
\qbezier(158.6364, -81.3026)(161.2342, -78.8012)(164.2457, -75.8725)
\qbezier(164.2457, -75.8725)(167.2571, -72.9438)(170.7524, -69.5195)
\qbezier(170.7524, -69.5195)(174.2477, -66.0953)(178.3085, -62.0955)
\qbezier(178.3085, -62.0955)(182.3693, -58.0956)(187.0903, -53.4269)
\qbezier(187.0903, -53.4269)(191.8114, -48.7582)(197.3029, -43.3116)
\qbezier(197.3029, -43.3116)(202.7944, -37.8651)(209.1846, -31.5135)
\qbezier(125.7955, -81.6591)(126.8910, -81.6591)(127.9922, -81.5470)
\qbezier(127.9922, -81.5470)(129.0934, -81.4350)(130.2119, -81.2097)
\qbezier(130.2119, -81.2097)(131.3303, -80.9844)(132.4776, -80.6436)
\qbezier(132.4776, -80.6436)(133.6250, -80.3027)(134.8131, -79.8428)
\qbezier(134.8131, -79.8428)(136.0013, -79.3828)(137.2427, -78.7989)
\qbezier(137.2427, -78.7989)(138.4841, -78.2150)(139.7917, -77.5010)
\qbezier(139.7917, -77.5010)(141.0994, -76.7871)(142.4868, -75.9357)
\qbezier(142.4868, -75.9357)(143.8743, -75.0843)(145.3561, -74.0866)
\qbezier(145.3561, -74.0866)(146.8378, -73.0888)(148.4294, -71.9343)
\qbezier(148.4294, -71.9343)(150.0210, -70.7798)(151.7389, -69.4564)
\qbezier(151.7389, -69.4564)(153.4569, -68.1331)(155.3192, -66.6272)
\qbezier(155.3192, -66.6272)(157.1814, -65.1212)(159.2075, -63.4169)
\qbezier(159.2075, -63.4169)(161.2335, -61.7126)(163.4444, -59.7922)
\qbezier(163.4444, -59.7922)(165.6553, -57.8718)(168.0742, -55.7152)
\qbezier(168.0742, -55.7152)(170.4930, -53.5587)(173.1451, -51.1434)
\qbezier(173.1451, -51.1434)(175.7972, -48.7282)(178.7101, -46.0291)
\qbezier(178.7101, -46.0291)(181.6231, -43.3300)(184.8273, -40.3189)
\qbezier(184.8273, -40.3189)(188.0315, -37.3078)(191.5604, -33.9533)
\qbezier(191.5604, -33.9533)(195.0893, -30.5987)(198.9798, -26.8657)
\qbezier(125.7955, -50.9773)(126.9384, -50.9773)(128.0835, -50.9062)
\qbezier(128.0835, -50.9062)(129.2287, -50.8352)(130.3804, -50.6929)
\qbezier(130.3804, -50.6929)(131.5322, -50.5506)(132.6950, -50.3364)
\qbezier(132.6950, -50.3364)(133.8578, -50.1222)(135.0362, -49.8353)
\qbezier(135.0362, -49.8353)(136.2146, -49.5485)(137.4131, -49.1878)
\qbezier(137.4131, -49.1878)(138.6116, -48.8272)(139.8348, -48.3914)
\qbezier(139.8348, -48.3914)(141.0580, -47.9556)(142.3106, -47.4429)
\qbezier(142.3106, -47.4429)(143.5633, -46.9302)(144.8502, -46.3387)
\qbezier(144.8502, -46.3387)(146.1372, -45.7472)(147.4633, -45.0746)
\qbezier(147.4633, -45.0746)(148.7895, -44.4019)(150.1600, -43.6456)
\qbezier(150.1600, -43.6456)(151.5306, -42.8892)(152.9507, -42.0463)
\qbezier(152.9507, -42.0463)(154.3709, -41.2033)(155.8462, -40.2704)
\qbezier(155.8462, -40.2704)(157.3215, -39.3376)(158.8576, -38.3113)
\qbezier(158.8576, -38.3113)(160.3937, -37.2849)(161.9966, -36.1611)
\qbezier(161.9966, -36.1611)(163.5995, -35.0374)(165.2753, -33.8118)
\qbezier(165.2753, -33.8118)(166.9511, -32.5862)(168.7063, -31.2542)
\qbezier(168.7063, -31.2542)(170.4615, -29.9221)(172.3028, -28.4784)
\qbezier(172.3028, -28.4784)(174.1442, -27.0347)(176.0788, -25.4738)
\qbezier(176.0788, -25.4738)(178.0134, -23.9128)(180.0488, -22.2287)
\qbezier(125.7955, -124.6136)(127.5804, -124.6136)(129.3716, -124.4650)
\qbezier(129.3716, -124.4650)(131.1627, -124.3164)(132.9725, -124.0182)
\qbezier(132.9725, -124.0182)(134.7822, -123.7199)(136.6230, -123.2700)
\qbezier(136.6230, -123.2700)(138.4638, -122.8201)(140.3484, -122.2153)
\qbezier(140.3484, -122.2153)(142.2331, -121.6106)(144.1746, -120.8468)
\qbezier(144.1746, -120.8468)(146.1161, -120.0831)(148.1279, -119.1551)
\qbezier(148.1279, -119.1551)(150.1397, -118.2270)(152.2357, -117.1283)
\qbezier(152.2357, -117.1283)(154.3318, -116.0296)(156.5265, -114.7526)
\qbezier(156.5265, -114.7526)(158.7213, -113.4755)(161.0299, -112.0114)
\qbezier(161.0299, -112.0114)(163.3386, -110.5472)(165.7771, -108.8857)
\qbezier(165.7771, -108.8857)(168.2157, -107.2243)(170.8010, -105.3541)
\qbezier(170.8010, -105.3541)(173.3863, -103.4839)(176.1363, -101.3920)
\qbezier(176.1363, -101.3920)(178.8862, -99.3001)(181.8198, -96.9720)
\qbezier(181.8198, -96.9720)(184.7535, -94.6439)(187.8911, -92.0635)
\qbezier(187.8911, -92.0635)(191.0287, -89.4831)(194.3919, -86.6326)
\qbezier(194.3919, -86.6326)(197.7552, -83.7821)(201.3674, -80.6417)
\qbezier(201.3674, -80.6417)(204.9797, -77.5013)(208.8659, -74.0493)
\qbezier(208.8659, -74.0493)(212.7520, -70.5973)(216.9390, -66.8098)
\qbezier(216.9390, -66.8098)(221.1261, -63.0223)(225.6429, -58.8732)
\qbezier(125.7955, -124.6136)(128.3046, -124.6136)(130.8156, -124.5159)
\qbezier(130.8156, -124.5159)(133.3267, -124.4181)(135.8434, -124.2225)
\qbezier(135.8434, -124.2225)(138.3602, -124.0268)(140.8864, -123.7330)
\qbezier(140.8864, -123.7330)(143.4127, -123.4392)(145.9524, -123.0467)
\qbezier(145.9524, -123.0467)(148.4920, -122.6542)(151.0489, -122.1625)
\qbezier(151.0489, -122.1625)(153.6057, -121.6708)(156.1837, -121.0791)
\qbezier(156.1837, -121.0791)(158.7616, -120.4874)(161.3646, -119.7948)
\qbezier(161.3646, -119.7948)(163.9676, -119.1022)(166.5995, -118.3077)
\qbezier(166.5995, -118.3077)(169.2314, -117.5132)(171.8962, -116.6156)
\qbezier(171.8962, -116.6156)(174.5611, -115.7180)(177.2630, -114.7158)
\qbezier(177.2630, -114.7158)(179.9648, -113.7137)(182.7078, -112.6055)
\qbezier(182.7078, -112.6055)(185.4507, -111.4974)(188.2389, -110.2815)
\qbezier(188.2389, -110.2815)(191.0271, -109.0657)(193.8648, -107.7403)
\qbezier(193.8648, -107.7403)(196.7025, -106.4149)(199.5939, -104.9779)
\qbezier(199.5939, -104.9779)(202.4854, -103.5410)(205.4350, -101.9903)
\qbezier(205.4350, -101.9903)(208.3847, -100.4396)(211.3969, -98.7729)
\qbezier(211.3969, -98.7729)(214.4092, -97.1061)(217.4887, -95.3207)
\qbezier(217.4887, -95.3207)(220.5682, -93.5353)(223.7196, -91.6286)
\qbezier(223.7196, -91.6286)(226.8710, -89.7219)(230.0991, -87.6910)
\qbezier(218.8956, -41.2245)(212.5440, -47.6147)(207.0974, -53.1062)
\qbezier(207.0974, -53.1062)(201.6509, -58.5977)(196.9822, -63.3187)
\qbezier(196.9822, -63.3187)(192.3134, -68.0398)(188.3136, -72.1006)
\qbezier(188.3136, -72.1006)(184.3138, -76.1614)(180.8895, -79.6567)
\qbezier(180.8895, -79.6567)(177.4653, -83.1520)(174.5366, -86.1634)
\qbezier(174.5366, -86.1634)(171.6079, -89.1749)(169.1065, -91.7727)
\qbezier(169.1065, -91.7727)(166.6051, -94.3705)(164.4725, -96.6154)
\qbezier(164.4725, -96.6154)(162.3399, -98.8602)(160.5264, -100.8045)
\qbezier(160.5264, -100.8045)(158.7129, -102.7488)(157.1762, -104.4379)
\qbezier(157.1762, -104.4379)(155.6394, -106.1270)(154.3436, -107.6004)
\qbezier(154.3436, -107.6004)(153.0477, -109.0738)(151.9625, -110.3658)
\qbezier(151.9625, -110.3658)(150.8772, -111.6578)(149.9773, -112.7985)
\qbezier(149.9773, -112.7985)(149.0774, -113.9393)(148.3417, -114.9556)
\qbezier(148.3417, -114.9556)(147.6061, -115.9718)(147.0176, -116.8871)
\qbezier(147.0176, -116.8871)(146.4291, -117.8025)(145.9740, -118.6383)
\qbezier(145.9740, -118.6383)(145.5188, -119.4742)(145.1865, -120.2501)
\qbezier(145.1865, -120.2501)(144.8541, -121.0260)(144.6368, -121.7600)
\qbezier(144.6368, -121.7600)(144.4194, -122.4940)(144.3120, -123.2033)
\qbezier(144.3120, -123.2033)(144.2045, -123.9125)(144.2045, -124.6136)
\qbezier(223.5434, -51.4293)(219.8104, -55.3198)(216.4558, -58.8487)
\qbezier(216.4558, -58.8487)(213.1013, -62.3776)(210.0902, -65.5818)
\qbezier(210.0902, -65.5818)(207.0790, -68.7860)(204.3800, -71.6990)
\qbezier(204.3800, -71.6990)(201.6809, -74.6119)(199.2656, -77.2640)
\qbezier(199.2656, -77.2640)(196.8504, -79.9160)(194.6939, -82.3349)
\qbezier(194.6939, -82.3349)(192.5373, -84.7538)(190.6169, -86.9647)
\qbezier(190.6169, -86.9647)(188.6965, -89.1756)(186.9922, -91.2016)
\qbezier(186.9922, -91.2016)(185.2879, -93.2277)(183.7819, -95.0899)
\qbezier(183.7819, -95.0899)(182.2760, -96.9522)(180.9526, -98.6702)
\qbezier(180.9526, -98.6702)(179.6293, -100.3881)(178.4748, -101.9797)
\qbezier(178.4748, -101.9797)(177.3203, -103.5713)(176.3225, -105.0530)
\qbezier(176.3225, -105.0530)(175.3248, -106.5348)(174.4734, -107.9223)
\qbezier(174.4734, -107.9223)(173.6220, -109.3097)(172.9080, -110.6173)
\qbezier(172.9080, -110.6173)(172.1941, -111.9250)(171.6102, -113.1664)
\qbezier(171.6102, -113.1664)(171.0263, -114.4078)(170.5663, -115.5960)
\qbezier(170.5663, -115.5960)(170.1064, -116.7841)(169.7655, -117.9314)
\qbezier(169.7655, -117.9314)(169.4247, -119.0788)(169.1994, -120.1972)
\qbezier(169.1994, -120.1972)(168.9741, -121.3156)(168.8621, -122.4169)
\qbezier(168.8621, -122.4169)(168.7500, -123.5181)(168.7500, -124.6136)
\qbezier(228.1804, -70.3603)(226.4963, -72.3957)(224.9353, -74.3303)
\qbezier(224.9353, -74.3303)(223.3744, -76.2649)(221.9307, -78.1063)
\qbezier(221.9307, -78.1063)(220.4870, -79.9476)(219.1549, -81.7028)
\qbezier(219.1549, -81.7028)(217.8228, -83.4580)(216.5973, -85.1338)
\qbezier(216.5973, -85.1338)(215.3717, -86.8096)(214.2480, -88.4125)
\qbezier(214.2480, -88.4125)(213.1242, -90.0153)(212.0978, -91.5515)
\qbezier(212.0978, -91.5515)(211.0715, -93.0876)(210.1386, -94.5629)
\qbezier(210.1386, -94.5629)(209.2058, -96.0382)(208.3628, -97.4584)
\qbezier(208.3628, -97.4584)(207.5198, -98.8785)(206.7635, -100.2491)
\qbezier(206.7635, -100.2491)(206.0072, -101.6196)(205.3345, -102.9458)
\qbezier(205.3345, -102.9458)(204.6619, -104.2719)(204.0704, -105.5589)
\qbezier(204.0704, -105.5589)(203.4789, -106.8458)(202.9662, -108.0984)
\qbezier(202.9662, -108.0984)(202.4535, -109.3511)(202.0177, -110.5743)
\qbezier(202.0177, -110.5743)(201.5819, -111.7975)(201.2213, -112.9960)
\qbezier(201.2213, -112.9960)(200.8606, -114.1945)(200.5738, -115.3729)
\qbezier(200.5738, -115.3729)(200.2869, -116.5512)(200.0727, -117.7141)
\qbezier(200.0727, -117.7141)(199.8585, -118.8769)(199.7162, -120.0287)
\qbezier(199.7162, -120.0287)(199.5739, -121.1804)(199.5028, -122.3256)
\qbezier(199.5028, -122.3256)(199.4318, -123.4707)(199.4318, -124.6136)
\put(125.7955, -124.6136){\line(1, 1){ 95.1136}}
\put( 88.9773, -124.6136){\vector(1, 0){144.2045}}
\put(125.7955, -161.4318){\vector(0, 1){144.2045}}
\put(242.3864, -129.2159){\makebox(0,0)[]{\footnotesize{$\chi$}}}
\put(116.5909, -15.6932){\makebox(0,0)[]{\footnotesize{$\eta$}}}
\put(165.6818, -11.0909){\makebox(0,0)[]{\footnotesize{$R = \mbox{const}$}}}
\put(227.0455, -20.2955){\makebox(0,0)[]{\footnotesize{$T = \mbox{const}$}}}
\put(259.2614, -87.7955){\makebox(0,0)[]{\footnotesize{$T = \mbox{const}$}}}
\put(248.5227, -40.2386){\makebox(0,0)[]{\footnotesize{$R = \mbox{const}$}}}
\end{picture}
\vspace{0mm} \newline
{\footnotesize \sf Figure 10. Minkowski diagram with reference to the
$(\eta,\chi)$-system for a flat universe model dominated by
matter and \mbox{radiation}. The diagram shows world lines
$R = \mbox{constant}$ and simultaneity curves
$T = \mbox{constant} \,$. In this universe model the
conformal system expands relative to the Hubble flow.}
\vspace{0mm} \newline
\itm The world lines of the reference particles with $R
= \mbox{constant}$ in the conformal coordinate system is given
by equation (36) in reference [1] which leads to
\begin{equation} \label{e_86}
\left( \frl{d \chi}{d \eta} \right)_{R = R_1}
= - \tanh \theta
= \frl{2 \eta \chi}{\eta^2 + \chi^2} > 0
\mbox{ .}
\end{equation}
This means that the $(T,R)$-system expands relative
to the $(\eta,\chi)$-system. Hence the
$(\eta,\chi)$-system contracts relative to the
$(T,R)$-system with a velocity
\begin{equation} \label{e_114}
\left( \frl{d R}{d T} \right)_{\chi = \chi_1}
= \tanh \theta = \frl{2 T R}
{T^2 + R^2} < 0
\end{equation}
as shown in Figure 9.
\itm In a flat LIVE dominated universe the scale factor is
\begin{equation} \label{e_115}
a(t) = e^{\widehat{H}_{\Lambda} (t - t_0)}
\mbox{ ,}
\end{equation}
where $\widehat{H}_{\Lambda} \,$ is given in equation \eqref{e_290}
and is constant. Then the parametric time $\eta$ is
\begin{equation} \label{e_91}
\eta = - \frl{1}{\widehat{H}_{\Lambda} \rule[-0mm]{0mm}{4.15mm}} \hs{0.7mm}
e^{- \widehat{H}_{\Lambda} (t - t_0)}
\mbox{ ,}
\end{equation}
which increases from $-\infty$ to $0$ as $t$ increases from $-\infty$ to
$\infty$. The world lines $\chi = \mbox{constant}$ and the simultaneity
curves $\eta = \mbox{constant}$ are shown in a Minkowski diagram with
reference to the conformal coordinate system for this universe model
in Figure 11. The corresponding world lines $R = \mbox{constant}$ and
simultaneity curves $T = \mbox{constant}$ are shown in Figure 12.
\itm In terms of the parametric time $\eta$ or of the conformal
coordinates $R$ and $T$, the scale factor may be
written
\begin{equation} \label{e_87}
a(\eta) = - \frl{1}{\widehat{H}_{\Lambda} \eta \rule[-0mm]{0mm}{4.15mm}}
= \frl{T^2 - R^2}{\widehat{H}_{\Lambda} T \rule[-0mm]{0mm}{4.15mm}}
\mbox{ .}
\end{equation}
In this case the line element in conformal coordinates takes the form [14]
\begin{equation} \label{e_88}
ds^2 = \frl{1}{\widehat{H}_{\Lambda}^2 T^2 \rule[-0mm]{0mm}{4.15mm}} ds_M^2
\end{equation}
as in the negatively curved case. There is no continual creation in such
a universe model.
\vspace*{3mm} \newline
\begin{picture}(50,232)(-96,-177)
\qbezier(185.7669,  28.4752)(182.0412,  24.4193)(178.6402,  20.6596)
\qbezier(178.6402,  20.6596)(175.2392,  16.8998)(172.1394,  13.4102)
\qbezier(172.1394,  13.4102)(169.0397,   9.9205)(166.2197,   6.6769)
\qbezier(166.2197,   6.6769)(163.3998,   3.4333)(160.8401,   0.4132)
\qbezier(160.8401,   0.4132)(158.2805,  -2.6069)(155.9634,  -5.4244)
\qbezier(155.9634,  -5.4244)(153.6463,  -8.2418)(151.5558, -10.8761)
\qbezier(151.5558, -10.8761)(149.4652, -13.5103)(147.5868, -15.9797)
\qbezier(147.5868, -15.9797)(145.7083, -18.4490)(144.0290, -20.7705)
\qbezier(144.0290, -20.7705)(142.3496, -23.0920)(140.8577, -25.2817)
\qbezier(140.8577, -25.2817)(139.3658, -27.4715)(138.0510, -29.5446)
\qbezier(138.0510, -29.5446)(136.7363, -31.6176)(135.5895, -33.5884)
\qbezier(135.5895, -33.5884)(134.4428, -35.5592)(133.4562, -37.4414)
\qbezier(133.4562, -37.4414)(132.4696, -39.3235)(131.6363, -41.1301)
\qbezier(131.6363, -41.1301)(130.8029, -42.9366)(130.1171, -44.6800)
\qbezier(130.1171, -44.6800)(129.4313, -46.4234)(128.8883, -48.1157)
\qbezier(128.8883, -48.1157)(128.3452, -49.8080)(127.9412, -51.4610)
\qbezier(127.9412, -51.4610)(127.5372, -53.1140)(127.2694, -54.7390)
\qbezier(127.2694, -54.7390)(127.0016, -56.3641)(126.8681, -57.9725)
\qbezier(126.8681, -57.9725)(126.7347, -59.5808)(126.7347, -61.1837)
\qbezier(159.8897,  32.4767)(158.0660,  29.5781)(156.3539,  26.7482)
\qbezier(156.3539,  26.7482)(154.6417,  23.9184)(153.0385,  21.1531)
\qbezier(153.0385,  21.1531)(151.4353,  18.3879)(149.9386,  15.6830)
\qbezier(149.9386,  15.6830)(148.4420,  12.9781)(147.0495,  10.3294)
\qbezier(147.0495,  10.3294)(145.6571,   7.6808)(144.3667,   5.0843)
\qbezier(144.3667,   5.0843)(143.0764,   2.4879)(141.8863,  -0.0602)
\qbezier(141.8863,  -0.0602)(140.6961,  -2.6083)(139.6043,  -5.1120)
\qbezier(139.6043,  -5.1120)(138.5126,  -7.6157)(137.5175, -10.0787)
\qbezier(137.5175, -10.0787)(136.5224, -12.5418)(135.6225, -14.9679)
\qbezier(135.6225, -14.9679)(134.7227, -17.3941)(133.9166, -19.7870)
\qbezier(133.9166, -19.7870)(133.1106, -22.1800)(132.3971, -24.5433)
\qbezier(132.3971, -24.5433)(131.6837, -26.9067)(131.0618, -29.2440)
\qbezier(131.0618, -29.2440)(130.4399, -31.5814)(129.9086, -33.8963)
\qbezier(129.9086, -33.8963)(129.3773, -36.2112)(128.9357, -38.5071)
\qbezier(128.9357, -38.5071)(128.4942, -40.8031)(128.1417, -43.0836)
\qbezier(128.1417, -43.0836)(127.7893, -45.3641)(127.5255, -47.6326)
\qbezier(127.5255, -47.6326)(127.2616, -49.9011)(127.0859, -52.1610)
\qbezier(127.0859, -52.1610)(126.9102, -54.4210)(126.8225, -56.6758)
\qbezier(126.8225, -56.6758)(126.7347, -58.9306)(126.7347, -61.1837)
\qbezier(126.7347, -44.6531)(127.3642, -44.6531)(128.0011, -44.5566)
\qbezier(128.0011, -44.5566)(128.6380, -44.4601)(129.2972, -44.2650)
\qbezier(129.2972, -44.2650)(129.9563, -44.0698)(130.6530, -43.7713)
\qbezier(130.6530, -43.7713)(131.3497, -43.4729)(132.1003, -43.0642)
\qbezier(132.1003, -43.0642)(132.8508, -42.6555)(133.6728, -42.1271)
\qbezier(133.6728, -42.1271)(134.4947, -41.5986)(135.4073, -40.9380)
\qbezier(135.4073, -40.9380)(136.3198, -40.2775)(137.3442, -39.4693)
\qbezier(137.3442, -39.4693)(138.3685, -38.6612)(139.5287, -37.6868)
\qbezier(139.5287, -37.6868)(140.6889, -36.7123)(142.0119, -35.5486)
\qbezier(142.0119, -35.5486)(143.3349, -34.3850)(144.8517, -33.0051)
\qbezier(144.8517, -33.0051)(146.3684, -31.6251)(148.1143, -29.9967)
\qbezier(148.1143, -29.9967)(149.8602, -28.3682)(151.8760, -26.4533)
\qbezier(151.8760, -26.4533)(153.8918, -24.5383)(156.2245, -22.2921)
\qbezier(156.2245, -22.2921)(158.5573, -20.0460)(161.2614, -17.4161)
\qbezier(161.2614, -17.4161)(163.9655, -14.7863)(167.1042, -11.7114)
\qbezier(167.1042, -11.7114)(170.2428,  -8.6366)(173.8893,  -5.0449)
\qbezier(173.8893,  -5.0449)(177.5357,  -1.4532)(181.7750,   2.7391)
\qbezier(181.7750,   2.7391)(186.0143,   6.9314)(190.9455,  11.8222)
\qbezier(190.9455,  11.8222)(195.8766,  16.7130)(201.6147,  22.4165)
\qbezier(126.7347, -22.6122)(127.7184, -22.6122)(128.7073, -22.5116)
\qbezier(128.7073, -22.5116)(129.6962, -22.4110)(130.7005, -22.2087)
\qbezier(130.7005, -22.2087)(131.7048, -22.0064)(132.7350, -21.7004)
\qbezier(132.7350, -21.7004)(133.7653, -21.3943)(134.8322, -20.9813)
\qbezier(134.8322, -20.9813)(135.8991, -20.5682)(137.0139, -20.0439)
\qbezier(137.0139, -20.0439)(138.1286, -19.5196)(139.3028, -18.8785)
\qbezier(139.3028, -18.8785)(140.4770, -18.2374)(141.7229, -17.4729)
\qbezier(141.7229, -17.4729)(142.9687, -16.7084)(144.2993, -15.8124)
\qbezier(144.2993, -15.8124)(145.6299, -14.9165)(147.0591, -13.8798)
\qbezier(147.0591, -13.8798)(148.4882, -12.8431)(150.0309, -11.6548)
\qbezier(150.0309, -11.6548)(151.5735, -10.4665)(153.2458,  -9.1142)
\qbezier(153.2458,  -9.1142)(154.9180,  -7.7619)(156.7373,  -6.2315)
\qbezier(156.7373,  -6.2315)(158.5566,  -4.7011)(160.5419,  -2.9767)
\qbezier(160.5419,  -2.9767)(162.5272,  -1.2522)(164.6993,   0.6843)
\qbezier(164.6993,   0.6843)(166.8713,   2.6208)(169.2528,   4.7896)
\qbezier(169.2528,   4.7896)(171.6342,   6.9583)(174.2499,   9.3820)
\qbezier(174.2499,   9.3820)(176.8656,  11.8057)(179.7429,  14.5095)
\qbezier(179.7429,  14.5095)(182.6201,  17.2134)(185.7889,  20.2257)
\qbezier(185.7889,  20.2257)(188.9578,  23.2379)(192.4512,  26.5900)
\qbezier(126.7347,   4.9388)(127.7610,   4.9388)(128.7893,   5.0026)
\qbezier(128.7893,   5.0026)(129.8176,   5.0663)(130.8518,   5.1941)
\qbezier(130.8518,   5.1941)(131.8860,   5.3219)(132.9302,   5.5143)
\qbezier(132.9302,   5.5143)(133.9744,   5.7066)(135.0325,   5.9642)
\qbezier(135.0325,   5.9642)(136.0907,   6.2218)(137.1669,   6.5456)
\qbezier(137.1669,   6.5456)(138.2431,   6.8695)(139.3414,   7.2608)
\qbezier(139.3414,   7.2608)(140.4398,   7.6522)(141.5647,   8.1125)
\qbezier(141.5647,   8.1125)(142.6895,   8.5729)(143.8451,   9.1040)
\qbezier(143.8451,   9.1040)(145.0007,   9.6352)(146.1916,  10.2392)
\qbezier(146.1916,  10.2392)(147.3824,  10.8432)(148.6131,  11.5223)
\qbezier(148.6131,  11.5223)(149.8438,  12.2015)(151.1190,  12.9584)
\qbezier(151.1190,  12.9584)(152.3943,  13.7154)(153.7190,  14.5531)
\qbezier(153.7190,  14.5531)(155.0438,  15.3907)(156.4232,  16.3123)
\qbezier(156.4232,  16.3123)(157.8026,  17.2340)(159.2419,  18.2431)
\qbezier(159.2419,  18.2431)(160.6812,  19.2522)(162.1860,  20.3527)
\qbezier(162.1860,  20.3527)(163.6908,  21.4532)(165.2668,  22.6493)
\qbezier(165.2668,  22.6493)(166.8429,  23.8455)(168.4964,  25.1418)
\qbezier(168.4964,  25.1418)(170.1499,  26.4382)(171.8871,  27.8399)
\qbezier(171.8871,  27.8399)(173.6243,  29.2415)(175.4520,  30.7539)
\qbezier(126.7347, -61.1837)(125.1319, -61.1837)(123.5235, -61.3171)
\qbezier(123.5235, -61.3171)(121.9151, -61.4506)(120.2900, -61.7184)
\qbezier(120.2900, -61.7184)(118.6650, -61.9862)(117.0120, -62.3902)
\qbezier(117.0120, -62.3902)(115.3590, -62.7942)(113.6667, -63.3373)
\qbezier(113.6667, -63.3373)(111.9744, -63.8803)(110.2310, -64.5661)
\qbezier(110.2310, -64.5661)(108.4876, -65.2519)(106.6811, -66.0853)
\qbezier(106.6811, -66.0853)(104.8746, -66.9186)(102.9924, -67.9052)
\qbezier(102.9924, -67.9052)(101.1103, -68.8918)( 99.1395, -70.0385)
\qbezier( 99.1395, -70.0385)( 97.1687, -71.1852)( 95.0956, -72.5000)
\qbezier( 95.0956, -72.5000)( 93.0225, -73.8148)( 90.8328, -75.3067)
\qbezier( 90.8328, -75.3067)( 88.6431, -76.7986)( 86.3216, -78.4780)
\qbezier( 86.3216, -78.4780)( 84.0001, -80.1573)( 81.5307, -82.0358)
\qbezier( 81.5307, -82.0358)( 79.0614, -83.9142)( 76.4271, -86.0048)
\qbezier( 76.4271, -86.0048)( 73.7928, -88.0953)( 70.9754, -90.4124)
\qbezier( 70.9754, -90.4124)( 68.1579, -92.7294)( 65.1379, -95.2891)
\qbezier( 65.1379, -95.2891)( 62.1178, -97.8488)( 58.8741, -100.6687)
\qbezier( 58.8741, -100.6687)( 55.6305, -103.4887)( 52.1409, -106.5884)
\qbezier( 52.1409, -106.5884)( 48.6512, -109.6881)( 44.8915, -113.0891)
\qbezier( 44.8915, -113.0891)( 41.1317, -116.4901)( 37.0758, -120.2159)
\qbezier(126.7347, -61.1837)(124.4816, -61.1837)(122.2268, -61.2714)
\qbezier(122.2268, -61.2714)(119.9720, -61.3592)(117.7120, -61.5349)
\qbezier(117.7120, -61.5349)(115.4521, -61.7106)(113.1836, -61.9744)
\qbezier(113.1836, -61.9744)(110.9151, -62.2383)(108.6346, -62.5907)
\qbezier(108.6346, -62.5907)(106.3541, -62.9432)(104.0582, -63.3847)
\qbezier(104.0582, -63.3847)(101.7622, -63.8262)( 99.4473, -64.3576)
\qbezier( 99.4473, -64.3576)( 97.1324, -64.8889)( 94.7951, -65.5108)
\qbezier( 94.7951, -65.5108)( 92.4577, -66.1327)( 90.0944, -66.8461)
\qbezier( 90.0944, -66.8461)( 87.7310, -67.5595)( 85.3381, -68.3656)
\qbezier( 85.3381, -68.3656)( 82.9451, -69.1716)( 80.5190, -70.0715)
\qbezier( 80.5190, -70.0715)( 78.0928, -70.9714)( 75.6298, -71.9665)
\qbezier( 75.6298, -71.9665)( 73.1667, -72.9615)( 70.6630, -74.0533)
\qbezier( 70.6630, -74.0533)( 68.1593, -75.1451)( 65.6112, -76.3353)
\qbezier( 65.6112, -76.3353)( 63.0631, -77.5254)( 60.4667, -78.8157)
\qbezier( 60.4667, -78.8157)( 57.8703, -80.1060)( 55.2216, -81.4985)
\qbezier( 55.2216, -81.4985)( 52.5730, -82.8909)( 49.8681, -84.3876)
\qbezier( 49.8681, -84.3876)( 47.1631, -85.8843)( 44.3979, -87.4875)
\qbezier( 44.3979, -87.4875)( 41.6326, -89.0907)( 38.8028, -90.8029)
\qbezier( 38.8028, -90.8029)( 35.9730, -92.5150)( 33.0743, -94.3387)
\qbezier( 43.1346, -136.0637)( 48.8380, -130.3256)( 53.7288, -125.3945)
\qbezier( 53.7288, -125.3945)( 58.6196, -120.4633)( 62.8119, -116.2240)
\qbezier( 62.8119, -116.2240)( 67.0043, -111.9847)( 70.5959, -108.3382)
\qbezier( 70.5959, -108.3382)( 74.1876, -104.6918)( 77.2624, -101.5532)
\qbezier( 77.2624, -101.5532)( 80.3373, -98.4145)( 82.9671, -95.7104)
\qbezier( 82.9671, -95.7104)( 85.5970, -93.0063)( 87.8432, -90.6735)
\qbezier( 87.8432, -90.6735)( 90.0893, -88.3408)( 92.0043, -86.3250)
\qbezier( 92.0043, -86.3250)( 93.9193, -84.3092)( 95.5477, -82.5633)
\qbezier( 95.5477, -82.5633)( 97.1761, -80.8174)( 98.5561, -79.3006)
\qbezier( 98.5561, -79.3006)( 99.9360, -77.7839)(101.0996, -76.4609)
\qbezier(101.0996, -76.4609)(102.2633, -75.1378)(103.2378, -73.9777)
\qbezier(103.2378, -73.9777)(104.2123, -72.8175)(105.0204, -71.7931)
\qbezier(105.0204, -71.7931)(105.8285, -70.7688)(106.4890, -69.8562)
\qbezier(106.4890, -69.8562)(107.1496, -68.9437)(107.6781, -68.1218)
\qbezier(107.6781, -68.1218)(108.2065, -67.2998)(108.6152, -66.5492)
\qbezier(108.6152, -66.5492)(109.0239, -65.7987)(109.3224, -65.1020)
\qbezier(109.3224, -65.1020)(109.6208, -64.4052)(109.8160, -63.7461)
\qbezier(109.8160, -63.7461)(110.0112, -63.0870)(110.1076, -62.4501)
\qbezier(110.1076, -62.4501)(110.2041, -61.8132)(110.2041, -61.1837)
\qbezier( 38.9610, -126.9002)( 42.3131, -123.4067)( 45.3254, -120.2379)
\qbezier( 45.3254, -120.2379)( 48.3376, -117.0691)( 51.0415, -114.1918)
\qbezier( 51.0415, -114.1918)( 53.7453, -111.3146)( 56.1690, -108.6989)
\qbezier( 56.1690, -108.6989)( 58.5927, -106.0832)( 60.7615, -103.7017)
\qbezier( 60.7615, -103.7017)( 62.9302, -101.3203)( 64.8667, -99.1482)
\qbezier( 64.8667, -99.1482)( 66.8033, -96.9762)( 68.5277, -94.9909)
\qbezier( 68.5277, -94.9909)( 70.2522, -93.0056)( 71.7825, -91.1863)
\qbezier( 71.7825, -91.1863)( 73.3129, -89.3670)( 74.6652, -87.6948)
\qbezier( 74.6652, -87.6948)( 76.0175, -86.0225)( 77.2058, -84.4799)
\qbezier( 77.2058, -84.4799)( 78.3941, -82.9372)( 79.4308, -81.5080)
\qbezier( 79.4308, -81.5080)( 80.4675, -80.0789)( 81.3635, -78.7483)
\qbezier( 81.3635, -78.7483)( 82.2594, -77.4177)( 83.0239, -76.1718)
\qbezier( 83.0239, -76.1718)( 83.7884, -74.9260)( 84.4295, -73.7518)
\qbezier( 84.4295, -73.7518)( 85.0706, -72.5776)( 85.5949, -71.4628)
\qbezier( 85.5949, -71.4628)( 86.1192, -70.3481)( 86.5323, -69.2812)
\qbezier( 86.5323, -69.2812)( 86.9453, -68.2143)( 87.2514, -67.1840)
\qbezier( 87.2514, -67.1840)( 87.5575, -66.1538)( 87.7597, -65.1494)
\qbezier( 87.7597, -65.1494)( 87.9620, -64.1451)( 88.0626, -63.1563)
\qbezier( 88.0626, -63.1563)( 88.1633, -62.1674)( 88.1633, -61.1837)
\qbezier( 34.7972, -109.9010)( 36.3095, -108.0733)( 37.7111, -106.3361)
\qbezier( 37.7111, -106.3361)( 39.1128, -104.5988)( 40.4092, -102.9454)
\qbezier( 40.4092, -102.9454)( 41.7056, -101.2919)( 42.9017, -99.7158)
\qbezier( 42.9017, -99.7158)( 44.0978, -98.1397)( 45.1984, -96.6349)
\qbezier( 45.1984, -96.6349)( 46.2989, -95.1301)( 47.3080, -93.6908)
\qbezier( 47.3080, -93.6908)( 48.3171, -92.2515)( 49.2387, -90.8722)
\qbezier( 49.2387, -90.8722)( 50.1603, -89.4928)( 50.9980, -88.1680)
\qbezier( 50.9980, -88.1680)( 51.8356, -86.8433)( 52.5926, -85.5680)
\qbezier( 52.5926, -85.5680)( 53.3495, -84.2927)( 54.0287, -83.0621)
\qbezier( 54.0287, -83.0621)( 54.7079, -81.8314)( 55.3118, -80.6405)
\qbezier( 55.3118, -80.6405)( 55.9158, -79.4497)( 56.4470, -78.2941)
\qbezier( 56.4470, -78.2941)( 56.9781, -77.1385)( 57.4385, -76.0136)
\qbezier( 57.4385, -76.0136)( 57.8989, -74.8888)( 58.2902, -73.7904)
\qbezier( 58.2902, -73.7904)( 58.6816, -72.6920)( 59.0054, -71.6158)
\qbezier( 59.0054, -71.6158)( 59.3292, -70.5397)( 59.5868, -69.4815)
\qbezier( 59.5868, -69.4815)( 59.8444, -68.4234)( 60.0367, -67.3792)
\qbezier( 60.0367, -67.3792)( 60.2291, -66.3350)( 60.3569, -65.3008)
\qbezier( 60.3569, -65.3008)( 60.4847, -64.2666)( 60.5485, -63.2383)
\qbezier( 60.5485, -63.2383)( 60.6122, -62.2100)( 60.6122, -61.1837)
\put( 41.3265, -146.5918){\line(1, 1){170.8163}}
\put( 30.3061, -61.1837){\vector(1, 0){192.8571}}
\put(126.7347, -157.6122){\vector(0, 1){192.8571}}
\put(231.4286, -65.3163){\makebox(0,0)[]{\footnotesize{$R$}}}
\put(118.4694,  36.6224){\makebox(0,0)[]{\footnotesize{$T$}}}
\put(162.5510,  42.1327){\makebox(0,0)[]{\footnotesize{$\chi = \mbox{const}$}}}
\put(217.6531,  33.8673){\makebox(0,0)[]{\footnotesize{$\eta = \mbox{const}$}}}
\put(  4.1327, -95.0714){\makebox(0,0)[]{\footnotesize{$\eta = \mbox{const}$}}}
\put( 12.3980, -136.9490){\makebox(0,0)[]{\footnotesize{$\chi = \mbox{const}$}}}
\end{picture}
\vspace{0mm} \newline
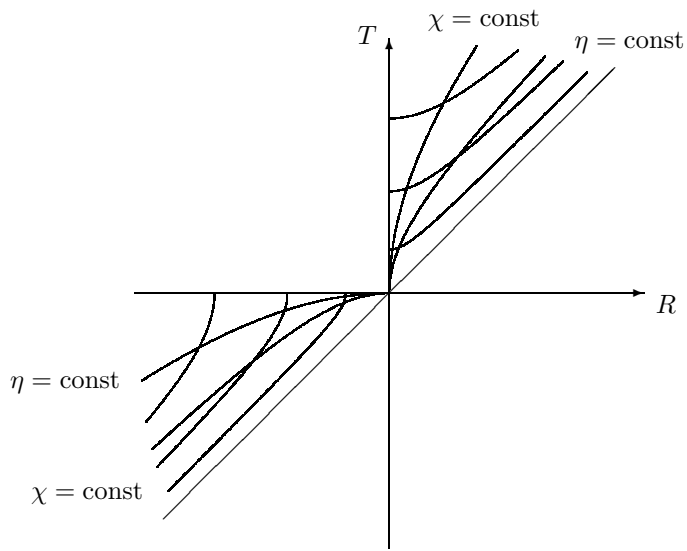
{\footnotesize \sf Figure 11. Minkowski diagram with reference to the
conformal coordinate system for a flat universe model dominated by vacuum
energy. The diagram shows world lines $\chi = \mbox{constant}$ and
simultaneity curves $\eta = \mbox{constant} \,$. For this universe model
the Hubble flow is expanding for $T > 0$ and contracting for $T < 0$
relative to the conformal system.}
\vspace{0mm} \newline
\vspace*{0mm} \newline
\begin{picture}(50,192)(-96,-182)
\qbezier(125.7955, -41.7727)(125.7955, -43.5577)(125.9441, -45.3489)
\qbezier(125.9441, -45.3489)(126.0927, -47.1400)(126.3909, -48.9497)
\qbezier(126.3909, -48.9497)(126.6891, -50.7594)(127.1391, -52.6003)
\qbezier(127.1391, -52.6003)(127.5890, -54.4411)(128.1938, -56.3257)
\qbezier(128.1938, -56.3257)(128.7985, -58.2103)(129.5623, -60.1519)
\qbezier(129.5623, -60.1519)(130.3260, -62.0934)(131.2540, -64.1052)
\qbezier(131.2540, -64.1052)(132.1821, -66.1170)(133.2808, -68.2130)
\qbezier(133.2808, -68.2130)(134.3795, -70.3090)(135.6565, -72.5038)
\qbezier(135.6565, -72.5038)(136.9336, -74.6985)(138.3977, -77.0072)
\qbezier(138.3977, -77.0072)(139.8619, -79.3159)(141.5233, -81.7544)
\qbezier(141.5233, -81.7544)(143.1848, -84.1930)(145.0550, -86.7783)
\qbezier(145.0550, -86.7783)(146.9252, -89.3636)(149.0171, -92.1135)
\qbezier(149.0171, -92.1135)(151.1090, -94.8635)(153.4371, -97.7971)
\qbezier(153.4371, -97.7971)(155.7652, -100.7307)(158.3456, -103.8683)
\qbezier(158.3456, -103.8683)(160.9260, -107.0059)(163.7765, -110.3692)
\qbezier(163.7765, -110.3692)(166.6270, -113.7325)(169.7674, -117.3447)
\qbezier(169.7674, -117.3447)(172.9078, -120.9569)(176.3598, -124.8431)
\qbezier(176.3598, -124.8431)(179.8118, -128.7293)(183.5993, -132.9163)
\qbezier(183.5993, -132.9163)(187.3867, -137.1033)(191.5359, -141.6202)
\qbezier(125.7955, -41.7727)(125.7955, -44.2819)(125.8932, -46.7929)
\qbezier(125.8932, -46.7929)(125.9910, -49.3039)(126.1866, -51.8207)
\qbezier(126.1866, -51.8207)(126.3822, -54.3374)(126.6761, -56.8637)
\qbezier(126.6761, -56.8637)(126.9699, -59.3900)(127.3624, -61.9296)
\qbezier(127.3624, -61.9296)(127.7549, -64.4693)(128.2466, -67.0261)
\qbezier(128.2466, -67.0261)(128.7383, -69.5830)(129.3300, -72.1609)
\qbezier(129.3300, -72.1609)(129.9217, -74.7389)(130.6143, -77.3419)
\qbezier(130.6143, -77.3419)(131.3069, -79.9448)(132.1014, -82.5767)
\qbezier(132.1014, -82.5767)(132.8959, -85.2087)(133.7935, -87.8735)
\qbezier(133.7935, -87.8735)(134.6911, -90.5384)(135.6933, -93.2402)
\qbezier(135.6933, -93.2402)(136.6954, -95.9421)(137.8036, -98.6850)
\qbezier(137.8036, -98.6850)(138.9117, -101.4280)(140.1276, -104.2162)
\qbezier(140.1276, -104.2162)(141.3434, -107.0044)(142.6688, -109.8421)
\qbezier(142.6688, -109.8421)(143.9942, -112.6797)(145.4311, -115.5712)
\qbezier(145.4311, -115.5712)(146.8681, -118.4627)(148.4188, -121.4123)
\qbezier(148.4188, -121.4123)(149.9695, -124.3619)(151.6362, -127.3742)
\qbezier(151.6362, -127.3742)(153.3030, -130.3865)(155.0884, -133.4660)
\qbezier(155.0884, -133.4660)(156.8737, -136.5455)(158.7804, -139.6969)
\qbezier(158.7804, -139.6969)(160.6872, -142.8483)(162.7181, -146.0763)
\qbezier(209.1846, -134.8729)(202.7944, -128.5213)(197.3029, -123.0747)
\qbezier(197.3029, -123.0747)(191.8114, -117.6281)(187.0903, -112.9594)
\qbezier(187.0903, -112.9594)(182.3693, -108.2907)(178.3085, -104.2909)
\qbezier(178.3085, -104.2909)(174.2477, -100.2911)(170.7524, -96.8668)
\qbezier(170.7524, -96.8668)(167.2571, -93.4425)(164.2457, -90.5139)
\qbezier(164.2457, -90.5139)(161.2342, -87.5852)(158.6364, -85.0838)
\qbezier(158.6364, -85.0838)(156.0386, -82.5823)(153.7937, -80.4498)
\qbezier(153.7937, -80.4498)(151.5489, -78.3172)(149.6046, -76.5037)
\qbezier(149.6046, -76.5037)(147.6603, -74.6902)(145.9712, -73.1535)
\qbezier(145.9712, -73.1535)(144.2821, -71.6167)(142.8087, -70.3208)
\qbezier(142.8087, -70.3208)(141.3353, -69.0250)(140.0433, -67.9398)
\qbezier(140.0433, -67.9398)(138.7513, -66.8545)(137.6105, -65.9546)
\qbezier(137.6105, -65.9546)(136.4698, -65.0547)(135.4535, -64.3190)
\qbezier(135.4535, -64.3190)(134.4373, -63.5834)(133.5220, -62.9949)
\qbezier(133.5220, -62.9949)(132.6066, -62.4064)(131.7707, -61.9512)
\qbezier(131.7707, -61.9512)(130.9349, -61.4961)(130.1590, -61.1637)
\qbezier(130.1590, -61.1637)(129.3831, -60.8314)(128.6491, -60.6140)
\qbezier(128.6491, -60.6140)(127.9151, -60.3967)(127.2058, -60.2892)
\qbezier(127.2058, -60.2892)(126.4965, -60.1818)(125.7955, -60.1818)
\qbezier(198.9798, -139.5207)(195.0893, -135.7877)(191.5604, -132.4331)
\qbezier(191.5604, -132.4331)(188.0315, -129.0786)(184.8273, -126.0674)
\qbezier(184.8273, -126.0674)(181.6231, -123.0563)(178.7101, -120.3572)
\qbezier(178.7101, -120.3572)(175.7972, -117.6581)(173.1451, -115.2429)
\qbezier(173.1451, -115.2429)(170.4930, -112.8277)(168.0742, -110.6711)
\qbezier(168.0742, -110.6711)(165.6553, -108.5146)(163.4444, -106.5941)
\qbezier(163.4444, -106.5941)(161.2335, -104.6737)(159.2075, -102.9695)
\qbezier(159.2075, -102.9695)(157.1814, -101.2652)(155.3192, -99.7592)
\qbezier(155.3192, -99.7592)(153.4569, -98.2533)(151.7389, -96.9299)
\qbezier(151.7389, -96.9299)(150.0210, -95.6066)(148.4294, -94.4521)
\qbezier(148.4294, -94.4521)(146.8378, -93.2975)(145.3561, -92.2998)
\qbezier(145.3561, -92.2998)(143.8743, -91.3020)(142.4868, -90.4506)
\qbezier(142.4868, -90.4506)(141.0994, -89.5992)(139.7917, -88.8853)
\qbezier(139.7917, -88.8853)(138.4841, -88.1714)(137.2427, -87.5875)
\qbezier(137.2427, -87.5875)(136.0013, -87.0036)(134.8131, -86.5436)
\qbezier(134.8131, -86.5436)(133.6250, -86.0836)(132.4776, -85.7428)
\qbezier(132.4776, -85.7428)(131.3303, -85.4019)(130.2119, -85.1767)
\qbezier(130.2119, -85.1767)(129.0934, -84.9514)(127.9922, -84.8393)
\qbezier(127.9922, -84.8393)(126.8910, -84.7273)(125.7955, -84.7273)
\qbezier(180.0488, -144.1577)(178.0134, -142.4735)(176.0788, -140.9126)
\qbezier(176.0788, -140.9126)(174.1442, -139.3517)(172.3028, -137.9080)
\qbezier(172.3028, -137.9080)(170.4615, -136.4643)(168.7063, -135.1322)
\qbezier(168.7063, -135.1322)(166.9511, -133.8001)(165.2753, -132.5746)
\qbezier(165.2753, -132.5746)(163.5995, -131.3490)(161.9966, -130.2252)
\qbezier(161.9966, -130.2252)(160.3937, -129.1014)(158.8576, -128.0751)
\qbezier(158.8576, -128.0751)(157.3215, -127.0488)(155.8462, -126.1159)
\qbezier(155.8462, -126.1159)(154.3709, -125.1831)(152.9507, -124.3401)
\qbezier(152.9507, -124.3401)(151.5306, -123.4971)(150.1600, -122.7408)
\qbezier(150.1600, -122.7408)(148.7895, -121.9844)(147.4633, -121.3118)
\qbezier(147.4633, -121.3118)(146.1372, -120.6392)(144.8502, -120.0477)
\qbezier(144.8502, -120.0477)(143.5633, -119.4562)(142.3106, -118.9435)
\qbezier(142.3106, -118.9435)(141.0580, -118.4308)(139.8348, -117.9950)
\qbezier(139.8348, -117.9950)(138.6116, -117.5592)(137.4131, -117.1985)
\qbezier(137.4131, -117.1985)(136.2146, -116.8379)(135.0362, -116.5510)
\qbezier(135.0362, -116.5510)(133.8578, -116.2642)(132.6950, -116.0500)
\qbezier(132.6950, -116.0500)(131.5322, -115.8358)(130.3804, -115.6935)
\qbezier(130.3804, -115.6935)(129.2287, -115.5511)(128.0835, -115.4801)
\qbezier(128.0835, -115.4801)(126.9384, -115.4091)(125.7955, -115.4091)
\qbezier(225.6429, -107.5132)(221.1261, -103.3640)(216.9390, -99.5765)
\qbezier(216.9390, -99.5765)(212.7520, -95.7891)(208.8659, -92.3371)
\qbezier(208.8659, -92.3371)(204.9797, -88.8851)(201.3674, -85.7447)
\qbezier(201.3674, -85.7447)(197.7552, -82.6043)(194.3919, -79.7538)
\qbezier(194.3919, -79.7538)(191.0287, -76.9032)(187.8911, -74.3229)
\qbezier(187.8911, -74.3229)(184.7535, -71.7425)(181.8198, -69.4144)
\qbezier(181.8198, -69.4144)(178.8862, -67.0863)(176.1363, -64.9944)
\qbezier(176.1363, -64.9944)(173.3863, -62.9025)(170.8010, -61.0323)
\qbezier(170.8010, -61.0323)(168.2157, -59.1621)(165.7771, -57.5006)
\qbezier(165.7771, -57.5006)(163.3386, -55.8392)(161.0299, -54.3750)
\qbezier(161.0299, -54.3750)(158.7213, -52.9108)(156.5265, -51.6338)
\qbezier(156.5265, -51.6338)(154.3318, -50.3568)(152.2357, -49.2581)
\qbezier(152.2357, -49.2581)(150.1397, -48.1593)(148.1279, -47.2313)
\qbezier(148.1279, -47.2313)(146.1161, -46.3033)(144.1746, -45.5395)
\qbezier(144.1746, -45.5395)(142.2331, -44.7758)(140.3484, -44.1710)
\qbezier(140.3484, -44.1710)(138.4638, -43.5663)(136.6230, -43.1164)
\qbezier(136.6230, -43.1164)(134.7822, -42.6664)(132.9725, -42.3682)
\qbezier(132.9725, -42.3682)(131.1627, -42.0699)(129.3716, -41.9213)
\qbezier(129.3716, -41.9213)(127.5804, -41.7727)(125.7955, -41.7727)
\qbezier(230.0991, -78.6953)(226.8710, -76.6644)(223.7196, -74.7577)
\qbezier(223.7196, -74.7577)(220.5682, -72.8510)(217.4887, -71.0656)
\qbezier(217.4887, -71.0656)(214.4092, -69.2803)(211.3969, -67.6135)
\qbezier(211.3969, -67.6135)(208.3847, -65.9467)(205.4350, -64.3960)
\qbezier(205.4350, -64.3960)(202.4854, -62.8454)(199.5939, -61.4084)
\qbezier(199.5939, -61.4084)(196.7025, -59.9715)(193.8648, -58.6461)
\qbezier(193.8648, -58.6461)(191.0271, -57.3207)(188.2389, -56.1048)
\qbezier(188.2389, -56.1048)(185.4507, -54.8890)(182.7078, -53.7808)
\qbezier(182.7078, -53.7808)(179.9648, -52.6727)(177.2630, -51.6705)
\qbezier(177.2630, -51.6705)(174.5611, -50.6684)(171.8962, -49.7708)
\qbezier(171.8962, -49.7708)(169.2314, -48.8731)(166.5995, -48.0786)
\qbezier(166.5995, -48.0786)(163.9676, -47.2841)(161.3646, -46.5916)
\qbezier(161.3646, -46.5916)(158.7616, -45.8990)(156.1837, -45.3073)
\qbezier(156.1837, -45.3073)(153.6057, -44.7156)(151.0489, -44.2239)
\qbezier(151.0489, -44.2239)(148.4920, -43.7322)(145.9524, -43.3397)
\qbezier(145.9524, -43.3397)(143.4127, -42.9472)(140.8864, -42.6534)
\qbezier(140.8864, -42.6534)(138.3602, -42.3595)(135.8434, -42.1639)
\qbezier(135.8434, -42.1639)(133.3267, -41.9682)(130.8156, -41.8705)
\qbezier(130.8156, -41.8705)(128.3046, -41.7727)(125.7955, -41.7727)
\qbezier(144.2045, -41.7727)(144.2045, -42.4738)(144.3120, -43.1831)
\qbezier(144.3120, -43.1831)(144.4194, -43.8924)(144.6368, -44.6264)
\qbezier(144.6368, -44.6264)(144.8541, -45.3604)(145.1865, -46.1363)
\qbezier(145.1865, -46.1363)(145.5188, -46.9122)(145.9740, -47.7480)
\qbezier(145.9740, -47.7480)(146.4291, -48.5839)(147.0176, -49.4992)
\qbezier(147.0176, -49.4992)(147.6061, -50.4146)(148.3417, -51.4308)
\qbezier(148.3417, -51.4308)(149.0774, -52.4470)(149.9773, -53.5878)
\qbezier(149.9773, -53.5878)(150.8772, -54.7286)(151.9625, -56.0206)
\qbezier(151.9625, -56.0206)(153.0477, -57.3126)(154.3436, -58.7860)
\qbezier(154.3436, -58.7860)(155.6394, -60.2593)(157.1762, -61.9484)
\qbezier(157.1762, -61.9484)(158.7129, -63.6376)(160.5264, -65.5819)
\qbezier(160.5264, -65.5819)(162.3399, -67.5261)(164.4725, -69.7710)
\qbezier(164.4725, -69.7710)(166.6051, -72.0159)(169.1065, -74.6137)
\qbezier(169.1065, -74.6137)(171.6079, -77.2115)(174.5366, -80.2229)
\qbezier(174.5366, -80.2229)(177.4653, -83.2344)(180.8895, -86.7297)
\qbezier(180.8895, -86.7297)(184.3138, -90.2250)(188.3136, -94.2858)
\qbezier(188.3136, -94.2858)(192.3134, -98.3466)(196.9822, -103.0676)
\qbezier(196.9822, -103.0676)(201.6509, -107.7887)(207.0974, -113.2802)
\qbezier(207.0974, -113.2802)(212.5440, -118.7717)(218.8956, -125.1619)
\qbezier(168.7500, -41.7727)(168.7500, -42.8682)(168.8621, -43.9695)
\qbezier(168.8621, -43.9695)(168.9741, -45.0707)(169.1994, -46.1892)
\qbezier(169.1994, -46.1892)(169.4247, -47.3076)(169.7655, -48.4549)
\qbezier(169.7655, -48.4549)(170.1064, -49.6022)(170.5663, -50.7904)
\qbezier(170.5663, -50.7904)(171.0263, -51.9786)(171.6102, -53.2200)
\qbezier(171.6102, -53.2200)(172.1941, -54.4614)(172.9080, -55.7690)
\qbezier(172.9080, -55.7690)(173.6220, -57.0766)(174.4734, -58.4641)
\qbezier(174.4734, -58.4641)(175.3248, -59.8515)(176.3225, -61.3333)
\qbezier(176.3225, -61.3333)(177.3203, -62.8151)(178.4748, -64.4067)
\qbezier(178.4748, -64.4067)(179.6293, -65.9982)(180.9526, -67.7162)
\qbezier(180.9526, -67.7162)(182.2760, -69.4342)(183.7819, -71.2964)
\qbezier(183.7819, -71.2964)(185.2879, -73.1587)(186.9922, -75.1847)
\qbezier(186.9922, -75.1847)(188.6965, -77.2107)(190.6169, -79.4217)
\qbezier(190.6169, -79.4217)(192.5373, -81.6326)(194.6939, -84.0514)
\qbezier(194.6939, -84.0514)(196.8504, -86.4703)(199.2656, -89.1224)
\qbezier(199.2656, -89.1224)(201.6809, -91.7745)(204.3800, -94.6874)
\qbezier(204.3800, -94.6874)(207.0790, -97.6003)(210.0902, -100.8046)
\qbezier(210.0902, -100.8046)(213.1013, -104.0088)(216.4558, -107.5377)
\qbezier(216.4558, -107.5377)(219.8104, -111.0666)(223.5434, -114.9570)
\qbezier(199.4318, -41.7727)(199.4318, -42.9157)(199.5028, -44.0608)
\qbezier(199.5028, -44.0608)(199.5739, -45.2059)(199.7162, -46.3577)
\qbezier(199.7162, -46.3577)(199.8585, -47.5095)(200.0727, -48.6723)
\qbezier(200.0727, -48.6723)(200.2869, -49.8351)(200.5738, -51.0135)
\qbezier(200.5738, -51.0135)(200.8606, -52.1919)(201.2213, -53.3904)
\qbezier(201.2213, -53.3904)(201.5819, -54.5889)(202.0177, -55.8121)
\qbezier(202.0177, -55.8121)(202.4535, -57.0353)(202.9662, -58.2879)
\qbezier(202.9662, -58.2879)(203.4789, -59.5406)(204.0704, -60.8275)
\qbezier(204.0704, -60.8275)(204.6619, -62.1144)(205.3345, -63.4406)
\qbezier(205.3345, -63.4406)(206.0072, -64.7668)(206.7635, -66.1373)
\qbezier(206.7635, -66.1373)(207.5198, -67.5078)(208.3628, -68.9280)
\qbezier(208.3628, -68.9280)(209.2058, -70.3482)(210.1386, -71.8235)
\qbezier(210.1386, -71.8235)(211.0715, -73.2988)(212.0978, -74.8349)
\qbezier(212.0978, -74.8349)(213.1242, -76.3710)(214.2480, -77.9739)
\qbezier(214.2480, -77.9739)(215.3717, -79.5768)(216.5973, -81.2526)
\qbezier(216.5973, -81.2526)(217.8228, -82.9283)(219.1549, -84.6835)
\qbezier(219.1549, -84.6835)(220.4870, -86.4387)(221.9307, -88.2801)
\qbezier(221.9307, -88.2801)(223.3744, -90.1214)(224.9353, -92.0561)
\qbezier(224.9353, -92.0561)(226.4963, -93.9907)(228.1804, -96.0261)
\put(125.7955, -41.7727){\line(1, -1){ 95.1136}}
\put( 88.9773, -41.7727){\vector(1, 0){144.2045}}
\put(125.7955, -149.1591){\vector(0, 1){144.2045}}
\put(242.3864, -46.3750){\makebox(0,0)[]{\footnotesize{$\chi$}}}
\put(116.5909,  -3.4205){\makebox(0,0)[]{\footnotesize{$\eta$}}}
\put(165.6818, -155.2955){\makebox(0,0)[]{\footnotesize{$R = \mbox{const}$}}}
\put(227.0455, -146.0909){\makebox(0,0)[]{\footnotesize{$T = \mbox{const}$}}}
\put(259.2614, -78.5909){\makebox(0,0)[]{\footnotesize{$T = \mbox{const}$}}}
\put(248.5227, -126.1477){\makebox(0,0)[]{\footnotesize{$R = \mbox{const}$}}}
\end{picture}
\vspace{0mm} \newline
{\footnotesize \sf Figure 12. Minkowski diagram with reference to
the $(\eta,\chi)$-system for a flat universe model dominated by
vacuum energy. The diagram shows world lines $R = \mbox{constant}$
and simultaneity curves $T = \mbox{constant} \,$. In this universe
model the conformal system contracts relative to the Hubble flow.}
\vspace{0mm} \newline
\itm The evolution of the flat LIVE dominated universe in the CFS-system
is as follows. The universe starts at $T = 0$ and expands from an initial
state with vanishing extension. In this era the conformal space has finite
extension, although it expands without limit. Again, as the conformal time
$T$ approaches infinity, the clocks are reset to come from minus infinity,
and the sign of the radial coordinate is changed. The conformal space then
has infinite extension and is contracting. The contraction slows down
to a final state at rest at $T = 0$, in which the conformal space still
has infinite extension.
%
%
\vspace{5mm} \newline
{\bf 9.2. Particle horizon in flat universe models using conformal time}
\vspace{3mm} \newline
We shall here discuss the particle horizon in flat universe models from
the perspective of the conformal coordinates $T$ and $R$. Let us first
consider a universe model dominated by vacuum energy. It extends backwards
in time to $t \rightarrow -\infty$. In such a universe model there is no
particle horizon. However, this universe model is not realistic. We
cannot expect that the general theory of relativity can give a realistic
description of spacetime before the Planck time. Hence we assume that the
inflationary era, which may be described classically, starts at the Planck
time ${\eta}_{Pl}$. Then there is a particle horizon around an observer at
$P$ with coordinates $T = T_0 > 0$ and $R = 0$. As in section 4.3 the
horizon $H$ is defined as the intersection between the past light cone at
$P$, $R = T_0 - T$, and the space at $\eta = {\eta}_{Pl}$. In the present
case this space is represented by the hyperbola $R^2 = T^2 + 2 b_{Pl} T$
where $b_{Pl} = (2 {\eta}_{Pl})^{-1}$. This gives
\begin{equation} \label{e_23}
T_H = \frac{T_0^2}{2 (T_0 + b_{Pl})}
\mbox{\hspace{2mm}, \hspace{2mm}}
R_H = \frac{(T_0 + 2 b_{Pl}) T_0}{2 (T_0 + b_{Pl})}
\mbox{ .}
\end{equation}
Inserting these expressions into the transformation \eqref{e_82} gives
\begin{equation} \label{e_24}
{\chi}_H = - \frl{1}{T_0} - \frl{1}{2 b_{Pl}} = - \frl{1}{T_0} - {\eta}_{Pl}
\mbox{ .}
\end{equation}
Since $P$ has coordinates $T = T_0$ and $R = 0$, the transformation
\eqref{e_82} also gives
\begin{equation} \label{e_25}
{\eta}_P = - \frl{1}{T_0}
\mbox{ .}
\end{equation}
Hence the equation of the particle horizon for a flat universe model
starting at the Planck time is
\begin{equation} \label{e_26}
{\chi}_H = {\eta}_P - {\eta}_{Pl}
\mbox{ .}
\end{equation}
This has here been deduced by using the spacetime diagram in
Figure 11 with respect to the conformal coordinates $T$ and $R$.
Equation \eqref{e_26} is in accordance with equation (10) in
reference [1] with $t_i = t_{Pl}$.
%
%
\vspace{5mm} \newline
{\bf 10. The CFS Hubble parameters of some universe models}
\vspace{3mm} \newline
The behaviour of the conformal space in different universe models
may be investigated by calculating the CFS Hubble parameter defined in
equation (48) in reference [1]. We first consider negatively curved
universe models with respectively dust and radiation as described in
CFS systems of type I. Using equation \eqref{e_76} we obtain
\begin{equation} \label{e_253}
H_R = \frac{2 T_i T}{(\sqrt{T^2 - R^2} - T_i)^3}
\end{equation}
for a dust dominated universe, and by means of equation \eqref{e_77}
we get
\begin{equation} \label{e_254}
H_R = \frac{2 T_i^2 T}{(T^2 - R^2 - T_i^2)^2}
\end{equation}
for a radiation dominated universe. As shown in section 4.4 there is
continual creation at the boundary of the conformal space given by
$T^2 - R^2 = T_i^2$. Hence the CFS Hubble parameter for both of these
universe models approaches infinity at the boundary with continual
creation.
\itm The CFS Hubble parameter for a negatively curved radiation dominated
universe with conformal coordinates of type II and III is
\begin{equation} \label{e_293}
H_R = \mbox{\small sgn} (T) \hs{0.5mm}
\frl{(1 + \widehat{T}^2 - \widehat{R}^2)^2 - 4 \widehat{T}^2
(\widehat{T}^2 - \widehat{R}^2)}{4 \beta \widehat{T}^2}
\mbox{ .}
\end{equation}
Correspondingly for a flat universe one obtains
\begin{equation} \label{e_294}
H_R = \frl{|T^2 - R^2| \hs{0.5mm} (3 T^2 + R^2)}{\beta T^2}
\mbox{ .}
\end{equation}
The Hubble parameter for a LIVE dominated universe model
with $k = -1$ and CFS coordinates of type I is
\begin{equation} \label{e_255}
H_R = \sqrt{\Omega_0} \hs{0.5mm} \widehat{H}_0 \hs{0.5mm} \frl{T}{T_f}
\mbox{ .}
\end{equation}
In this case the conformal Hubble parameter in independent of the
position $R$.
The Hubble parameter for the same universe model but
with CFS coordinates of type II and III is
\begin{equation} \label{e_291}
H_R = - \widehat{H}_{\Lambda} \hs{0.9mm} \mbox{\small sgn} (T)
\end{equation}
where we have omitted the hat and the tilde on $T$.
This formula is also valid for a flat LIVE dominated universe.
The conformal space therefore expands for $T < 0$ and contracts
for $T > 0$. This behaviour is a result of two competing motions. The
Hubble flow expands exponentially. But as seen in Figure 12, the conformal
reference frame contracts relative to the Hubble flow. The sign of the
conformal Hubble parameter shows that the expansion dominates for
$T < 0$ corresponding to the region $- \chi < \eta < 0$ in the
$(\eta,\chi)$-plane, while the contaction of the conformal system
dominates for $T > 0$ corresponding to the region $0 < \chi < - \eta$.
%
%
\vspace{5mm} \newline
{\bf 11. Conclusion}
\vspace{3mm} \newline
In the $(\eta,\chi)$-system, time and space are created at the
Big Bang singularity at $t = 0$. In the $(T,R)$-system, new space is
also created at later times. Big Bang happens continually at the
boundary of space, i.e. on the hyperbola $T^2 - R^2 = T_i^2$
in Figure 1. Just as our universe is said not to exist before
$t = \eta = 0$, our universe did not exist below this hyperbola.
Or, alternatively if cosmic time and space was created at the
Planck time, conformal space is continually created at the Planck
boundary.
\itm What then about continual creation of new space? Is this only a
coordinate effect? In a way it is. As made clear in the introduction
conformal space is a coordinate space. But cosmic space is also a
coordinate space. The first one is defined by $T = \mbox{constant}$
and the second one by $t = \mbox{constant}$. Is the one more fundamental
than the other?
\itm A difference between the cosmic and the conformal space is that
the reference particles of the first one are freely moving, and those
of the second ones must be acted upon by non-gravitational forces. The
reference particles of the cosmic space constitute an inertial flow,
while those of the conformal space do not. In this sense the cosmic
space is more fundamental than the conformal space.
\itm Continual creation of conformal space, matter and energy is physical
and real. But it belongs to a CFS picture of the universe which does not
have the same physical significance as the picture of our universe based
upon cosmic time and freely moving reference particles, because these
particles which define the Hubble flow, are in fact the clusters of
galaxes.
\itm In the present paper we have described FRW universe models with
negative and vanishing spatial curvature using different types of
conformally flat spacetime coordinates. They are comoving in reference
frames that move relative to each other. Due to the relativity of
simultaneity they therefore define different spatial sections of
spacetime. Hence the picture that they give of the creation and the
evolution of the universe are different, and also rather strange when
we are used to the standard picture of Big Bang happening everywhere at
a certain moment, $t = 0$, provided by using cosmic coordinate time and
spatial coordinates comoving with free particles.
\itm In the picture associated with the first type of CFS coordinates
the universe first appears at the spatial origin of the coordinate system
and then expands with superluminal velocity. The universe is inhomogeneous,
and the energy density and temperature increase towards infinity at the
expanding front which defines the boundary of the universe. New space,
matter and energy are created at this front. Inside the front space
expands with a constant, subluminal velocity. This is not a spaces defined
by free particles, but by the reference particles of the CFS coordinate
system.
\itm The motion of the CFS reference particles relative to those of the
standard cosmic coordinate system, defining the inertial Hubble flow,
has been depicted for all three types of CFS coordinates.
\itm The picture of the universe with reference to CFS coordinates of the
second type depends upon the contents of the universe. The matter and
radiation dominated universe contracts and has finite age and spatial
extension, while the corresponding universe dominated by vacuum energy
expands.
\itm The CFS picture of type III of the evolution of a universe with
radiation and dust is rather pathological.
Looking at Figure 7, one might expect that the universe starts at an
infinitely past conformal time. This is, however, no so. A dust and
radiation dominated universe starts at a cosmic time $t = 0$. From
equation \eqref{e_73} it follows that this corresponds to a parametric
time $\eta = 0$. Equation \eqref{e_133} therefore implies that the
universe starts at a conformal time $\widetilde{T} = 0$. Figure 7 shows
that the universe then starts with a hole in a finite region around
$\widetilde{R} = 0$ where the conformal space does not exist, surrounded
by a region with radiation and dust of infinitely great extension. The
conformal space then expands and so does the hole. As $\widetilde{T}$
approaches infinity, the clocks are reset to come from minus infinity,
and $\widetilde{R}$ changes from negative to positive values. A negative
sign of $\widetilde{R}$ can be removed by changing the coordinates
$\theta$ and $\phi$ corresponding to a reflection through the origin
in space.
This means that the expansion is replaced by a contraction. The conformal
space then contracts to a vanishing extension at $\widetilde{T} = -1$.
\itm The corresponding picture for a LIVE dominated universe with negative
spatial curvature is as follows. Equation \eqref{e_79} implies that
such a universe start at a cosmic time $t = 0$. From equation \eqref{e_80}
it follows that this corresponds to a parametric time
$\eta \rightarrow -\infty$. Equation \eqref{e_133} therefore implies that
this universe starts at a conformal time $\widetilde{T} = 1$. Figure 8
now shows that the conformal space then appears with vanishing extension
and expands towards infinite extension as the conformal time approaches
infinity. At this point the clocks are reset to come from minus infinity,
and $\widetilde{R}$ changes from positive to negative values, so that the
expansion is replaced by a contraction. The conformal space then contracts
to a final state at $\widetilde{T} = 0$ similar to the initial state
with radiation and dust.
\itm The evolution of the conformally flat space of a universe with
critical density dominated by dust and radiation, corresponding to the
evolution using cosmic time, is rather peculiar. The Big Bang is replaced
by the appearance of a hole in space which expands with a velocity of
light with center at the observer, even if the space as defined by
constant cosmic time is homogeneous, and the observer has an arbitrary
position. Then, at an infinitely far future, the CFS clocks are reset
to minus infinity, and the universe contracts relative to the conformal
frame.
\itm To the CFS observer the resetting of the clocks may seem rather
artificial. It would therefore be tempting to reinterpret Figure 9
saying that the universe started by contracting from an infinitely
remote past. Such a reinterpretation is however not permitted. We
know from the usual description of the universe in terms of cosmic time
that objects in the universe grow older with increasing cosmic time.
Hence obsevations would show that at $T < 0$ an object in the universe
would be older than it was at $T > 0$.
\itm By applying the rule in Appendix B in reference [1] for composing
generating functions one may obtain a deeper understanding of the
conformal coordinate transformations and find new ones.
Note the similarity between the transformations to CFS coordinates
of type II and III in a universe models with negative spatial
curvature given by the equations \eqref{e_17} and \eqref{e_133}.
One may wonder whether this points to a relationship between these
trasnformations. The answer is that the generating function in
\eqref{e_205} may be obtained as a composition of the generating
function \eqref{e_204} for the CFS coordinates of type II in a universe
with negative spatial curvature and the generating function \eqref{e_206}
for the CFS coordinates of type II in a flat universe.
\itm It remains to describe universe models with positive spatial
curvature in terms of CFS coordinates. This will be done in the
third and final paper in this series, where we will also consider
Penrose diagrams and compositions of transformations between CFS
coordinates for universe models with different spatial curvature.
\vspace{5mm} \newline
{\bf References}
%
\begin{enumerate}
\item \O .Gr\o n and S.Johannesen, \textit{FRW Universe Models in
Conformally Flat Spacetime Coordinates. I: General Formalism},
Eur.Phys.J.Plus \textbf{126}, 28 (2011).

\item L.Infield and A.Schild, \textit{A New Approach to Kinematic
Cosmology}, Phys.Rev. \textbf{68}, 250 - 272, (1945).

\item G.E.Tauber, \textit{Expanding Universe in Conformally Flat
Coordinates}, J.Math.Phys. \textbf{8}, 118 - 123 (1967).

\item G.Endean, \textit{Cosmology in Conformally Flat Spacetime}, The
Astrophysical Journal \textbf{479}, 40 - 45 (1997).

\item  M.Ibison, \textit{On the conformal forms of the
Robertson-Walker metric}, J.Math.Phys. \textbf{48},
122501-1 -- 122501-23 (2007).

\item M.Iihoshi, S.V.Ketov and A.Morishita, \textit{Conformally Flat
FRW Metrics}, Prog.Theor. Phys. \textbf{118},
475 - 489 (2007).

\item K.Shankar and B.F.Whiting, \textit{Conformal coordinates for a
constant density star}, arXiv:0706.4324.

\item J.Garecki, \textit{On Energy of the Friedmann Universes in
Conformally Flat Coordinates}, arXiv:0708.2783.

\item G.U.Varieschi, \textit{A Kinematical Approach to Conformal Cosmology},
Gen.Rel.Grav. \textbf{42}, 929 - 974 (2010).

\item M.J.Chodorowski, \textit{A direct consequence of the
expansion of space?} Astro-ph: 0610590.

\item G.F.Lewis, M.J.Francis, L.A.Barnes and J.B.James, {\it Coordinate
Confusion in Conformal Cosmology}, Mon.Not.R.Astron.Soc. {\bf 381},
L50 - L54 (2007).

\item V.F.Mukhanov, \textit{Physical foundations of cosmology},
Cambridge University Press, (2005).

\item \O .Gr\o n and S.Hervik, Einstein's General Theory of Relativity,
Springer, (2007), ch.11.

\item E.Eriksen and \O .Gr\o n, {\it The De Sitter Universe Models},
Int.J.Mod.Phys. {\bf D4}, 115 - 159 (1995).

\end{enumerate}

\end{document}